\documentclass[reprint,amsmath,amssymb,aps,prd,floatfix,showkeys,nofootinbib]{revtex4-2}

\usepackage{graphicx}
\usepackage{dcolumn}
\usepackage{bm}
\usepackage{mathtools}
\usepackage{taghere}
\usepackage[squaren,Gray]{SIunits}
\usepackage{hyperref}
\usepackage{pifont}

\renewcommand{\vec}[1]{\bm{#1}}
\newcommand{\bra}[1]{\langle #1 |}
\newcommand{\ket}[1]{| #1 \rangle}
\DeclareMathOperator{\Tr}{Tr}
\newcommand{\cprime}{{\prime\!}}
\newcommand{\overbar}[1]{\mkern 1.5mu\overline{\mkern-1.5mu#1\mkern-1.5mu}\mkern 1.5mu}
\newcommand{\MSbar}{\overbar{\text{MS}}}
\newcommand{\RI}{\text{RI}^\prime\mkern-3mu\raisebox{1pt}{$/$}\text{SMOM}}
\newcommand{\greencheck}{{\color{green}\ding{51}}}
\newcommand{\redxcheck}{{\color{red}\ding{55}}}

\begin{document}

\title{Mellin moments of spin dependent and independent PDFs of the pion and rho meson}%

\author{Marius~L{\"o}ffler}
  \email{marius.loeffler@ur.de}
\author{Philipp~Wein}%
\author{Thomas~Wurm}%
\author{Simon~Weish{\"a}upl}%
\author{Daniel~Jenkins}%
\author{Rudolf~R{\"o}dl}%
\author{Andreas~Sch{\"a}fer}%
\author{Lisa~Walter}%
\affiliation{Institut f{\"u}r Theoretische Physik, Universit{\"a}t Regensburg, 93040 Regensburg, Germany}%

\collaboration{RQCD Collaboration}\noaffiliation

\date{\today}

\begin{abstract}
We compute the second moments of pion and rho parton distribution functions (PDFs) in lattice QCD with $N_f = 2+1$ flavors of improved Wilson fermions. We determine both singlet and non-singlet flavor combinations and, for the first time, take disconnected contributions fully into account. In the case of the rho, we also calculate the additional contribution arising from the $b_1$ structure function. The numerical analysis includes 26 ensembles, mainly generated by the CLS effort, with pion masses ranging from \unit{420}{\mega\electronvolt} down to \unit{214}{\mega\electronvolt} and with 5 different lattice spacings in the range of \unit{0.1}{\femto\meter} to \unit{0.05}{\femto\meter}. This enables us to take the continuum limit, as well as to resolve the quark mass dependencies reliably. Additionally we discuss the contaminations of rho correlation functions by two-pion states.
%
%
\end{abstract}
\maketitle
%
%
%
\section{\label{sec_Introduction}Introduction}%
The pion is routinely investigated on the lattice, although, to the best of our knowledge, disconnected quark contributions, e.g., to quark momentum fractions, were not included so far. Since the pion is the pseudo-Goldstone boson of dynamical chiral symmetry breaking (DCSB) its quark structure could differ substantially from that of other mesons and if so, the flavor singlet sea quark contribution is a natural place for such a difference to show up. In contrast to the pion, the quark structure of the $\rho$ is only rarely analyzed on the lattice, primarily due to the complications caused by its resonance nature.\par
For the pion experimental data exists primarily from two classes of experiments, namely Drell-Yan reactions with (secondary) pion beams, e.g., $\pi+N\to \mu^++\mu^-+X$, which are sensitive to the pion PDF at large~$x\gtrsim 0.15$, and semi-inclusive (tagged) deep inelastic scattering (DIS), e.g., $e+N \to e^\prime+N+X$, which is sensitive to small $x$ and exploits the fact that the electron can scatter off the nucleon pion cloud via the Sullivan process~\cite{Sullivan:1971kd}.
Experiments of the first type were performed by NA10~\cite{Betev:1985pf}, E326~\cite{Greenlee:1985gd}, E615~\cite{Conway:1989fs}, and, more recently, by COMPASS~\cite{Aghasyan:2017jop}. This will be continued by \href{https://nqf-m2.web.cern.ch/}{AMBER} at CERN~\cite{Denisov:2018unj,Amber:2021}.
Experiments of the second type were performed at HERA~\cite{Chekanov:2002pf,Aaron:2010ab} (see also~\cite{Holtmann:1994rs,McKenney:2015xis,Barry:2018ort}) and are currently pursued at JLab~Hall~A~\cite{Montgomery:2017hab} (cf.\ ~the conditionally approved proposal~\cite{Keppel:2015}). They are also under consideration for the physics program at the EIC~\cite{Aguilar:2019teb}.\par%
In contrast to the pion case, there exists very little relevant experimental data for the $\rho$ quark PDFs. The rho meson is the lightest strongly decaying particle with a branching fraction of $>99.9\%$ into 2 pions~\cite{Zyla:2020zbs}. It is spin-1 which implies the existence of novel polarization dependent structure functions~\cite{Hoodbhoy:1988am}. The unstable nature of the $\rho$ complicates the analysis of its structure, both, on the lattice -- we will discuss some of the implications in this article -- and in experiment. However, as the goal of hadron physics must be to also determine the quark-gluon structure of resonances rather than only ground states, the $\rho$ is one of the most attractive light mesons to explore.
To the best of our knowledge, no existing or planned experiment will investigate the spin structure of the $\rho$, and so lattice calculations may offer the best, if not only, chance to determine it. In ref.~\cite{Best:1997qp} it was speculated whether one could analyse the spin structure of the $\rho$ in the meson cloud of a nucleon in a (polarized) Sullivan process (see also ref.~\cite{Holtmann:1994rs}), but the interpretation of such measurements would be very non-trivial in view of the required analytic continuation from the $t$ to the $s$ channel~\cite{Ji:private}.
However, the $b_1$ structure function of the deuteron was measured by HERMES~\cite{Airapetian:2005cb} (using DIS on tensor-polarized deuteron gas with negligible vector polarization) and turned out to be surprisingly large for such a loosly bound system. Also, while the data for a limited $x$ range cannot really test the Close--Kumano sum rule~for the first moment of $b_1(x) $\cite{Close:1990zw}, an unexpected behavior outside of the measured $x$ range is needed to fulfill it. Overall the results differ from the expectation that the deuteron is in an $S$ wave with only a small $D$ wave admixture (cf., e.g., ref.~\cite{Kumano:2016cqs}). Also, there are efforts to measure the deuteron $b_1$ via the proton-deuteron Drell-Yan process (conditionally approved proposal at JLab~Hall~C~\cite{Alleda:2013} and feasibility studies for Fermilab~\cite{Kumano:2016cqs,Song:2019awx}), as well as discussions of a measurement via DIS at the EIC~\cite{Boer:2019pzo}.\par%
The study of mesonic structure using lattice QCD has, by now, a history of over three decades. Traditionally, such calculations focused on moments on PDFs and distribution amplitudes (DAs). While earlier simulations~\cite{Martinelli:1987zd,Martinelli:1987bh,Best:1997qp,Best:1997ab,Best:1997rm,Guagnelli:2004ga,Capitani:2005jp} used quenched fermion representations, more recent simulations~\cite{Brommel:2006zz,Bali:2013gya,Abdel-Rehim:2015owa,Oehm:2018jvm,Alexandrou:2020gxs,Alexandrou:2021mmi} use, for example, (clover-improved) dynamical Wilson fermions, where the fermion determinant, and thus the quark sea, is taken into account. However, what all these studies have in common is that they neglect disconnected contributions, because the latter are notoriously difficult to calculate and usually come with a large statistical error. During our analysis we found that the noise on the light and strange quark disconnected loops is highly correlated. We can use this to our advantage by looking at the non-singlet ($\bar{u} u + \bar{d} d - 2 \bar{s} s$) and singlet ($\bar{u} u + \bar{d} d + \bar{s} s$) flavor combinations instead of the light and strange loops themselves. While the large statistical errors persist for the singlet flavor combination, they are reduced by over an order of magnitude for the non-singlet operator, which allows us to obtain quite precise results in this case even though we take the disconnected contributions fully into account.\par%
The reach of such calculations of moments of PDFs and DAs is limited, primarily because for higher moments the problems caused by operator mixing become untraceable. Therefore, in recent years ever more attention has focused on coordinate space methods~\cite{Braun:2007wv,Ji:2013dva,Ma:2014jla,Radyushkin:2017cyf,Ji:2020ect} which allow to calculate the full functional forms of DAs and PDFs. As far as we know, no lattice results for $b_1(x,Q^2)$ of the $\rho$ have been published so far using these methods, but some work exists for DAs of vector mesons~\cite{Hua:2020gnw}, for which there also exist results for some Mellin moments~\cite{Braun:2016wnx}, and other mesons~\cite{Cichy:2021lih}. DAs and PDFs probe independent aspects of the meson quark structure, and thus provide valuable complementary information.\par%
In this article we directly calculate the second moments of the pion and rho PDFs by evaluating operators that contain a covariant derivative. The same method would not be directly applicable to higher moments, since one would face the problem of mixing with lower-dimensional operators. Sparked by the presentation in ref.~\cite{Ji:2013dva}, position space methods have recently fueled a lot of excitement, since they, in principle, allow for a resolution of the complete PDF. There are recent studies on the pion PDF exploring possible methods, such as the current-current method~\cite{Sufian:2019bol,Sufian:2020vzb}, large momentum effective theory~\cite{Chen:2018fwa,Izubuchi:2019lyk} (using quasi PDFs), or Ioffe time distributions~\cite{Joo:2019bzr} (using pseudo PDFs).
Similar to experiment and in contrast to the pion, the rho meson structure has only been studied once to our knowledge, in the work presented in ref.~\cite{Best:1997qp} (also discussed in refs.~\cite{Best:1997ab,Best:1997rm}) based on a quenched simulation at large quark masses. The use of large quark masses is probably due to the additional difficulty raised by the instability of the rho meson at physical quark masses. Actually, due to the finite volume, there is no continuum of two-pion final states in a lattice simulation, and therefore the rho meson cannot decay dynamically into two pions. Nevertheless, the discretized set of two-pion (or, at higher energies, even multi-pion) finite volume states is present and might overshadow the contribution from the rho in the correlation function. We discuss this issue in some detail in Sec.~\ref{sec_TwoPionStates}.\par%
This article is structured as follows: We set the stage with a general discussion of PDF properties and their connection to DIS structure functions in Sec.~\ref{sec_GeneralProperties}. Next, in Sec.~\ref{sec_LatticeComputation}, we will present the details of the lattice calculation, including simulation parameters, the analysis of correlation functions, and possible two-pion contributions. We describe the extrapolation strategy and our final results in Sec.~\ref{sec_Results}, and summarize in Sec.~\ref{sec_Summary}.
\section{\label{sec_GeneralProperties}General properties of PDFs}
The cross-section of deep inelastic scattering can be written as a product of a leptonic and a hadronic part. The hadronic tensor is given by%
\begin{align}%
    W^{\mu \nu}(\vec p, \lambda) &= \int \frac{ \mathrm{d}^4z}{4 \pi} \,\, e^{iq \cdot z} \bra{ \vec p, \lambda} \left[j^{\mu}(z), j^{\nu}(0) \right] \ket{\vec p, \lambda} \,, \label{eq_scattering-amplitude}
\end{align}%
where $\vec p$ is the three-momentum and $\lambda$ labels the spin of the target hadron along a quantization direction~\cite{Hoodbhoy:1988am,Manohar:1992tz}. Using parity and time-reversal invariance it is straight forward to show that the most general hadronic tensor for polarized DIS from targets with spin-1 or less can be decomposed into eight structure functions%
\begin{align}%
    \begin{split}
        W^{\mu \nu} = &-F_1 g_{\mu \nu} + F_2 \frac{p_{\mu} p_{\nu}}{p \cdot q} + i \frac{g_1}{p \cdot q} \epsilon_{\mu \nu \lambda \sigma} q^{\lambda} s^{\sigma} \\
        &+ i \frac{g_2}{(p \cdot q)^2} \epsilon_{\mu \nu \lambda \sigma} q^{\lambda} \left( p \cdot q s^{\sigma} - s \cdot q p^{\sigma} \right)  \\
        &- b_1 r_{\mu \nu} + \frac{1}{6} b_2 \left(s_{\mu \nu} + t_{\mu \nu} + u_{\mu \nu} \right) \\
        &+ \frac{1}{2} b_3 \left(s_{\mu \nu} - u_{\mu \nu} \right) +\frac{1}{2} b_4 \left( s_{\mu \nu} - t_{\mu \nu} \right)  \,, \taghere
        \label{eq_scattering-amplitude-decomposition}
    \end{split}
\end{align}
with the kinematic factors $r_{\mu \nu}$, $s_{\mu \nu}$, $t_{\mu \nu}$, and $u_{\mu \nu}$, which depend on the momentum transfer $q$, the target momentum $p$ and the target polarization vector $\epsilon$ (cf.\ App.~\ref{app_coordinates_and_polarization_vectors} for our conventions). The quantities $r_{\mu \nu}$, $s_{\mu \nu}$, $t_{\mu \nu}$, and $u_{\mu \nu}$ are constructed such that they vanish upon averaging over the target spin, see ref.~\cite{Hoodbhoy:1988am}. For spin-$\frac{1}{2}$ targets $s^{\mu}$ corresponds to the spin four-vector. For spin-$1$ targets it corresponds to $s^\mu=-i\epsilon^{\mu\nu\rho\sigma} e^*_\nu e_\rho p_\sigma$. Note that, due to current conservation any term proportional to $q_{\mu}$ or $q_{\nu}$ in Eq.~\eqref{eq_scattering-amplitude-decomposition} would vanish. Which of the structure functions can contribute depends on the target spin: In case of spin-0 only $F_1$ and $F_2$ do. For spin-$\frac{1}{2}$ targets one has $F_1$, $F_2$, $g_1$, and $g_2$, where the measurement of $g_1$ and $g_2$ requires a polarized beam. In case of spin-1 targets the full set of eight structure functions can contribute. Notably, as argued in ref.~\cite{Hoodbhoy:1988am}, the additional structure functions $b_{1-4}$ can be measured using an unpolarized electron beam.\par%
The hadronic tensor can be factorized into a hard scattering kernel, which can be calculated perturbatively, and in PDFs containing the nonperturbative information. The PDFs related to the structure functions in Eq.~\eqref{eq_scattering-amplitude-decomposition} are defined as\footnote{In general one finds three quark and three gluon PDFs analogous to Eqs.~\eqref{eq_def_qPDF} and~\eqref{eq_def_qtildePDF}, see, e.g.,~\cite{Ji:1998pc}.}%
\begin{align}
    q^\lambda_H &= \int \limits_{-\infty}^\infty \frac{dz^-}{4 \pi} e^{-i x p^+ z^-} \bra{\vec p, \lambda} \bar{q}(z) \gamma^+ q(0) \ket{\vec p, \lambda} \,,
    \label{eq_def_qPDF} \\
    \Delta q^\lambda_H &=  \int \limits_{-\infty}^\infty \frac{dz^-}{4 \pi} e^{-i x p^+ z^-} \bra{\vec p, \lambda} \bar{q}(z) \gamma^+ \gamma_5 q(0)  \ket{\vec p, \lambda} \,,    \label{eq_def_qtildePDF}
\end{align}
where $z$ is a lightlike vector with vanishing plus component, $z^\mu=z^- n^\mu_-$, where $n^\mu_-$ is dimensionless and can be used to project vectors onto their plus component, e.g., $n_- \cdot p = p^+$. PDF evolution with respect to the factorization scale, which delineates long- from short-distance physics, is governed by the well-known DGLAP equations~\cite{Dokshitzer:1977sg,Gribov:1972ri,Altarelli:1977zs}. To assure gauge invariance the fields in the nonlocal operators are connected by Wilson lines which we do not write out explicitly. The PDF in Eq.~\eqref{eq_def_qPDF} corresponds to the sum $q=q_\uparrow + q_\downarrow$, while the PDF in Eq.~\eqref{eq_def_qtildePDF} corresponds to the difference $\Delta  q=q_\uparrow - q_\downarrow$ of the densities for quarks with opposite helicity. For spin-$1$ hadrons symmetry implies that distributions for different polarizations, $\lambda=+,0,-$, are related~\cite{Hoodbhoy:1988am,Best:1997qp}%
\begin{align}
    q^+ &= q^- \,, &
    \Delta q^+&=-\Delta q^- \,, &
    \Delta q^0&=0 \,,
\end{align}
such that only three independent quark PDFs remain. The quark PDFs defined above support $-1<x<1$, where the values at negative $x$ have to be interpreted as momentum fractions of anti-quarks%
\begin{align}%
    q^\lambda(x)&=-\bar q^\lambda(-x) \,, &
    \Delta q^\lambda(x)&=\Delta \bar q^\lambda(-x) \,, &
    \text{ for $x<0$} \,.
\end{align}\par%
In order to see the connection between PDFs and structure functions let us consider the operator product expansion%
\begin{align}%
    \mathcal{O}_a(z) \, \mathcal{O}_b(0) &= \sum_k c_{abk}(z) \, \mathcal{O}_k(0) \label{eq_operator-product} \,,
\end{align}%
which allows us to rewrite the product of two operators as a sum over local operators assuming that the momentum components of the external states under consideration are small compared to the inverse separation $1/z$. Using this concept allows us to expand the product of the electromagnetic currents in Eq.~\eqref{eq_scattering-amplitude} into a series of local operators multiplied by coefficient functions depending solely on the momentum transfer $q$. However, this is only valid for target matrix elements provided that the momentum transfer $q$ is much larger than the typical hadronic mass scale $\Lambda_{\rm{QCD}}$.\par%
For any general operator $\mathcal{O}_{d,n}^{\mu_1 \dots \mu_n}$ of dimension $d$ and spin $n$ one can show that the terms in the expansion have the structure%
\begin{align}%
    c_{\mu_1 \dots \mu_n} \mathcal{O}_{d,n}^{\mu_1 \dots \mu_n} &\rightarrow  \omega^n \left( \frac{Q}{M} \right)^{2-t} \,,
\end{align}%
where $M$ is the target hadron mass, $\omega = (2 p \cdot q)/(-q^2)$, and the twist $t = d - n$. Taking into account that QCD operators contain at least two quark fields ($t = 1$ each) and an arbitrary number of covariant, symmetrized derivatives $\overleftrightarrow{D}^{\mu} = \overrightarrow{D}^{\mu} - \overleftarrow{D}^{\mu}$ ($t = 0$ each) a conventional basis for lowest twist $t = 2$ quark operators can be written in terms of two towers of operators\footnote{In general one finds six towers of twist $t = 2$ operators, see, e.g.,~\cite{Ji:1998pc}.}%
\begin{align}%
    \mathcal{O}^{\mu_1 \dots \mu_n} &= \frac{1}{2^{n-1}}  \, \mathcal{S} \, \bar{q} \, \gamma^{\mu_1} i \overleftrightarrow{D}^{\mu_2} \dots i \overleftrightarrow{D}^{\mu_n} \, q  \,,
    \label{eq_ope-operator-vector}\\
    \mathcal{O}_5^{\mu_1 \dots \mu_n} &= \frac{1}{2^{n-1}}\, \mathcal{S} \, \bar{q} \, \gamma^{\mu_1} \, \gamma_5 i \overleftrightarrow{D}^{\mu_2} \dots i \overleftrightarrow{D}^{\mu_n} \, q \,,
    \label{eq_ope-operator-axial}
\end{align}%
where $\mathcal{S}$ projects out the completely symmetrized and traceless components of the r.h.s.\ tensor. It is straight forward to confirm that the matrix elements of these operators correspond to Mellin moments of the PDFs~\cite{Diehl:2003ny},\footnote{See Eq.~(33) in ref.~\cite{Diehl:2003ny} and note that the forward matrix elements in Eqs.~\eqref{eq_def_qPDF} and~\eqref{eq_def_qtildePDF} are invariant under translation.} e.g.,%
\begin{align}%
    \begin{split}
        \MoveEqLeft[2] \int_{-1}^1 dx \, x^{n-1} q(x) = \langle x^{n-1} \rangle_q  + (-1)^n  \langle x^{n-1} \rangle_{\bar q} \\
        &= \frac{1}{2 p_+^n} n^-_{\mu_1} \cdots n^-_{\mu_n}  \bra{\vec p} \mathcal{O}^{\mu_1 \dots \mu_n} \ket{\vec p}  \,,
        \label{eq_moments_vector}\taghere
    \end{split} \\
    \begin{split}
        \MoveEqLeft[2] \int_{-1}^1 dx \, x^{n-1} \Delta q(x) = \langle x^{n-1} \rangle_{\Delta q}  - (-1)^n  \langle x^{n-1} \rangle_{\Delta \bar q}\\
        &= \frac{1}{2 p_+^n} n^-_{\mu_1} \cdots n^-_{\mu_n}   \bra{\vec p} \mathcal{O}^{\mu_1 \dots \mu_n}_5 \ket{\vec p}  \,,
        \label{eq_moments_axial}\taghere
    \end{split}
\end{align}%
where we define the $n$-th moment of a function as%
\begin{align}%
    M_n(f) &= \int_0^1 \mathrm{d}x \, x^{n-1} \, f(x) = \langle x^{n-1} \rangle_f \,.
\end{align}\par%
In perturbation theory and to leading twist accuracy the structure functions are directly related to the PDFs, see, e.g., refs.~\cite{Hoodbhoy:1988am,Best:1997qp}, where a generic structure function $F$ is always obtained as the sum over the contributions from quarks and antiquarks for the individual quark flavors weighted by the square of their electric charge $e_q$:%
\begin{align}\label{eq_structure_function}%
    F &= \sum_q e_q^2 \bigl( F^q + F^{\bar q}\bigr) \,.
\end{align}%
In the following we will only write down the quark contribution. The antiquark contribution is obtained by simply substituting $q\rightarrow \bar q$. For spin-$0$ targets one obtains%
\begin{align}
    F_1^q(x) &= \frac{1}{2}q(x) + \mathcal O(\alpha_s) \,, \\
    F_2^q(x) &= x q(x) + \mathcal O(\alpha_s) \,,
\end{align}
satisfying the Callan-Gross relation~\cite{Callan:1969uq}, $F_2^q = 2 x F_1^q + \mathcal O(\alpha_s)$. The gluon PDF does not appear at leading order, since the gluons do not carry electric charge, and thus can only couple through a quark loop. For spin-$1$ targets the hadronic tensor depends on the hadron spin. Taking an average over the target spins one finds
\begin{align} \label{eq_spin1_relation_structure_function_to_pdf}
    F_1^q(x) &= \frac{1}{6}\Bigl( q^+(x) + q^0(x) + q^-(x) \Bigr) + \mathcal O(\alpha_s) \,, \\
    F_2^q(x) &= \frac{x}{3}\Bigl( q^+(x) + q^0(x) + q^-(x) \Bigr) + \mathcal O(\alpha_s) \,.
\end{align}
Considering the difference between targets with polarization $\lambda = \pm$ and $\lambda = 0$ one finds%
\begin{align}
    g_1^q(x) &= \frac{1}{2} \Delta q^+(x) + \mathcal O(\alpha_s) \,, \\
    b_1^q(x) &= \frac{1}{2}\Bigl(q^0(x) - q^+(x)\Bigr) + \mathcal O(\alpha_s) \,, \\
    b_2^q(x) &= x \Bigl(q^0(x) - q^+(x)\Bigr) + \mathcal O(\alpha_s) \,,
\end{align}
which means that $b_1$ and $b_2$ are sensitive to a possible dependence of the quark densities on the hadron polarization. The structure functions $g_2$, $b_3$, and $b_4$ do not contribute at leading twist.\par
Next, we perform a Lorentz decomposition for the forward matrix elements of the operators Eqs.~\eqref{eq_ope-operator-vector} and~\eqref{eq_ope-operator-axial}. For a spin-0 particle this yields %
\begin{align}
	\bra{\vec p} \mathcal{O}^{\mu_1 \dots \mu_n} \ket{\vec p} = 2 \, \mathcal{S} \left[ v^q_n p^{\mu_1} \cdots p^{\mu_n} \right],
	\label{eq_pseudoscalar-matrix-element}
\end{align}%
with the so-called reduced matrix element~$v_n$. Operators containing~$\gamma_5$ do not contribute because of symmetry relations. For a spin-$1$ particle we find three independent structures%
\begin{widetext}%
    \begin{align}%
        \bra{\vec p, \lambda} \mathcal{O}^{\mu_1 \dots \mu_n} \ket{\vec  p, \lambda} &= 2 \, \mathcal{S} \left[a^q_n p^{\mu_1} \cdots p^{\mu_n} + d^q_n \left( m^2 \epsilon^{* \mu_1}(\vec{p}, \lambda) \, \epsilon^{\mu_2}(\vec{p}, \lambda) - \frac{1}{3} p^{\mu_1} p^{\mu_2} \right) p^{\mu_3} \cdots p^{\mu_n} \right] \,,
        \label{eq_vector-matrix-element} \\
        \bra{\vec  p, \lambda} \mathcal{O}^{\mu_1 \dots \mu_n}_5 \ket{\vec  p, \lambda} &= 2 i \, \mathcal{S} \left[r^q_n \epsilon^{\rho \sigma \tau \mu_1} \epsilon^{*}_{\rho}(\vec{p}, \lambda) \epsilon_{\sigma}(\vec{p}, \lambda) p_{\tau} \, p^{\mu_2} \cdots p^{\mu_n} \right] \,,
        \label{eq_axial-vector-matrix-element}
    \end{align}%
\end{widetext}%
where we use the convention that $\varepsilon^{0123}=-1$. Here, $a_n$ is related to the polarization averaged contribution and $d_n$ to the polarized contribution of the quark PDF~$q$. The reduced matrix element $r_n$ in Eq.~\eqref{eq_axial-vector-matrix-element} is related to the (quark-)spin dependent PDF~$\Delta q$.\par%
In the structure functions always the sum of quark and antiquark contributions $F^{q+\bar q}= F^q + F^{\bar q}$ is relevant, see Eq.~\eqref{eq_structure_function}. Comparing this to Eqs.~\eqref{eq_moments_vector} and~\eqref{eq_moments_axial} one notices that the matrix elements given above yield information about either the even or (exclusive) the odd moments of a given structure function. For spin-$0$ targets one finds%
\begin{align}
    \begin{split}
        2 M_n(F_1^{q+\bar q})   &= C_n^{(1)} v^q_n \,, \qquad n \text{ even} \,, \\
        M_{n-1}(F_2^{q+\bar q}) &= C_n^{(2)} v^q_n \,, \qquad n \text{ even} \,, \taghere
        \label{eq_moments-of-structure-functions-spin0}
    \end{split}
\end{align}
while, for spin-$1$ targets,%
\begin{align}%
    \begin{split}
        2 M_n(F_1^{q+\bar q})   &= C_n^{(1)} a^q_n \,, \qquad n \text{ even} \,, \\
        M_{n-1}(F_2^{q+\bar q}) &= C_n^{(2)} a^q_n \,, \qquad n \text{ even} \,, \\
        2 M_n(b_1^{q+\bar q})   &= C_n^{(1)} d^q_n \,, \qquad n \text{ even} \,, \\
        M_{n-1}(b_2^{q+\bar q}) &= C_n^{(2)} d^q_n \,, \qquad n \text{ even} \,, \\
        2 M_n(g_1^{q+\bar q})   &= C_n^{(3)} r^q_n \,, \qquad n \text{ odd}  \,. \taghere
        \label{eq_moments-of-structure-functions-spin1}
    \end{split}
\end{align}%
The $C^{(k)}_n = 1 + \mathcal{O}(\alpha_s)$ are the Wilson coefficients of the OPE.\par%
We can also relate the moments of the PDFs to the reduced matrix elements. By substituting Eq.~\eqref{eq_pseudoscalar-matrix-element} into Eq.~\eqref{eq_moments_vector}, we find for spin-$0$%
\begin{align}%
    \begin{split}
        v^q_n &= \langle x^{n-1} \rangle_{q+\bar q}  \,, \qquad n \text{ even} \,, \\
        v^q_n &= \langle x^{n-1} \rangle_{q-\bar q}  \,, \qquad n \text{ odd} \,. \taghere
        \label{eq_red_matrixelement_pdf_relation_spin0}
    \end{split}
\end{align}%
For spin-$1$ hadrons we find (by substituting Eq.~\eqref{eq_vector-matrix-element} into Eq.~\eqref{eq_moments_vector}) that%
\begin{align}%
    \begin{split}
        a^q_n &= \frac{1}{3} \sum_{\lambda=\pm,0} \langle x^{n-1} \rangle_{q^\lambda+\bar q^\lambda}  \,, \qquad n \text{ even} \,, \\
        a^q_n &= \frac{1}{3} \sum_{\lambda=\pm,0} \langle x^{n-1} \rangle_{q^\lambda-\bar q^\lambda}  \,, \qquad n \text{ odd}  \,, \\
        d^q_n &= \langle x^{n-1} \rangle_{q^0+\bar q^0} - \langle x^{n-1} \rangle_{q^+ +\bar q^+}  \,, \qquad n \text{ even} \,, \\
        d^q_n &= \langle x^{n-1} \rangle_{q^0-\bar q^0} - \langle x^{n-1} \rangle_{q^+ -\bar q^+}  \,, \qquad n \text{ odd}  \,, \taghere
        \label{eq_red_matrixelement_pdf_relation_spin1}
    \end{split}
\end{align}%
i.e., $a^q_n$ yields the polarization average, while $d^q_n$ corresponds to the difference between hadrons with polarization $\lambda = \pm$ and $\lambda = 0$. In the following we will be particularly interested in the second moments, since the corresponding operator (cf.\ Eq.~\eqref{eq_ope-operator-vector} with $n=2$) is equivalent to the quark part of the energy-momentum tensor~\cite{Ji:1996ek}, and describes the distribution of the momentum within the hadron. For instance, in the spin-$1$ case a non-zero value of $d^q_2$ would indicate that the portion of the momentum carried by quarks of flavor $q$ depends on the polarization direction of the hadron.\par%
In analogy to the relations in Eqs.~\eqref{eq_red_matrixelement_pdf_relation_spin0} and~\eqref{eq_red_matrixelement_pdf_relation_spin1} the reduced matrix element $r_n$ in Eq.~\eqref{eq_axial-vector-matrix-element} can be related to the moments of $\Delta q$ defined in Eq.~\eqref{eq_moments_axial}. We will, however, restrict ourselves to the computation of the second moments of the vector PDF Eq.~\eqref{eq_def_qPDF} for the rest of this work.
\section{\label{sec_LatticeComputation}Computation on the lattice}%
\subsection{\label{sec_LatticeSetup}Lattice setup and numerical methods}%
To calculate the second moment of the structure functions introduced in the last section we analyzed a subset of the lattice gauge ensembles generated within the Coordinated Lattice Simulations (CLS) effort~\cite{Bruno:2014jqa}. The ensembles have been generated using a tree-level Symanzik improved gauge action with $N_f = 2+1$ flavors of nonperturbatively order $a$ improved Wilson (clover) fermions. Stable Monte-Carlo sampling is achieved by applying twisted-mass determinant reweighting~\cite{Luscher:2012av} to avoid near zero modes of the Wilson-Dirac operator.\par%
To avoid freezing of the topological charge and large autocorrelation times for the very fine lattices we use open boundary conditions for most of our simulations~\cite{Luscher:2012av, Luscher:2011kk}. Only some of the coarser lattices are simulated using periodic boundaries. The CLS gauge ensembles are generated along three different trajectories in the renormalized quark mass plane%
\begin{itemize}%
    \item ${\rm Tr M} = \text{const.}$: The trace of the quark mass matrix is kept constant near its physical value~\cite{Bruno:2014jqa}
    \item $m_s = \text{const.}$: The strange quark mass is kept constant close to its physical value~\cite{Bali:2016umi}
    \item $m_l = m_s$: The symmetric line.
\end{itemize}%
This strategy is explained in~\cite{Bali:2016umi} while an additional graphical illustration can be found in~\cite{Bali:2019yiy}. A complete list of the gauge ensembles used in this work is shown in Tab.~\ref{tab_Cls}. We use five different lattice spacings from $\unit{0.0497}{\femto\meter}$ up to $\unit{0.0984}{\femto\meter}$ and $m_{\pi}$ covers a range from $\sim \unit{420}{\mega\electronvolt}$ down to $\sim \unit{220}{\mega\electronvolt}$ with volumes $L m_{\pi}$ between $3.8$ and $6.4$, see Tab.~\ref{tab_Cls}.\par%
\begin{table*}
    \caption{\label{tab_Cls}CLS gauge ensembles analyzed in this work labeled by their identifier and sorted by the inverse coupling~$\beta$ and~$m_{\pi}$. We also label lattice volume in spatial and temporal direction, the boundary condition in time, open (o) or periodic (p), the lattice spacing, the pion mass, the volume in terms of $L m_{\pi}$, and the rho meson mass $m_{\rho}$ computed in Sec.~\ref{sec_RhoMass}. The list of source-sink distances analyzed for the connected three-point function of the ensemble is labeled by $t$ and if more than one measurement for a set of source-sink separations was performed we denote the corresponding number of measurements in parenthesis. In physical units these distances roughly correspond to $\unit{[0.7, 0.9, 1.0, 1.2]}{\femto\meter}$. The number of configurations analyzed for the ensemble is denoted as $n_{\rm{cnfgs}}$ and traj. specifies the trajectory in the quark mass plane of the ensemble~\cite{Bali:2016umi}. An in-depth description of the ensemble generation can be found in~\cite{Bruno:2014jqa}.}
    \begin{ruledtabular}
        \begin{tabular}{ccccccccccc}
            Ens. & $\beta$ & $N_s \times N_t $ & bc & $a$[fm] & $m_{\pi}$ [MeV] & $L m_{\pi}$ & $m_{\rho}$ [MeV] & $t/a$ & $n_{\rm{cnfgs}}$ & traj. \\ \hline
            A653 & 3.34 & $24 \times 48$ & p & 0.0984 & 426 & 5.1 & 870 & [7, 9, 11, 13]  & 2525 & sym \\
            A650 & 3.34 & $24 \times 48$ & p & 0.0984 & 368 & 4.4 & 813 & [7, 9, 11, 13]  & 2216 & sym \\
            H101 & 3.4 & $32 \times 96$ & o & 0.0859 & 420 & 5.9 & 860 & [8, 10, 12, 14]  & 2016 & trm/sym \\
            H102r001 & 3.4 & $32 \times 96$ & o & 0.0859 & 352 & 4.9 & 828 & [8, 10, 12, 14] (2) & 1992 & trm \\
            H102r002 & 3.4 & $32 \times 96$ & o & 0.0859 & 356 & 5.0 & 820 & [8, 10, 12, 14] (2) & 2016 & trm \\
            H105 & 3.4 & $32 \times 96$ & o & 0.0859 & 279 & 3.9 & 793 & [8, 10, 12, 14]  & 2052 & trm \\
            H106 & 3.4 & $32 \times 96$ & o & 0.0859 & 272 & 3.8 & 805 & [8, 10, 12, 14]  & 1543 & ms \\
            H107 & 3.4 & $32 \times 96$ & o & 0.0859 & 366 & 5.1 & 860 & [8, 10, 12, 14]  & 1564 & ms \\
            C101 & 3.4 & $48 \times 96$ & o & 0.0859 & 220 & 4.6 & 753 & [8, 10, 12, 14]  & 1997 & trm \\
            C102 & 3.4 & $48 \times 96$ & o & 0.0859 & 222 & 4.6 & 756 & [8, 10, 12, 14]  & 1465 & ms \\
            B450 & 3.46 & $32 \times 64$ & p & 0.0760 & 418 & 5.2 & 869 & [9, 11, 14, 16]  & 1612 & trm/sym \\
            B452 & 3.46 & $32 \times 64$ & p & 0.0760 & 350 & 4.3 & 856 & [9, 11, 14, 16]  & 1944 & ms \\
            D450 & 3.46 & $64 \times 128$ & p & 0.0760 & 214 & 5.3 & 741 & [9, 11, 14, 16]  & 617 & trm \\
            N450 & 3.46 & $48 \times 128$ & p & 0.0760 & 285 & 5.3 & 812 & [9, 11, 14, 16]  & 1131 & ms \\
            S400 & 3.46 & $32 \times 128$ & o & 0.0760 & 352 & 4.3 & 839 & [9, 11, 14, 16]  & 2001 & ms \\
            X450 & 3.46 & $48 \times 64$ & p & 0.0760 & 263 & 4.9 & 737 & [9, 11, 14, 16]  & 400 & sym \\
            rqcd030 & 3.46 & $32 \times 64$ & p & 0.0760 & 317 & 3.9 & 795 & [9, 11, 14, 16]  & 1222 & sym \\
            N201 & 3.55 & $48 \times 128$ & o & 0.0643 & 285 & 4.5 & 822 & [11, 14, 16, 19]  & 1522 & ms \\
            N202 & 3.55 & $48 \times 128$ & o & 0.0643 & 411 & 6.4 & 860 & [11, 14, 16, 19]  & 899 & trm/sym \\
            N203 & 3.55 & $48 \times 128$ & o & 0.0643 & 345 & 5.4 & 833 & [11, 14, 16, 19] (2) & 3086 & trm \\
            N204 & 3.55 & $48 \times 128$ & o & 0.0643 & 351 & 5.5 & 859 & [11, 14, 16, 19]  & 1500 & ms \\
            N200 & 3.55 & $48 \times 128$ & o & 0.0643 & 284 & 4.4 & 804 & [11, 14, 16, 19] (2) & 3424 & trm \\
            X250 & 3.55 & $48 \times 64$ & p & 0.0643 & 348 & 5.4 & 816 & [11, 14, 16, 19]  & 345 & sym \\
            X251 & 3.55 & $48 \times 64$ & p & 0.0643 & 267 & 4.2 & 757 & [11, 14, 16, 19]  & 434 & sym \\
            J303 & 3.7 & $64 \times 192$ & o & 0.0497 & 257 & 4.2 & 802 & [14, 17, 21, 24]  & 1068 & trm \\
            N300 & 3.7 & $48 \times 128$ & o & 0.0497 & 422 & 5.1 & 891 & [14, 17, 21, 24]  & 1539 & trm/sym \\
            N304 & 3.7 & $48 \times 128$ & o & 0.0497 & 351 & 4.3 & 884 & [14, 17, 21, 24]  & 1726 & ms
        \end{tabular}%
    \end{ruledtabular}%
\end{table*}%
The two- and three-point functions introduced in Sec.~\ref{sec_CorrelationFunctions} are computed on the lattice using the gauge configurations in Tab.~\ref{tab_Cls}. While we get the two-point functions by an inversion of the lattice Dirac operator using common numerical solvers (in particular we use a modified version of the Wuppertal adaptive algebraic multigrid code DD-$\alpha$AMG~\cite{Babich:2010qb, Frommer:2013fsa} on SIMD architectures~\cite{Heybrock:2015kpy, Richtmann:2016kcq, Georg:2017diz, Georg:2017zua} and the IDFLS solver~\cite{Luscher:2007se, Luscher:2007es} on other architectures) the computation of the three-point functions is more involved. The three-point function connected parts of all ensembles are computed using stochastic estimators as described in App.~\ref{sec_StochasticPropagator}. The computation of the three-point function disconnected contributions is described in App.~\ref{sec_DisconnPropagator}. To improve the overlap of the interpolating currents at the source and the sink timeslice we use Wuppertal smeared~\cite{Gusken:1989qx} quarks in the source and sink interpolators employing APE-smoothed gauge links~\cite{Falcioni:1984ei}. All the computations are performed using the Chroma software package~\cite{Edwards:2004sx} and additional libraries implemented by our group.\par%
\subsection{\label{sec_Renormalization}Renormalization}%
In order to obtain physically meaningful results, the bare operators introduced in Eqs.~\eqref{eq_ope-operator-vector} and~\eqref{eq_ope-operator-axial} have to be renormalized. In this context, one faces the additional difficulty that the isosinglet quark operators will mix under renormalization with the gluonic operators, schematically,%
\begin{align}%
    \mathcal O^{\rm ren} &= Z^{qq} \mathcal O + Z^{qg} \mathcal O_g \,,
\end{align}%
where we have suppressed the Lorentz indices for better readability. For the opposite direction (i.e., admixture of quark operators into gluon operators) it has been shown in ref.~\cite{Alexandrou:2016ekb}, using one-loop perturbative renormalization, that this admixture is a few percent effect (see also the discussion in ref.~\cite{Shanahan:2018pib}). We will assume that the same is true for the admixture of gluonic operators into quark operators, and that, as a consequence, its effect is negligible within the statistical accuracy of this work. Still, this caveat has to be kept in mind and needs to be addressed in future work. Note, however, that operators without an isosinglet part (e.g., with flavor structure $\bar u u+\bar d d-2\bar s s$) are not affected. Furthermore, we will approximate the isosinglet renormalization factor by the (non-perturbatively calculated) renormalization factor for isovector currents. This is exact to NLO accuracy (within a perturbative renormalization procedure).\par%
On the lattice, the continuous Euclidean $\mathrm{O}(4)$ symmetry is reduced to that of its finite hypercubic subgroup $\mathrm{H}(4)$. Therefore, symmetry imposes much weaker constraints on the mixing of operators under renormalization. In order to avoid mixing as far as possible, in particular mixing with lower-dimensional operators, we use operators from suitably chosen multiplets that possess a definite C-parity and transform according to irreducible representations of~$\mathrm{H}(4)$, cf.\ refs.~\cite{Gockeler:1996mu,Best:1997qp}. To be specific, we will use the operators $\mathcal O_{\rm v2a} = \mathcal O^{0i}$ and $\mathcal O_{\rm v2b} = \frac{4}{3} \mathcal O^{00}$, cf.\ App.~\ref{app_Operators}, where also an explicit definition of the operators is provided.\par%
Our final results will be given in the $\MSbar$ scheme at a scale of $\unit{2}{\giga\electronvolt}$. To this end, we adopt a two step procedure: First, we calculate the renormalization factors nonperturbatively in the $\RI$ scheme. These are then converted to the $\MSbar$ scheme using perturbative QCD. The whole procedure is described in great detail in ref.~\cite{Bali:2020lwx}, including subtleties due to the use of open boundaries in time direction and details of the perturbative subtraction of lattice artifacts. To be specific, we use the values for $Z_{\rm v2a}$ and $Z_{\rm v2b}$ based on $\RI$ as intermediate scheme, with the perturbative subtraction of lattice artifacts (without the use of the so-called fixed-scale method). The explicit values of $Z^{qq}$ for the operator combinations $\mathcal O_{\rm v2a}$ and $\mathcal O_{\rm v2b}$ used in this work are given in Tab.~\ref{tab_renormalization}.
\subsection{\label{sec_CorrelationFunctions}Correlation functions}%
In order to calculate the DIS structure functions on the lattice one has to compute two- and three-point correlation functions in the forward limit:%
\begin{align}
    C_{{\rm 2}, \vec p, t}^{(\mu \nu)} &= a^3 \sum_{\vec{x}} e^{-i \vec{p} \cdot \vec{x}} \langle O_M^{(\mu)}(\vec{x},t) \Bar{O}_M^{(\nu)}(\vec{0}, 0) \rangle \,,
    \label{eq_Twopt} \\
    C_{{\rm 3}, \vec p, t, \tau}^{(\mu \nu)} &= a^6 \mathop{\smash{\sum_{\vec{x}, \vec{y}}}} e^{-i \vec{p} \cdot \vec{x}}
    \langle O_M^{(\mu)}(\vec{x},t) \mathcal{O}(\vec{y}, \tau) \Bar{O}_M^{(\nu)}(\vec{0}, 0) \rangle \,.
    \label{eq_Threept}
\end{align}
We will consider pion \mbox{($M=\pi$)} and rho mesons \mbox{($M=\rho$)}, where the Lorentz indices are only necessary in the latter case. The interpolating current $\smash{\Bar{O}_M^{(\mu)}}$ creates a meson state with matching quantum numbers at the source timeslice $t_{\rm{src}}$ while $\smash{O_M^{(\nu)}}$ annihilates the meson at the sink time slice $t_{\rm{snk}}$. They read%
\begin{align}%
    O_\pi &= \bar q_{f} \gamma_5 q_{g} \,, &
    O_\rho^i &= \bar q_{f} \gamma^i q_{g} \,,
\end{align}%
with appropriately chosen quark flavors $f$ and $g$ and $i=1,2,3$. The quark fields in the interpolating currents are spatially smeared (see Sec.~\ref{sec_LatticeSetup}) to enhance the ground state overlap. In addition to the two interpolating currents the three-point function contains an insertion current $\mathcal{O}$ at timeslice $\tau$ with $0 < \tau < t$. The extraction of the ground-state matrix element of $\mathcal{O}$ is the key task in the subsequent calculations. In this work we set $t = t_{\rm{snk}} - t_{\rm{src}}$, $\tau = t_{\rm{ins}} - t_{\rm{src}}$ and hence $t_{\rm{src}} = 0$ without loss of generality.\par%
\subsubsection{\label{sec_PseudoScalarMesons}The pion}%
For the pion case we first define the matrix elements%
\begin{align}%
    \bra{0} O_\pi \ket{\vec{p}} &= \sqrt{\vphantom{Z_{\vec{p}}}\smash{Z^\pi_{\vec{p}}}} \,,
    \label{eq_pseudoscalarMatrixElement1}
\end{align}%
where $Z^\pi_{\vec{p}}$ is smearing dependent and encodes the overlap of the ground state with the interpolating currents at the source and the sink. Inserting a complete set of states into Eq.~\eqref{eq_Twopt} allows us to expand the two-point correlation function in terms of hadronic matrix elements. At large Euclidian times the correlation function can be approximated by the ground state contribution%
\begin{align}%
   C_{{\rm 2}, \vec p, t} &= Z^\pi_{\vec{p}} \frac{e^{-E^\pi_{\vec{p}} t}}{2 E^\pi_{\vec{p}}} + \dots \,,
    \label{eq_pseudoscalarSpecDecomp}
\end{align}%
where we assume that the same smearing setup is used at the source and the sink. For the ground state energies $E^\pi_{\vec{p}}$ we impose the continuum dispersion relation ${E^\pi_{\vec{p}} = \sqrt{m_\pi^2 + \vec{p}^2}}$.\par%
Similarly, one can show that the spectral decomposition of the three-point function in Eq.~\eqref{eq_Threept} for large Euclidean times reads%
\begin{align}%
    \begin{split}
        C_{{\rm 3}, \vec p, t, \tau} &=  Z^\pi_{\vec{p}} \frac{e^{-E^\pi_{\vec{p}} t}}{(2 E^\pi_{\vec{p}})^2} \bra{\vec{p}} \mathcal{O} \ket{\vec{p}} + \dots \,. \taghere
        \label{eq_pseudoscalarThreeptSpecDecompLimit}
        \end{split}
\end{align}%
In practice it turns out that especially for the three-point functions the signal-to-noise ratio at large Euclidean time distances $t$ and $\tau$ does not only contain the ground state contribution. How to exclude further excited state contributions is explained in Sec.~\ref{sec_ExcitedStates}.\par%
\subsubsection{\label{sec_VectorMesons}The rho meson}%
For the rho mesons we define, in analogy to the pseudoscalar case Eq.~\eqref{eq_pseudoscalarMatrixElement1},
\begin{align}%
    \bra{0} O_\rho^{\mu} \ket{\vec{p},\lambda} &= \sqrt{\vphantom{Z_{\vec{p}}}\smash{Z^\rho_{\vec{p}}}} \epsilon^{\mu}(\vec{p}, \lambda) \,,
    \label{eq_vectorMatrixElement1}
\end{align}%
where the polarization vector $\epsilon^{\mu}$ obeys the general transversality condition Eq.~\eqref{eq_trans-condition}. Therefore, the insertion of a complete set of states into Eq.~\eqref{eq_Twopt} (including a sum over all possible polarizations), yields%
\begin{align}%
    C_{{\rm 2}, \vec p, t}^{\mu \nu} &= - Z^\rho_{\vec{p}} \frac{e^{-E^\rho_{\vec{p}} t}}{2 E^\rho_{\vec{p}}} \left( g^{\mu \nu} - \frac{p^{\mu} p^{\nu}}{m_\rho^2}\right) + \dots \,,
    \label{eq_vectorSpecDecomp}
\end{align}%
where we have only written out the contribution from the rho meson, which is the leading one-particle state at large Euclidean times. Note, however, that in this case also two-pion states can occur, which, depending on the simulation parameters, can have smaller energies than the rho meson. This problem will be discussed in Sec.~\ref{sec_TwoPionStates}.\par%
Inserting two complete sets of states into the three-point function Eq.~\eqref{eq_Threept} we find%
\begin{align}%
    C_{{\rm 3}, \vec p, t, \tau}^{\mu \nu} &= Z^\rho_{\vec{p}} \frac{e^{-E^\rho_{\vec{p}} t}}{(2 E^\rho_{\vec{p}})^2} \mathop{\smash{\sum_{\lambda^\prime, \lambda}}}
    \epsilon^\mu(\vec{p}, \lambda^\prime) \epsilon^{\nu *}(\vec{p}, \lambda) \, \bra{\vec{p}, \lambda^\prime} \mathcal O \ket{\vec{p}, \lambda} \,.
    \label{eq_vectorThreeptSpecDecompLimit}
\end{align}%
Excited state contributions will be treated in Sec.~\ref{sec_ExcitedStates}.\par%
\begin{table}[t]
    \renewcommand{\arraystretch}{1.5}
    \centering
    \caption{\label{tab_renormalization}Renormalization factors $Z^{qq}$ for the operator combinations $\mathcal O_{\rm v2a}$ and $\mathcal O_{\rm v2b}$ used in this work. All values taken from Tab. XIII in~\cite{Bali:2020lwx}.}%
    \begin{ruledtabular}
        \begin{tabular}{cccccc}
            $\beta$ & 3.34 & 3.4 & 3.46 & 3.55 & 3.7 \\
            \hline
            $Z^{qq}_{\rm v2a}$ & 1.0731 & 1.1010 & 1.1251 & 1.1578 & 1.2053 \\
            $Z^{qq}_{\rm v2b}$ & 1.0672 & 1.0938 & 1.1170 & 1.1485 & 1.1949
        \end{tabular}
    \end{ruledtabular}
\end{table}
\subsubsection{\label{sec_ExcitedStates}Excited states analysis}%
In the three-point functions Eqs.~\eqref{eq_pseudoscalarSpecDecomp} and~\eqref{eq_vectorSpecDecomp} the signal-to-noise ratio decreases exponentially with the source-sink separation in time. At small time distances between the operators, however, there are still noticeable excited state effects. We take these into account by allowing for a generic excited state contribution in the spectral decomposition of the correlation functions. For the pseudoscalar correlation functions Eqs.~\eqref{eq_pseudoscalarSpecDecomp} and~\eqref{eq_pseudoscalarThreeptSpecDecompLimit} our ansatz reads%
\begin{align}%
    C_{{\rm 2}, \vec p, t} &= Z^\pi_{\vec{p}} \frac{e^{-E^\pi_{\vec{p}} t}}{2 E^\pi_{\vec{p}}} \Bigl( 1 + A e^{-\Delta E_{\vec{p}} t} \Bigr) \,,\label{eq_specDecomposition_excitedStates}\\
    \begin{split}
        C_{{\rm 3}, \vec p, t, \tau} &=  Z^\pi_{\vec{p}} \frac{e^{-E^\pi_{\vec{p}} t}}{(2 E^\pi_{\vec{p}})^2} \bra{\vec{p}} \mathcal{O} \ket{\vec{p}} \taghere \\
        \MoveEqLeft \times \Bigl(1 + B_{10} e^{- \Delta E_{\vec{p}} (t - \tau)} +  B_{01} e^{- \Delta E_{\vec{p}} \tau} + B_{11} e^{- \Delta E_{\vec{p}} t} \Bigr) \,, \label{eq_specDecompositionThreept_excitedStates}
    \end{split}
\end{align}%
where $\Delta E_{\vec{p}}$ denotes the energy difference to the first excited state. The excited state amplitude in the two-point function, $A$, depends on the interpolating currents at the source and the sink, their smearing, and the momentum $\vec{p}$, while the amplitudes in the three-point function, $B_{10}$, $B_{01}$, and $B_{11}$, also depend on the operator insertion $\mathcal{O}$.\par%
For the rho meson case we perform the analysis analogously. However, in particular for ensembles with small quark masses and large volumes, one would in this situation expect a contribution from (possibly multiple) two-pion states, which can have even smaller energy than the ``ground state'' rho meson itself. Despite the fact that we do not find any trace of these two-pion states in our numerical analysis, we cannot claim to have this problem fully under control; cf.\ the discussion of this delicate issue in Sec.~\ref{sec_TwoPionStates}.\par%
\subsubsection{\label{sec_Ratios}Ratios}
Instead of performing a fit to three-point functions, one can equivalently fit to ratios of two- and three-point functions. As discussed in ref.~\cite{Bali:2019yiy}, this can be advantageous due to a cancellation of unwanted correlations between two- and three-point functions. Furthermore, the ratio can be chosen in such a way that contributions from the ground state directly corresponds to the matrix element we are interested in. For the pseudoscalar correlation functions we define%
\begin{align}%
    R_{\vec{p}} &= \frac{C_{3, \vec{p}, t, \tau}}{C_{2, \vec{p}, t}}
    \xrightarrow{t \gg \tau \gg 0} \frac{\bra{\vec{p}} \mathcal{O} \ket{\vec{p}}}{2 E^\pi_{\vec{p}} } \,,
    \label{eq_ratio_pseudoscalar}
\end{align}%
which holds for any operator insertion $\mathcal O$ in the three-point function. For the vector meson case we will consider the diagonal case with the same Lorentz indices at the sink and at the source (i.e., $\mu=\nu=i$ in Eqs.~\eqref{eq_Twopt} and~\eqref{eq_Threept}). Defining $J^{\vec{p}}_{\lambda^\prime \lambda} \equiv  \langle \vec{p}, \lambda^\prime | \mathcal O | \vec{p}, \lambda \rangle/(2 E^\rho_{\vec{p}})$, one obtains%
\begin{align}%
    \begin{split}
        \MoveEqLeft[1]  R^i_{\vec{p}} = \frac{C^{ii}_{3,\vec{p}, t, \tau}}{C^{ii}_{2,\vec{p}, t}} \\
        &\xrightarrow{t \gg \tau \gg 0} \frac{m_{\rho}^2}{\left(E^{\rho}_{\vec{p}}\right)^2}  \sum_{\lambda, \lambda^\prime} \epsilon^i(\vec{p}, \lambda^\prime) \epsilon^{i *}(\vec{p}, \lambda) J^{\vec{p}}_{\lambda^\prime \lambda}  \,, \taghere
    \end{split}
    \label{eq_ratio_vector}
\end{align}%
where $i$ is fixed (no summation). On the right-hand side a sum over multiple matrix elements occurs, which can be evaluated explicitly for the chosen three-momentum. For on-axis momenta $\hat{\vec{p}} = \pm \vec{e}_i$ one finds the simple formulas
\begin{align}
    J^{\vec{p}}_{00} &= R^i_{\vec{p}} \,, &
    J^{\vec{p}}_{++} +  J^{\vec{p}}_{--} &=\sum_{j\neq i}  R^j_{\vec{p}} \,. \label{eq_onaxis_ratio_to_me}
\end{align}
for the extraction of the polarization-conserving matrix elements.
\par%
\subsection{\label{sec_TwoPionStates}Two-pion state contribution in the vector meson case}
In an infinite volume, above the particle creation threshold, a continuum of states would contribute to the spectral decomposition of the rho meson. In particular in Eq.~\eqref{eq_vectorSpecDecomp}, a continuum of two-pion states would contribute above the $2m_\pi$ threshold. In the non-interacting case, their center of mass energies are given by%
\begin{align}
    E_{\rm cm} &= 2 \sqrt{m_\pi^2 + \vec{k}^2} \,,
\end{align}
where $\vec{k}$ and $-\vec{k}$ are the momenta of the two pions in the center of mass frame. In a finite volume momenta are quantized such that one gets a sum over a discrete set of states that contribute.\par%

\begin{table}
    \centering
    \caption{\label{tab_little_groups}Little groups and decomposition of angular momentum $1$ in irreducible representations for all momentum sectors $\vec{d}^2\leq4$, where $\vec{p} = \frac{2\pi}{L} \vec{d}$. The groups are isomorphic for each representative of a sector. Table taken from ref.~\cite{Werner:2019hxc}.}
    \begin{ruledtabular}
        \begin{tabular}{lcc}
            $\vec{d}^2$   & $\rm{LG}(\vec{p})$ & $\Gamma$ \\
            \hline
            $0$ & $O_h$     & $\mathrm{T{1u}}$ \\
            $1$ & $C_{4v}$  & $\mathrm{A1} \oplus \mathrm{E}$ \\
            $2$ & $C_{2v}$  & $\mathrm{A1} \oplus \mathrm{B1} \oplus \mathrm{B2}$ \\
            $3$ & $C_{3v}$  & $\mathrm{A1} \oplus \mathrm{E}$ \\
            $4$ & $C_{4v}$  & $\mathrm{A1} \oplus \mathrm{E}$ \\
        \end{tabular}
    \end{ruledtabular}
\end{table}
\begin{table}
    \centering
    \caption{\label{tab_phase_shifts}Scattering phase shifts (assuming that only the $P$-wave contributes) for momentum sectors $\vec{d}^2$ and irreducible representations $\Gamma$. See refs.~\cite{Werner:2019hxc,Gockeler:2012yj} for more details.}
    \begin{ruledtabular}
        \begin{tabular}{lll}
            $\vec{d}^2$ & $\Gamma$ & $\phi^{\vec{d}}_\Gamma$ \\
            \hline
            $0$	& $\mathrm{T1u}$ & $w_{0,0} - w_{2,0} - \sqrt{6} w_{2,2}$ \\
            $1$ & $\mathrm{A1}$ & $w_{0,0} + 2 w_{2,0}$ \\
            $1$ & $\mathrm{E}$ & $w_{0,0} - w_{2,0} $\\
            \dots & \dots & \dots
        \end{tabular}
    \end{ruledtabular}
\end{table}
\begin{figure*}[tb]
    \includegraphics[width=\textwidth]{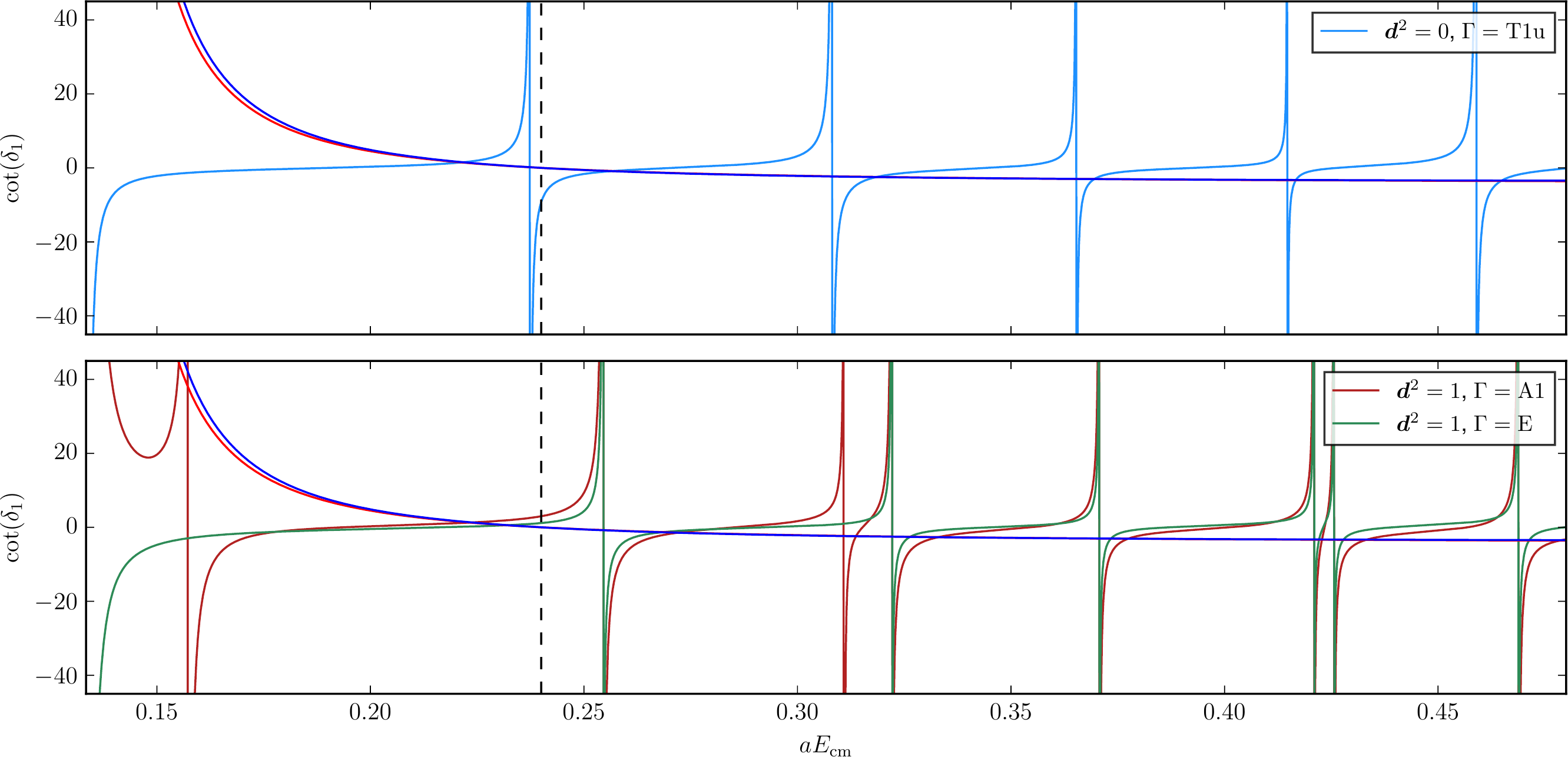}
    \caption{\label{fig_phase_shifts}Scattering phase shifts $\phi^{\vec{d}}_\Gamma$ for different total momenta $\vec{p} = \frac{2 \pi}{L} \vec{d}$ and irreducible representations $\Gamma$ calculated based on the relations given in Tab.~\ref{tab_phase_shifts}. As input we have used the pion mass, the rho mass and the volume of the ensemble D200. The dashed vertical line indicates the rho mass we have measured on this ensemble. The poles occur at the positions of the noninteracting center of mass energies. The red and blue curves correspond to the Breit-Wigner ansatz Eq.~\eqref{eq_delta_BW} and to the Gounaris-Sakurai ansatz Eq.~\eqref{eq_delta_GS}, respectively, see also Fig.~\ref{fig_phase_shifts_2}. The energy levels are at the intersections of the curves.}
\end{figure*}%
\begin{figure}[t]
    \includegraphics[width=\columnwidth]{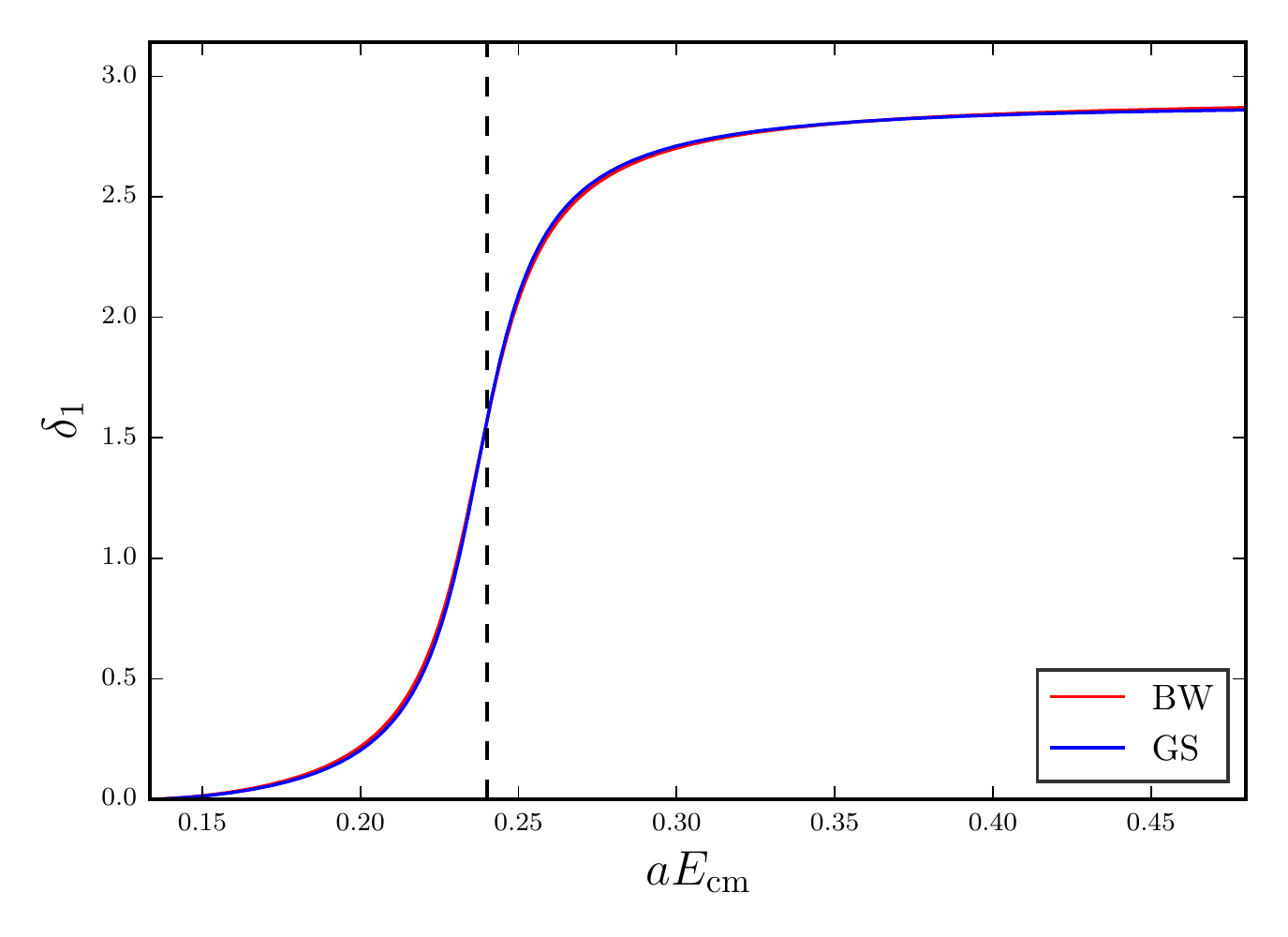}
    \caption{\label{fig_phase_shifts_2}Comparison of the Breit-Wigner (red) and Gounaris-Sakurai (blue) parametrizations of the phase shift. Here we use the pion mass and (naively measured) rho mass from D200 as input. The rho-pi-pi coupling constant is set to the phenomenological value $g_{\rho\pi\pi}=5.96$.}
\end{figure}
For particles of integer spin and at zero momentum (i.e., in the center of mass frame), the full symmetry group on the lattice is the octahedral group $O_h=O \otimes I$ defined as the direct product of the cubic group $O$ (consisting of $24$ rotations) and the group of space inversions~$I$.\footnote{For half-integer spin one would have to consider the corresponding double covers of $O$ and $O_h$.} In a moving frame, however, the symmetry is reduced to the so-called little groups (for details see, e.g., refs.~\cite{Gockeler:2012yj,Erben:2019nmx,Werner:2019hxc}) shown in Tab.~\ref{tab_little_groups} together with the decomposition into irreducible representations. The four-momenta in the laboratory frame are related to those in the center-of-mass frame by a Lorentz boost, such that the states that obey the quantization condition on the lattice in the different moving frames will in general correspond to different center-of-mass energies.\par%
The connection between the finite volume energy spectrum of two-pion states and infinite volume scattering phase shifts has been established by M.~L\"uscher in his seminal articles~\cite{Luscher:1986pf,Luscher:1990ux}. Recent discussions of this topic are also found in refs.~\cite{Gockeler:2012yj,Erben:2019nmx,Werner:2019hxc}. Being interested in the rho resonance in the vector channel, we may restrict ourselves to the $P$-wave ($l=1$) contribution, since it is usually found that nonzero phase shifts in higher odd partial waves are not required to describe the two-pion spectrum~\cite{Dudek:2012xn,Wilson:2015dqa}. In this simplified situation the $P$-wave phase shift $\delta_1$ is directly related to the quantized two-pion energy levels in finite volume. The latter appear when the condition%
\begin{align}\label{eq_quantization_condition}
    \cot{\delta_1} \overset{!}{=} \cot{ \phi^{\vec{d}}_\Gamma}
\end{align}
is satisfied, see Fig.~\ref{fig_phase_shifts}, which will be discussed in more detail below. The scattering phase shifts $\phi^{\vec{d}}_\Gamma$ can be taken from Tab.~\ref{tab_phase_shifts}, using
\begin{align}
    w_{lm} &= \frac{Z^{\vec{d}}_{lm}(1,q^2)}{\pi^{3/2} \, \sqrt{2l+1} \, \gamma \, q^{l+1}} \,, &
    q &= \frac{L k}{2 \pi} \,, \\
    k &= \sqrt{\frac{E_{\rm cm}^2}{4} - m_\pi^2 } \,. \label{eq_k}
\end{align}
For the numeric evaluation of the generalized zeta function~$Z^{\vec{d}}_{lm}(1,q^2)$ we use the representation derived in ref.~\cite{Gockeler:2012yj}.\par%
Using Eq.~\eqref{eq_quantization_condition} we obtain the energy levels in the interactive case via equating the phase shifts given in Tab.~\ref{tab_phase_shifts} with a phenomenological parametrization, where, for any given parametrization, we define the rho mass and width as~\cite{Gounaris:1968mw}%
\begin{align}
    \cot{\delta_1} \biggr|_{s=m_\rho^2} &= 0 \,, &
    m_\rho \Gamma_\rho &= \biggl(\frac{d \delta_1}{d s} \biggr)^{-1}_{s=m_\rho^2} \,,
\end{align}
using the Mandelstam variable $s=E_{\rm cm}^2$. For instance one can use a relativistic Breit-Wigner (BW) ansatz%
\begin{align} \label{eq_delta_BW}
    \frac{k^3}{\sqrt{s}}  \cot{\delta_1^{\rm BW}} &= \frac{6 \pi \bigl( m_\rho^2 - s\bigr)}{g_{\rho\pi\pi}^2 }
    = \frac{k_\rho^3 \bigl( m_\rho^2 - s \bigr)}{\Gamma_\rho m_\rho^2} \,,
    \intertext{with}
    \Gamma_\rho &= \frac{g_{\rho\pi\pi}^2 k^3_\rho} {6 \pi m_\rho^2} \,,
    \label{eq_width}
\end{align}
where $k=\sqrt{s/4 - m_\pi^2}$, as defined in Eq.~\eqref{eq_k}, and
\begin{align}
 k_\rho&=k\biggr|_{s= m_\rho^2}=\sqrt{m_\rho^2/4 - m_\pi^2} \,.
\end{align}
Alternatively, one can use a Gounaris-Sakurai (GS) parametrization~\cite{Gounaris:1968mw}, where
\begin{align} \label{eq_delta_GS}
\frac{k^3}{\sqrt{s}}  \cot{\delta_1^{\rm GS}} &= k^2 (h(s) - h(m_\rho^2)) + (k_\rho^2 - k^2) c \,,
\intertext{with}
 h(s)&= \frac{2}{\pi} \frac{k}{\sqrt{s}} \ln{ \biggl( \frac{\sqrt{s}+2k}{2 m_\pi} \biggr)} \,, \\
 c &= \frac{4 k_\rho^3}{m_\rho^2 \Gamma_\rho} + 4 k_\rho^2 h^\prime(m_\rho^2) \,.
\end{align}
If one chooses to apply the Kawarabayashi-Suzuki-Riazuddin-Fayyazuddin relation \cite{Kawarabayashi:1966kd,Riazuddin:1966sw}, $m_\rho^2=2g_{\rho\pi\pi}^2F_\pi^2$, (which, as argued in ref.~\cite{Djukanovic:2004mm}, is a consequence of chiral symmetry and the requirement of consistency of the effective field theory with respect to renormalizability), both the BW and the GS parametrizations are determined solely by the rho mass (given that the pion decay constant $F_\pi$ is well-known and the width of the rho is linked to the rho-pi-pi coupling constant~$g_{\rho\pi\pi}$ via Eq.~\eqref{eq_width}).\par%
In Fig.~\ref{fig_phase_shifts_2} we plot the $P$-wave phase shifts for both, the BW (red) and the GS (blue) parametrization, using the properties of our ensemble D200 (${N_s \times N_t = 64 \times 128, m_{\pi} = \unit{201}{\mega\electronvolt}, m_{\rho} = \unit{746}{\mega\electronvolt}}$) as input. As one can see, the two parametrizations yield quite similar results. In Fig.~\ref{fig_phase_shifts}, we illustrate the quantization condition Eq.~\eqref{eq_quantization_condition}: the energy levels are situated at the intersections between the phase shift parametrizations and the curves for~$\cot{ \phi^{\vec{d}}_\Gamma}$. The pole positions correspond to the center of mass energies of the noninteracting system.\par%
As pointed out in ref.~\cite{Gounaris:1968mw}, the phase shifts are linked to the pion form factor via
\begin{align}
    F_\pi(s) &= \frac{f(0)}{f(s)} \,,&
    &\text{with} &
    f(s) &= \frac{k^3}{\sqrt{s}} \bigl( \cot{\delta_1} - i \bigr)   \,.
\end{align}
Note that this formula only works for the GS ansatz, which we will use in the following, and not for the BW ansatz, because in the latter case $f(s)$ diverges at $s=0$. Using the GS parametrization one finds%
\begin{figure}[t]
    \includegraphics[width=\columnwidth]{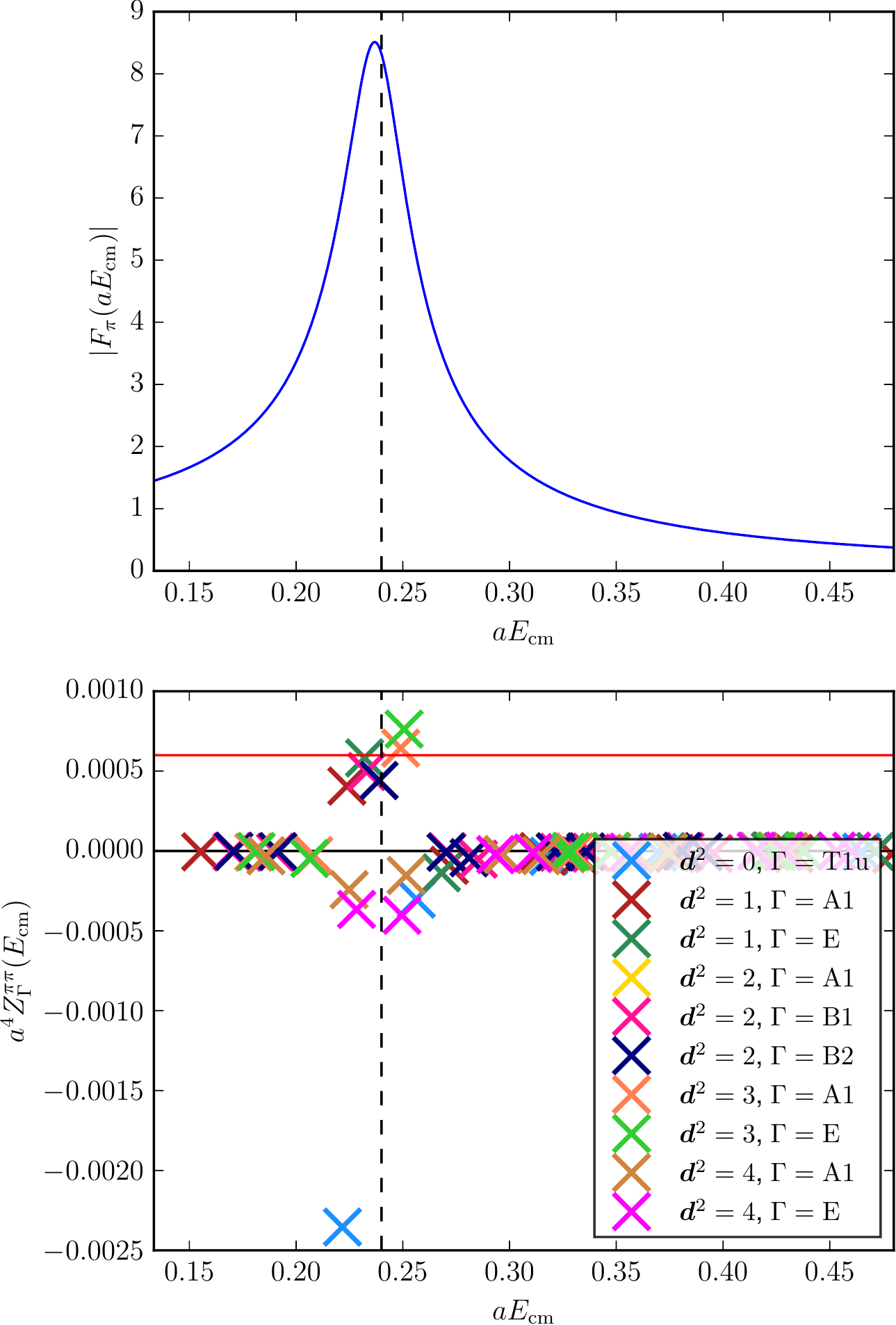}
    \caption{\label{fig_ff_and_overlap}The pion form factor obtained using Eq.~\eqref{eq_PiFF} and the corresponding overlap factors $Z^{\pi\pi}_\Gamma$ (again using D200 input parameters) for the local currents from Eq.~\eqref{eq_overlap_pipi}. The red line on the lower panel is plotted for comparison and corresponds to the estimated $Z^\rho = 2 f_\rho^2 m_\rho^2$ for the local currents using the phenomenological value $f_\rho = \unit{222}{\mega\electronvolt}$ as input.}
\end{figure}
\begin{align}
    F_\pi(s) &= \frac{m_\pi^2 h(m_\rho^2) + \frac{c}{4} m_\rho^2 -\frac{1}{\pi} m_\pi^2 }{\frac{k^3}{\sqrt{s}} \bigl( \cot{\delta_1^{\rm GS}} - i \bigr)} \,.
    \label{eq_PiFF}
\end{align}
As shown in ref.~\cite{Feng:2014gba} (which is a generalization of the original derivation given in ref.~\cite{Meyer:2011um} for moving frames), the form factor can be determined from the overlap factor of the two-pion states with a local (unsmeared) vector current. By inverting this relation (and adapting it to our conventions) we obtain the overlap factors
\begin{align} \label{eq_overlap_pipi}
    Z^{\pi\pi}_{\vec{p}}(s)  &= \biggl( q \frac{\partial \phi^{\vec{d}}_\Gamma}{\partial q} + k \frac{\partial \delta_1(k)}{\partial k} \biggr)^{-1} \frac{k^5}{6 \pi \sqrt{s}} | F_\pi(s) |^2 \,,
\end{align}
from a given form factor, which, in turn, can be determined from the phase shift.\par%
In Fig.~\ref{fig_ff_and_overlap} we show the form factor and the corresponding estimate for the overlap factor of local (i.e., unsmeared) vector currents at the source and the sink with the two-pion states at a given center of mass energy. The first thing to notice is, that two-pion states whose center of mass energy is much smaller (or larger) than the rho mass are strongly suppressed. In the example shown here the overlap of these states is (roughly) smaller by a factor of 100 compared to our estimate for overlap of the rho meson itself (horizontal red line). This means that these states will not yield large contributions to the correlation functions at the intermediate time distances available in our simulation, despite being energetically favored, which may explain why we do not see these states in our numerical analysis. More problematic is the possible contribution of states that have a center of mass energy close to the rho mass. As one can see in Fig.~\ref{fig_ff_and_overlap}, the overlap of these states is strongly enhanced, and can be of the same size or even larger than the overlap of the rho meson. This is particularly concerning, because we would not be able to distinguish such a state in the spectral decomposition within our numerical analysis. It is important to keep this caveat in mind when interpreting our results.\par%
That being said, we want to stress that the analysis provided above is actually only valid for unsmeared currents. Obviously, the situation might be less critical for the smeared currents that we use in our simulation. A posteriori, the trustworthiness of the numerical results presented in the following could be enhanced significantly, if future studies (e.g., by using the generalized eigenvalue method with two-pion interpolating currents, cf.\ refs.~\cite{Luscher:1990ck,Bali:2015gji,Erben:2019nmx,Fischer:2020fvl,Fischer:2020bgv}) can show that the overlap of smeared vector interpolating currents with the two-pion states is much smaller than for the local currents.\par%
\section{Analysis and results\label{sec_Results}}%
\subsection{Pion and rho mass}\label{sec_RhoMass}
To compute the reduced matrix elements introduced, e.g., in Eq.~\eqref{eq_vector-matrix-element}, we need the mass (energy) of the meson in the rest (boosted) frame. While the values for $m_{\pi}$ are taken from~\cite{Bali:2023} the values for $m_{\rho}$ are obtained by a direct fit to the correlation function using the spectral decomposition presented in Eq.~\eqref{eq_specDecomposition_excitedStates}. Beside the mass (energy) itself two additional amplitudes ($Z$ and $A$) and also the energy gap to the first excited state $\Delta E$ enter the fit as free parameters. However, using the ratio method introduced in Sec.~\ref{sec_Ratios} the additional amplitudes and also $\Delta E$ will not enter the results presented in this work.\par
\begin{figure}[tb]
    \includegraphics[]{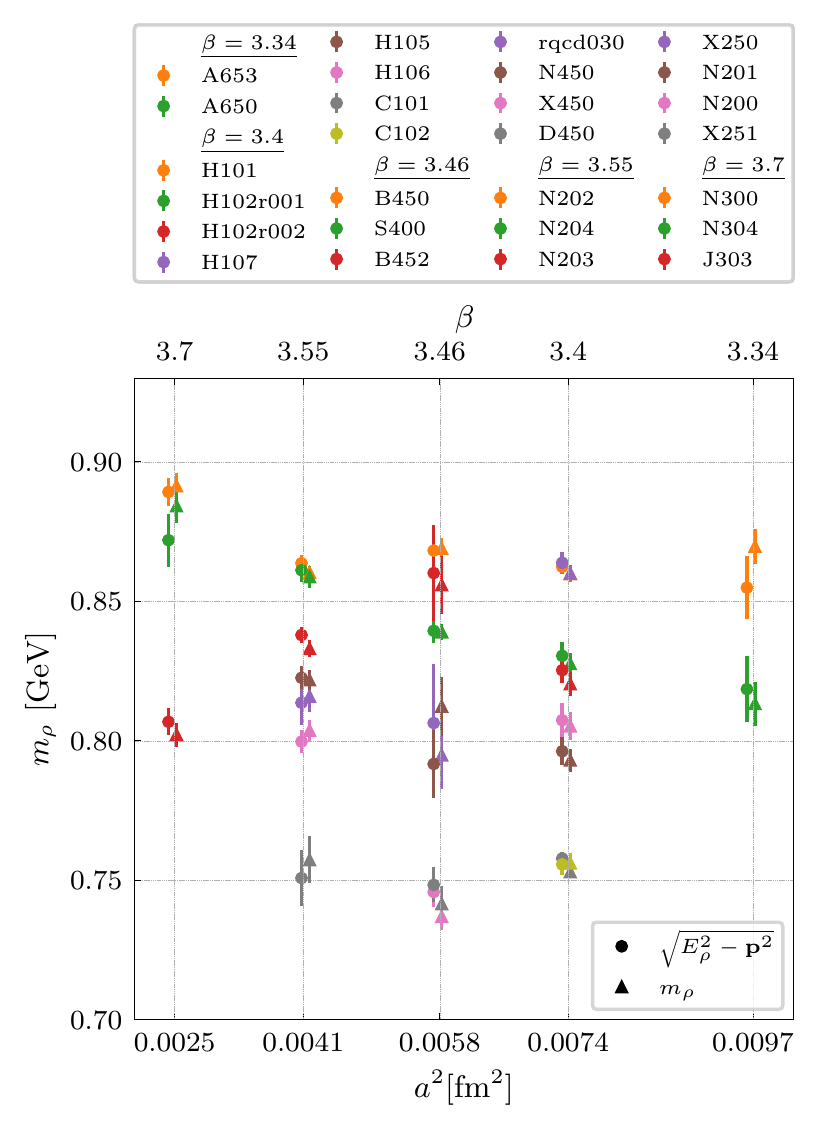}
    \caption{\label{fig_rho_mass}Rho masses for all ensembles analyzed from a double exponential fit (open boundaries) or a single exponential fit (periodic boundaries) to the two-point function correlator. The triangles depict the results in the rest frame ($\vec{p}^2 = 0$) while the circles correspond to fits in the boosted frame ($\vec{p}^2 = \frac{4 \pi^2}{L^2}$) projected to the rest frame using the continuum dispersion relation.}
\end{figure}

%
The fits are performed using a constant fit window of ${\sim} \unit{2}{\femto\meter}$ for all ensembles with open boundary conditions and we start 1 or 2 timeslices (${\sim} \, \unit{0.1}{\femto\meter}$) away from the source for the coarser or finer lattices. Due to the structure of Eq.~\eqref{eq_specDecomposition_excitedStates} the values obtained by the fit for the ground state and excited state energy can be interchanged. To overcome this technical issue we have introduced a cutoff for the double exponential fit at $t_{\mathrm{cut}} \approx \unit{0.65}{\femto\meter}$ and fit only the single exponential $Z^\rho_{\vec{p}} (2 E^\rho_{\vec{p}})^{-1} e^{-E^\rho_{\vec{p}} t}$ for larger times. In case of periodic boundary conditions we choose a symmetric ansatz of the form $Z^\rho_{\vec{p}} (2 E^\rho_{\vec{p}})^{-1} (e^{-E^\rho_{\vec{p}} t} + e^{-E^\rho_{\vec{p}} (T-t)})$ with $t_{\mathrm{cut}} \leq t \leq T-t_{\mathrm{cut}}$ where we only fit the amplitude and the ground state energy. The final results of these fits are shown in Fig.~\ref{fig_rho_mass}. We depict the rest frame results by triangles and the boosted frame results by circles. Note that the continuum dispertion relation $E_{\vec{p}}^\rho = \sqrt{m_\rho^2 + \vec{p}^2}$ is used to project the energies onto their corresponding mass values and to check that the dispersion relation is well satisfied for the momenta in use.
\subsection{Extraction of ground state matrix elements\label{sec_GSmatrixelements}}
The observables studied in this article are affected by disconnected quark loops. Often one can circumvent this problem by considering isovector current insertions, where the up and down quark disconnected loops cancel each other identically in the limit of exact isospin symmetry. This is not a viable solution in this case, since also the connected part vanishes for the isovector currents. Unfortunately, the disconnected contributions are notorious for having a large statistical error. However, as will be discussed later in this section, this is not true in general.\par%
\begin{figure*}[tp]
    \includegraphics[]{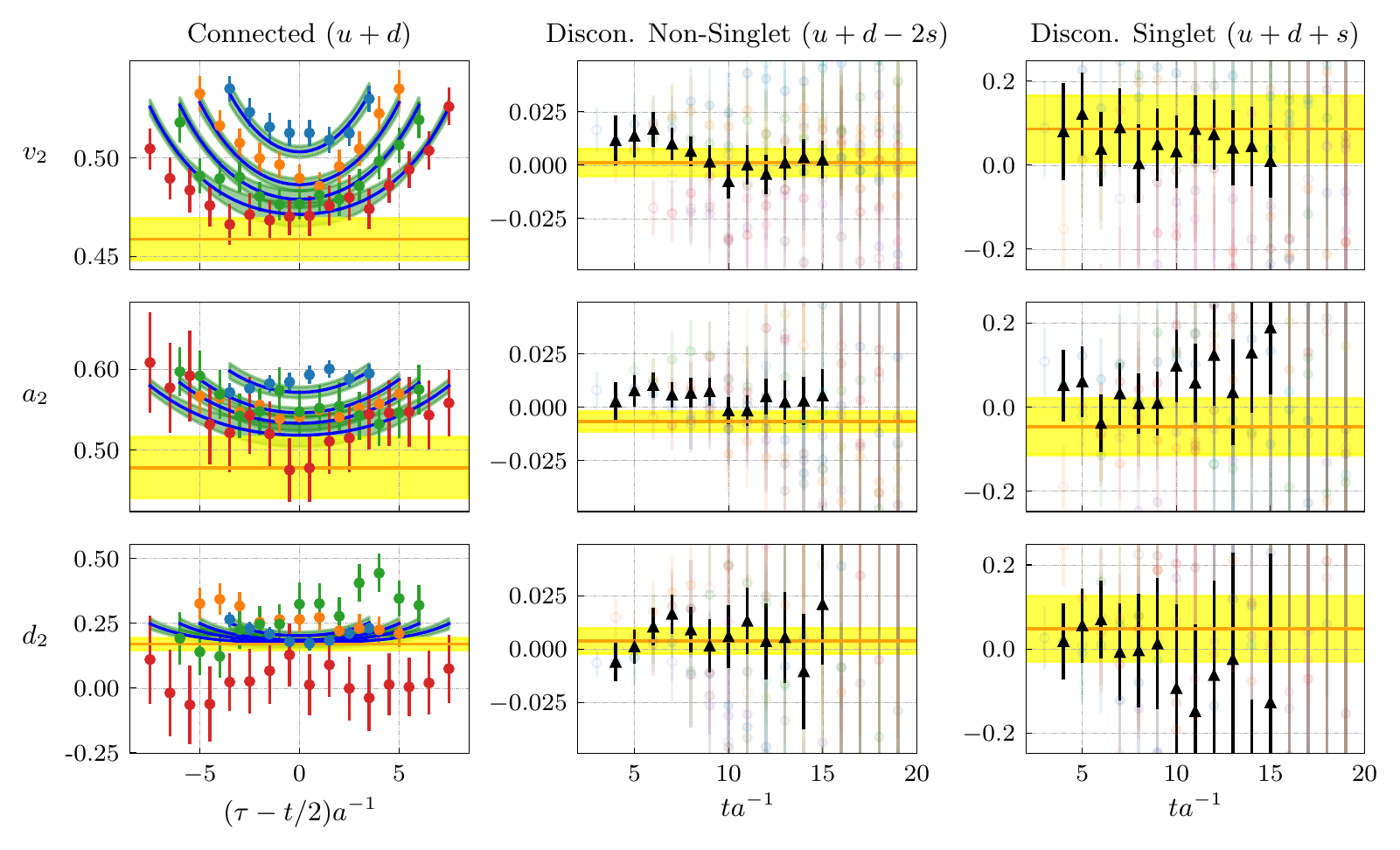}%
    \caption{\label{fig_N204_ratio_v2a}Extraction of renormalized values for $v_2$, $a_2$, and $d_2$ from the ratios obtained in Eqs.~\eqref{eq_v2a}--\eqref{eq_d2a} using the operator combination $\mathcal O_{\rm v2a}$. For illustrative purposes we only show the data points and individual fits (solid blue lines in the first column) for all momentum combinations with $\vec{n}^2 = 1$, however, the ground state results (orange lines) are obtained by a simultaneous fit to the operator combinations $\mathcal O_{\rm v2a}$ and $\mathcal O_{\rm v2b}$ using all possible momenta for the corresponding matrix element with $\vec{n}^2 \leq 1$ for the connected, disconnected non-singlet, and disconnected singlet contributions respectively. The analysis shown in this plot has been performed on the ensemble N204. The solid blue lines in the first column correspond to a simultaneous fit to the four source-sink separations of the ensemble, cf.\ Tab.~\ref{tab_Cls}, for the insertion current $(\bar{u} u + \bar{d} d)$ needed to construct the flavor (non-)singlet operator contributions. In case of $v_2$ and $a_2$, the fits allow for a generic excited state on top of the ground state, while in case of $d_2$ the excited state energy $\Delta E$ is fixed by the two-point function, see Eq.~\eqref{eq_specDecomposition_excitedStates}. The orange line depicts the extracted ground state contribution and directly correspond to the values of the reduced matrix elements. In the second and third column we show the disconnected contributions for the non-singlet ($\bar{u} u + \bar{d} d - 2 \bar{s} s$) and singlet ($\bar{u} u + \bar{d} d + \bar{s} s$) operators as function of the final timeslice $t$. In addition to the original data points (grayed out) we also show an average over all insertion times for every final timeslice $t$ (black triangle markers).}
\end{figure*}%
\begin{figure*}[tp]
    \includegraphics[]{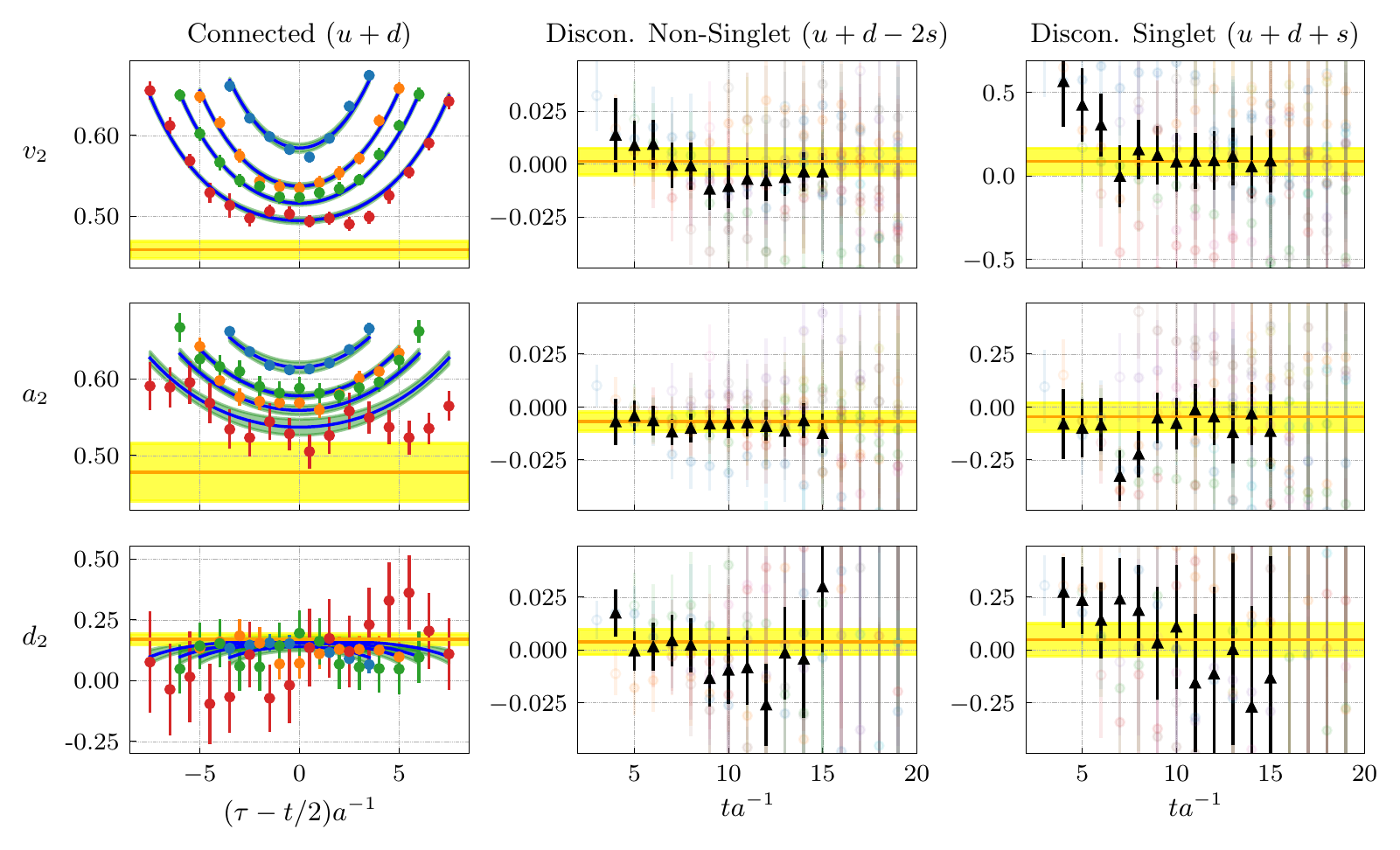}
    \caption{\label{fig_N204_ratio_v2b}Extraction of renormalized values for $v_2$, $a_2$, and $d_2$ from the ratios obtained in Eqs.~\eqref{eq_v2b}--\eqref{eq_d2b} using the operator combination $\mathcal O_{\rm v2b}$. The data is visualized in the same way as in Fig.~\ref{fig_N204_ratio_v2a}, i.e., we only plot the data points and individual fits for $\vec{n}^2 = 1$. Also here we want to stress, that the ground state results (orange line) are obtained by a simultaneous fit to the operator combinations $\mathcal O_{\rm v2a}$ and $\mathcal O_{\rm v2b}$ using all possible momenta for the corresponding matrix element with $\vec{n}^2 \leq 1$ for the connected, disconnected non-singlet, and disconnected singlet contributions respectively.}
\end{figure*}
In cases where the disconnected contributions are zero within the error one might be tempted to simply drop them. However, in situations where the statistical error is large (for instance the flavor singlet operators), their inclusion can shift the mean and, even more important, can increase the error for the final result substantially. I.e., they have to be included, if one wants to provide reliable error estimates for phenomenological applications. Nevertheless, we perform a second analysis in these cases, where we solely use the connected part, which allows us to compare to other lattice results for connected contributions.\par%
In principle, one would like to add the connected and the disconnected contributions already at the correlation function level. However, this is not feasible since the connected and the disconnected parts are calculated in different ways. The connected part is calculated with the stochastic propagator estimation presented in App.~\ref{sec_StochasticPropagator} using a setup with a fixed sink timeslice to obtain the result for all possible insertion times. The disconnected loops are calculated on fixed timeslices also using stochastic propagators (see App.~\ref{sec_DisconnPropagator} for details). For any computed two-point function we can therefore obtain the disconnected three-point function at a fixed insertion time but for arbitrary final times. In order to sum up the contributions at the correlation function level, we would have to throw away a large part of our data, since we could only use the insertion and final times where we have data for both. This would be prohibitively wasteful. Therefore, we will perform the extraction of the ground state matrix elements separately for the connected and the disconnected contribution.\par%
Next, we extract the ground state contribution from the ratios defined in Sec.~\ref{sec_Ratios}. We use kinematic prefactors and, in case of the rho, take appropriate linear combinations such that the ground state contributions directly correspond to the reduced matrix elements $v_2^q$ (for the pion) as well as $a_2^q$ and $d_2^q$ for the rho meson. For on-axis momenta $\hat{\vec{p}} = \pm \vec{e}_i$ we obtain (using Eqs.~\eqref{eq_ov2a_v2} and~\eqref{eq_ov2a_a2d2})
\begin{align}
    v_2 &= \frac{1}{p^i} R_{\vec{p}}\bigl( \mathcal O_{\rm v2a}^i \bigr) \,, \label{eq_v2a}
\end{align}
\begin{align}
    a_2 &= \frac{1}{3 p^i} \sum_j R^j_{\vec{p}} \bigl( \mathcal O_{\rm v2a}^i \bigr)  \,, \label{eq_a2a} \\
    d_2 &= \frac{3}{4 p^i} \bigl( 2 R^i_{\vec{p}}\bigl( \mathcal O_{\rm v2a}^i \bigr) - \sum_{j \neq i} R^j_{\vec{p}}\bigl( \mathcal O_{\rm v2a}^i \bigr) \bigr) \,, \label{eq_d2a}
\end{align}
with the operator $\mathcal O_{\rm v2a}^i$ as inserted current. For $\mathcal O_{\rm v2b}$ we find
\begin{align}
    v_2 &= \frac{3 E}{4 E^2 - m^2} R_{\vec{p}} \bigl( \mathcal O_{\rm v2b} \bigr) \,, \label{eq_v2b} \\
    a_2 &= \frac{E}{4 E^2 - m^2} \sum_j R^j_{\vec{p}} \bigl( \mathcal O_{\rm v2b} \bigr) \,, \label{eq_a2b} \\
    d_2 &= \frac{3 E}{8 (E^2 - m^2)} \bigl( 2 R^i_{\vec{p}}\bigl( \mathcal O_{\rm v2b} \bigr) - \sum_{j \neq i} R^j_{\vec{p}}\bigl( \mathcal O_{\rm v2b} \bigr) \bigr)  \,.\label{eq_d2b}
\end{align}
Here we hide the superscripts $\pi$, $\rho$, and the subscript $\vec{p}$ for the mass and energy. Measurements using $\mathcal O_{\rm v2a}^i$ always require nonzero momentum in direction $i$. If one uses $\mathcal O_{\rm v2b}$ the reduced matrix elements $v_2$ and $a_2$ can be measured for vanishing three-momentum. The extraction of $d_2$, however, always requires nonzero momentum. The reason for this is that $d_2$ corresponds to the difference of the PDF moment between longitudinally and transversally polarized rho mesons, which is no useful concept for mesons in their rest frame. Explicit operator definitions can be found in App.~\ref{app_Operators}.\par%
In Figs.~\ref{fig_N204_ratio_v2a} and~\ref{fig_N204_ratio_v2b}, we show examples of the ratio fits for the ensemble N204. For the statistical analysis we generate 500 bootstrap samples per ensemble using a bin size of 40 molecular dynamics units to eliminate autocorrelations. To visualize the fits and corresponding data points we only present plots for $\vec{n}^2 = 1$, where $\vec{n}$ is defined by $\vec{p} = \frac{2 \pi}{L} \, \vec{n}$, using the operator combinations Eqs.~\eqref{eq_v2a}--\eqref{eq_d2b}. However, the results for the reduced matrix elements $v_2$, $a_2$, and $d_2$ in the left column are obtained by combined fits to all ratios using an ansatz similar to Eq.~\eqref{eq_specDecompositionThreept_excitedStates} with the definitions Eqs.~\eqref{eq_ratio_pseudoscalar} and~\eqref{eq_ratio_vector}. In case of pseudoscalar mesons this reads
\begin{align}\label{eq_fitAnsatz}
    R(\mathcal{O}, \vec{p}^2, t, \tau) = B_0 &+ B_1\left(\mathcal{O}, \vec{p}^2\right) \,\, e^{- \Delta E_{\vec{p}^2} \,\, \left(t - \tau\right)} \nonumber \\
    &+ B_1\left(\mathcal{O}, \vec{p}^2\right) \,\, e^{- \Delta E_{\vec{p}^2} \,\, \tau},
\end{align}
where the ratio $R$ explicitly depends on $\vec{p}^2$, the operator $\mathcal{O} \in \left\{ \mathcal{O}_{\rm v2a}, \mathcal{O}_{\rm v2b} \right\}$, the sink timeslice $t$, and the insertion timeslice $\tau$. For the operator combination $\mathcal{O}_{\rm v2a}^i$, with $i=1,2,3$, we average the three spatial directions and fit to the data using the parameters $B_0$, $B_1\left(\mathcal{O}_{\rm v2a}, \vec{n}^2 = 1\right)$ and the excited state energy is given by $\Delta E_{\vec{n}^2 = 1}$, independent of the operator combination. Note that we require nonzero momentum in direction $i$ for $\mathcal{O}_{\rm v2a}$. Additionally the operator combination $\mathcal{O}_{\rm v2b}$ gives rise to the further excited state amplitudes $B_1\left(\mathcal{O}_{\rm v2b}, \vec{n}^2 \right)$ and the excited state energy $\Delta E_{\vec{n}^2 = 0}$. All in all this yields a simultaneous fit to three operator combinations for each source sink separation, to resolve the individual parameters. The actual fit is performed simultaneously to all source sink separations, cf.\ Tab.~\ref{tab_Cls}. A summary of the individual contributions is given in Tab.~\ref{tab_fits}. One can easily deduce that the ground state contribution $B_0$ is present in all operator combinations while the excited state amplitudes and energies depend on the operator combination and $\vec{n}^2$ respectively. A similar approach holds for the extraction of $a_2$ and $d_2$. However, for $d_2$ the statistical error is much larger and we are not able to resolve reliable excited state energies from the ratios. Therefore we fix the excited state energies $\Delta E_{\vec{n}^2 = 1}$ by an additional, simultaneous fit to the two-point function Eq.~\eqref{eq_specDecomposition_excitedStates} in case of the reduced matrix element $d_2$.
\begin{table}
    \renewcommand{\arraystretch}{1.5}
    \centering
    \caption{\label{tab_fits}Summary of the occurrence of the individual fit parameters in the ansatz Eq.~\eqref{eq_fitAnsatz} for the extraction of $v_2$ (pion) and $a_2$ (rho). A green check mark indicates that the fit parameter is present in the corresponding operator combination, whereas the red crosses indicate that the fit parameter is not present.}%
    \begin{ruledtabular}
        \begin{tabular}{cccc}
            Fit Parameter& $\mathcal{O}_{\rm v2a}(\vec{n}^2 = 1)$ & $\mathcal{O}_{\rm v2b}(\vec{n}^2 = 1)$ & $\mathcal{O}_{\rm v2b}(\vec{n}^2 = 0)$ \\
            \hline
            $B_0$ & \greencheck & \greencheck & \greencheck \\
            $B_1\left(\mathcal{O}_{\rm v2a}, \vec{n}^2 = 1 \right)$ & \greencheck & \redxcheck & \redxcheck \\
            $B_1\left(\mathcal{O}_{\rm v2b}, \vec{n}^2 = 1 \right)$ & \redxcheck & \greencheck & \redxcheck \\
            $B_1\left(\mathcal{O}_{\rm v2b}, \vec{n}^2 = 0\right)$ & \redxcheck & \redxcheck & \greencheck \\
            $\Delta E_{\vec{n}^2 = \, 1}$ & \greencheck & \greencheck & \redxcheck \\
            $\Delta E_{\vec{n}^2 = \, 0}$ & \redxcheck & \redxcheck & \greencheck
        \end{tabular}
    \end{ruledtabular}
\end{table}\par%
In summary this implies that the fits do not only take into account the data points shown in Figs.~\ref{fig_N204_ratio_v2a} and~\ref{fig_N204_ratio_v2b}, but are actually based on a larger data set stemming from various operator and momentum configurations. However, the prefactors in Eqs.~\eqref{eq_v2a} -- \eqref{eq_d2b} are solely determined by the meson masses and corresponding momentum contributions.

As discussed above, we perform the extraction of the ground state matrix elements separately for the connected (left column) and the disconnected (middle and right column) contributions. While analyzing the disconnected contribution we found that the considerable noise on the light and strange quark loops is highly correlated for all included ensembles, cf.\ Tab.~\ref{tab_Cls}. We can use this to our advantage by looking at the non-singlet ($\bar{u} u + \bar{d} d - 2 \bar{s} s$) and singlet ($\bar{u} u + \bar{d} d + \bar{s} s$) flavor combinations instead of the light and strange loops themselves. As depicted impressively in the middle and the right column of Figs.~\ref{fig_N204_ratio_v2a} and~\ref{fig_N204_ratio_v2b} (note the difference in scale), the statistical error is smaller by more than one order of magnitude for the non-singlet operator. For the disconnected contributions we do not see an indication for a significant excited state contribution and, consequently, content ourselves with a constant fit to extract the ground state signal employing the fit strategy described above.\par%
For the disconnected contribution we have data points for a large number of combinations of final times $t$ and insertion times $\tau$. If one plots the data for various insertions times in one plot (cf.\ the grayed out points in Figs.~\ref{fig_N204_ratio_v2a} and~\ref{fig_N204_ratio_v2b}), the statistical scattering of these data points alone can lead to the misconception that the statistical error of the extracted ground state (yellow band) is underestimated. In order to convince the viewer of the plots that this is not the case we also plot the black points, which are obtained by taking the average over data at all insertion times $\tau$ for the given final time $t$.\par%
\subsection{Quark mass dependence and continuum extrapolation\label{sec_Extrapolation}}
As described in Sec.~\ref{sec_LatticeSetup}, the ensembles we analyze have been generated along multiple trajectories in the quark mass plane and at different lattice spacings $a$. We obtain the final results by extrapolating to the continuum limit (at $a=0$) and to physical masses. To this end we employ the parametrization%
\begin{align}
    f(a,m_\pi^2,m_K^2) &= c_0 + c_1 a + c_2 m_\pi^2 + c_3 m_K^2 \,.
\end{align}
Note that we have to use a linear term in the lattice spacing as leading contribution despite the fact that our lattice action is order $a$ improved, because we lack the order $a$ improvement for the inserted currents.\par
\begin{figure*}[tb]
    \includegraphics[]{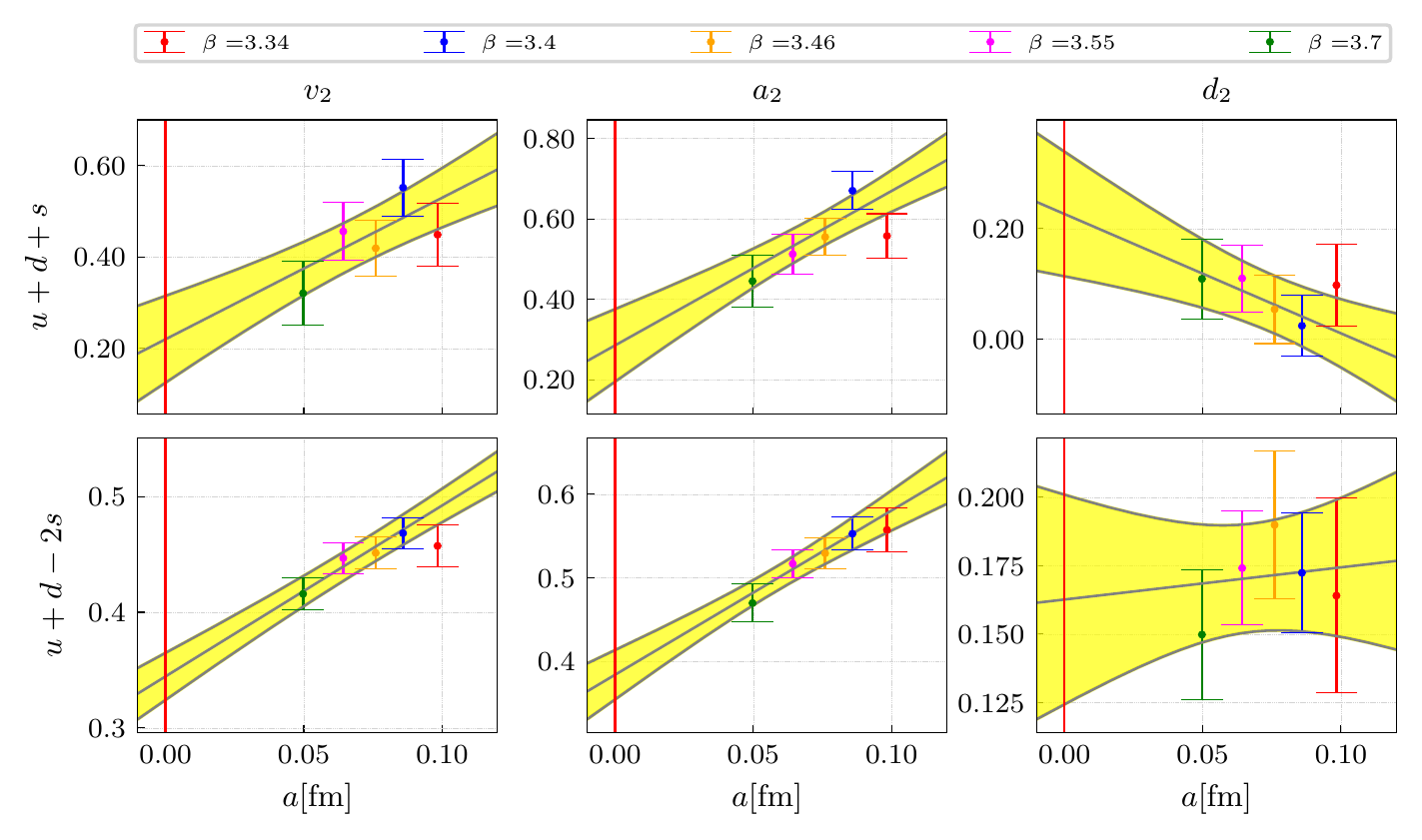}%
    \caption{\label{fig_extrapolation_average_a}
    Lattice spacing dependence of the extrapolations for the flavor singlet ($\bar{u} u + \bar{d} d + \bar{s} s$) and flavor non-singlet ($\bar{u} u + \bar{d} d - 2 \bar{s} s$) operator combinations using the fits shown in, e.g., Figs.~\ref{fig_N204_ratio_v2a} and~\ref{fig_N204_ratio_v2b}. The reduced matrix elements $v_2$ (pion), $a_2$, and $d_2$ (both rho) have been obtained by a translation to physical quark masses and averaging measurements with the same values of $\beta$ using the weighted average given in Eq.~\eqref{eq_weighted_avg}. From coarsest to finest lattice spacing, this corresponds to averaging the data of 2, 8, 7, 7, and 3 independent ensembles, cf.\ Tab.~\ref{tab_Cls}. The non-averaged plots for the individual trajectories can be found in appendix~\ref{sec_AdditionalPlots}.}
\end{figure*}

\begin{figure*}[tbp]
    \centering
    \includegraphics[width=0.9\textwidth,keepaspectratio]{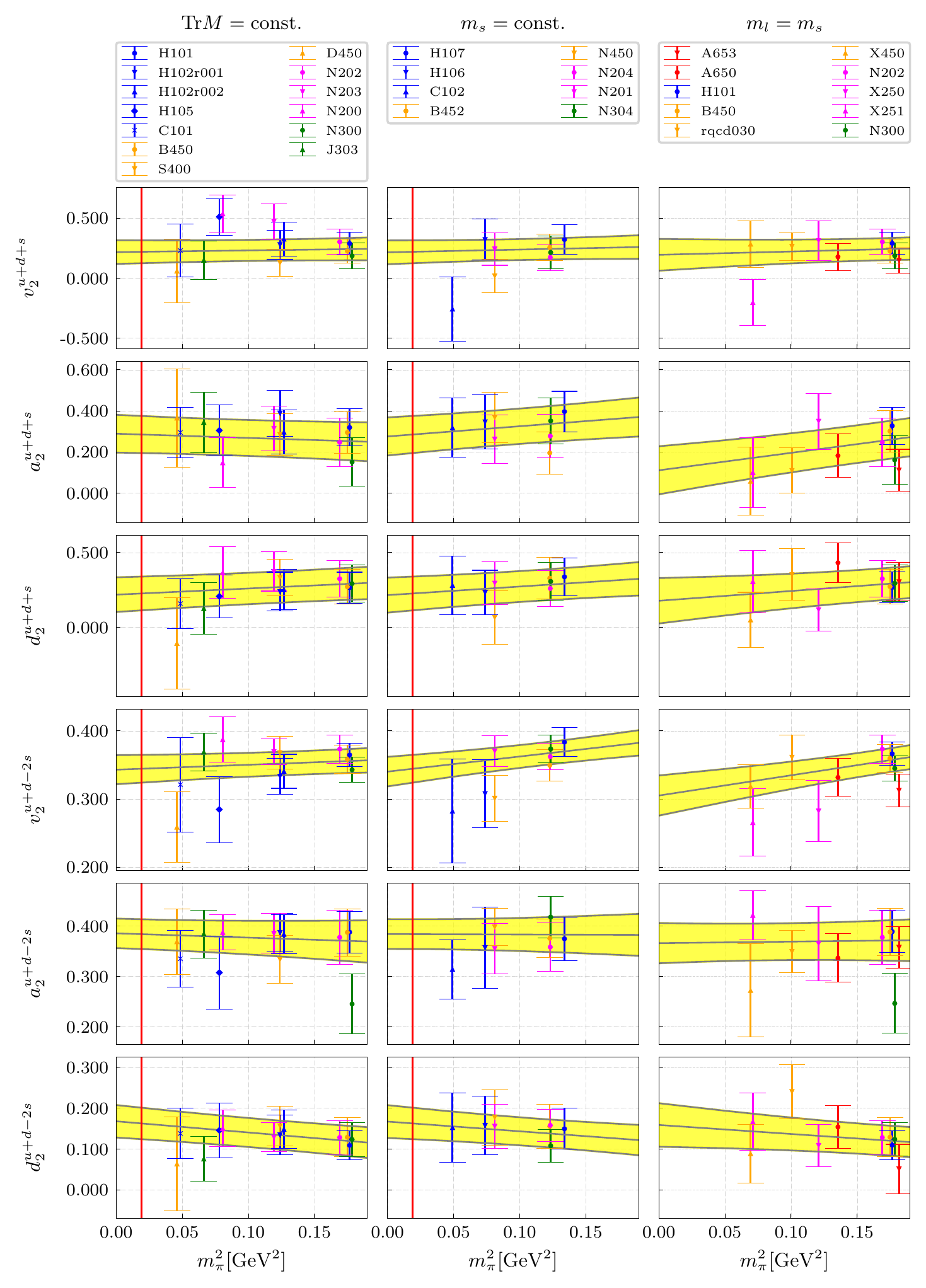}
    \caption{\label{fig_extrapolation_only_mpi}Extrapolation for the flavor singlet ($\bar{u} u + \bar{d} d + \bar{s} s$) and flavor non-singlet operator combinations ($\bar{u} u + \bar{d} d - 2 \bar{s} s$) using the fits shown in, e.g., Fig.~\ref{fig_N204_ratio_v2a}. From left to right we show the $m_{\pi}^2$ dependence of the reduced matrix elements $v_2$ (pion), $a_2$, and $d_2$ (both rho) for the three different trajectories we use in our analysis. Note that the symmetric trajectory $m_l = m_s$ with exact $\rm{SU}(3)$ flavor symmetry approaches the chiral limit in the quark mass plane and not the physical point.}
\end{figure*}

%
In Figs.~\ref{fig_extrapolation_average_a} and~\ref{fig_extrapolation_only_mpi}, the yellow bands show the extrapolations for the flavor non-singlet and flavor singlet operators as a function of $a$ and $m_{\pi}^2$, respectively. As discussed in Sec.~\ref{sec_GSmatrixelements}, the statistical error of the flavor singlet operator combinations is much larger than in the flavor non-singlet case, which makes it hard to draw any convincing conclusions. Nevertheless, the results for the flavor singlet combinations will allow us to give at least an upper bound for the reduced matrix elements $v_2$, $a_2$, and $d_2$. However, for the flavor non-singlet combinations, the situation is much better and we get meaningful results and errors. Fig.~\ref{fig_extrapolation_only_mpi} shows the extrapolations for the quark mass dependence along the ${\rm Tr M} = \text{const.}$, $m_s = \text{const.}$, and $m_l = m_s$ trajectories, from left to right. The data points are corrected for lattice spacing effects and are shifted to the corresponding trajectories. For, e.g., the ${\rm Tr M} = \text{const.}$ trajectory this keeps the average quark mass fixed (again using the fitted model shown above) and thus allows us to see the effect of flavor symmetry breaking. The lattice spacing dependence is depicted in Fig.~\ref{fig_extrapolation_average_a}. To visualize solely the discretization effects, the data points in Fig.~\ref{fig_extrapolation_average_a} are corrected for mass effects (using the fitted model), i.e., they are translated to physical masses along the fitted curve, and finally averaged for all ensembles with the same values of $\beta$ using the weighted average
\begin{align}
    \bar{A}_i &= \sum_i^{N_{\beta}} w_i \, A_i, & & \text{with} & w_i &= \frac{1 / \sigma_i^2}{\sum_i^{N_{\beta}} 1 / \sigma_i^2} \label{eq_weighted_avg},
\end{align}
where $N_{\beta}$ is the number of data points $A_i$ (ground state matrix elements) per $\beta$ with corresponding errors $\sigma_i$. Note that this procedure is only applied to the points in these plots for illustrative purposes, while the bands are obtained from the actual fit performed using the original data points, cf.\ App.~\ref{sec_AdditionalPlots}.\par%
The final results for the reduced matrix elements are given in Tab.~\ref{tab_results}. In addition to the statistical error $()_s$ we provide estimates for the systematic uncertainties due to the quark mass extrapolation $()_m$ and the continuum extrapolation $()_a$. To this end, we have performed additional fits with cuts in the mass range $\bar{m} = \sqrt{(2 m_K^2 + m_{\pi}^2) / 3} < \unit{450}{\mega\electronvolt}$ and $a < \unit{0.09}{\femto\meter}$, respectively. We then take the difference between the results from these fits and our main result as an estimate of the corresponding systematic uncertainties.
\subsection{Discussion\label{sec_Discussion}}
Using Eq.~\eqref{eq_structure_function} in combination with Eqs.~\eqref{eq_moments-of-structure-functions-spin0} and~\eqref{eq_moments-of-structure-functions-spin1}, one can write the ratio of the moments of the structure functions (at leading twist accuracy) with the corresponding Wilson coefficient $C^{(k)}_n = 1 + \mathcal{O}(\alpha_s)$ as a sum over the related reduced matrix element%
\begin{align}
    \frac{2 M_2(F_1)_\pi}{C_2^{(1)}}  = \frac{M_1(F_2)_\pi}{C_2^{(2)}}  &= \sum_q e_q^2 v_2^q = \frac{2}{9} \Bigl(v_2^{\rm fs} + \frac{1}{4} v_2^{\rm fns} \Bigr) \,, \label{eq_f1pi} \\
    \frac{2 M_2(F_1)_\rho}{C_2^{(1)}} = \frac{M_1(F_2)_\rho}{C_2^{(2)}} &= \sum_q e_q^2 a_2^q = \frac{2}{9} \Bigl(a_2^{\rm fs} + \frac{1}{4} a_2^{\rm fns} \Bigr) \,, \label{eq_f1rho}\\
    \frac{2 M_2(b_1)_\rho}{C_2^{(1)}} = \frac{M_1(b_2)_\rho}{C_2^{(2)}} &= \sum_q e_q^2 d_2^q = \frac{2}{9} \Bigl(d_2^{\rm fs} + \frac{1}{4} d_2^{\rm fns} \Bigr) \,, \label{eq_b1rho}
\end{align}
where we assume exact isospin symmetry and ${\rm fs} \equiv u+d+s$ is the flavor singlet while ${\rm fns} \equiv u+d-2s$ is the flavor non-singlet combination. If we only consider the connected part, the strange quark contribution drops out entirely. The result can be written in terms of the light quark connected contribution as
\begin{align}
    \frac{2M_2(F_1)_\pi^{\rm conn.}}{C_2^{(1)}}  = \frac{M_1(F_2)_\pi^{\rm conn.}}{C_2^{(2)}}  &=\frac{5}{9} v_2^{\ell, {\rm conn.}}\,, \label{eq_f1pi_conn} \\
    \frac{2M_2(F_1)_\rho^{\rm conn.}}{C_2^{(1)}} = \frac{M_1(F_2)_\rho^{\rm conn.}}{C_2^{(2)}} &=\frac{5}{9} a_2^{\ell, {\rm conn.}}\,, \label{eq_f1rho_conn} \\
    \frac{2M_2(b_1)_\rho^{\rm conn.}}{C_2^{(1)}} = \frac{M_1(b_2)_\rho^{\rm conn.}}{C_2^{(2)}} &=\frac{5}{9} d_2^{\ell, {\rm conn.}}\,. \label{eq_b1rho_conn}
\end{align}
\begin{table}
    \renewcommand{\arraystretch}{1.5}
    \centering
    \caption{\label{tab_results}Results obtained from the extrapolations in Figs.~\ref{fig_extrapolation_average_a} and~\ref{fig_extrapolation_only_mpi} and the coresponding connected-only contributions for the flavor combination $(u+d)$, all at $\mu = \unit{2}{\giga\electronvolt}$. The final statistical error is given by $()_s$ and estimates of the systematic uncertainties due to the quark mass by $()_m$, and due to the continuum extrapolation by $()_a$. The values of $\chi^2$ per degrees of freedom are obtained from the corresponding extrapolations.}%
    \begin{ruledtabular}
        \begin{tabular}{clc}
            Matrix element & Final result & $\chi^2 / \rm{d.o.f.}$\\
            \hline
            $v_2^{(u+d+s)}$ & 0.220 $( 95 )_s ( 98 )_m ( 155 )_a$ & 1.71 \\
            $a_2^{(u+d+s)}$ & 0.285 $( 90 )_s ( 76 )_m ( 271 )_a$ & 1.78 \\
            $d_2^{(u+d+s)}$ & 0.226 $( 112 )_s ( 6 )_m ( 54 )_a$ & 0.72 \\
            $v_2^{(u+d-2s)}$ & 0.344 $( 20 )_s ( 3 )_m ( 19 )_a$ & 1.72 \\
            $a_2^{(u+d-2s)}$ & 0.384 $( 29 )_s ( 11 )_m ( 42 )_a$ & 1.33 \\
            $d_2^{(u+d-2s)}$ & 0.163 $( 38 )_s ( 5 )_m ( 7 )_a$ & 0.58 \\
            $v_2^{(u+d),\rm{conn.}}$ & 0.357 $( 16 )_s ( 2 )_m ( 15 )_a$ & 1.74 \\
            $a_2^{(u+d),\rm{conn.}}$ & 0.393 $( 29 )_s ( 10 )_m ( 35 )_a$ & 1.47 \\
            $d_2^{(u+d),\rm{conn.}}$ & 0.180 $( 38 )_s ( 5 )_m ( 7 )_a$ & 0.59 \\
        \end{tabular}
    \end{ruledtabular}
\end{table}
In Tab.~\ref{tab_structure_function_results} we give our final results for these linear combinations. As shown in the last section the flavor singlet contributions contain relatively large errors which affect Eqs.~\eqref{eq_f1pi}--\eqref{eq_b1rho}. However, treating the connected part only\footnote{The connected only results are presented as a comparison option for other studies neglecting disconnected contributions.} reduces the errors significantly. The reader should be aware of the fact that we use (in both cases) the flavor non-singlet renormalization constants, which is only an approximation, cf.\ the discussion in Sec.~\ref{sec_Renormalization}.\par%
For the flavor non-singlet contributions given in Tab.~\ref{tab_results} we obtain very precise results, despite the fact that all disconnected quark loops are fully taken into account. The non-singlet operators do not mix with gluonic operators under renormalization, and the necessary renormalization factors have been calculated non-perturbatively, cf.\ Sec.~\ref{sec_Renormalization}. As a first step one can compare the connected results in Tab.~\ref{tab_structure_function_results} to the connected-only and the flavor non-singlet results in Tab.~\ref{tab_results}. Multiplying the latter two by the prefactors given in Eqs.~\eqref{eq_f1pi_conn}--\eqref{eq_b1rho_conn} one finds that both the first moments of the unpolarized structure functions $F_1^{\pi}$ and $F_1^{\rho}$ and the first moment of the structure function $b_1$ are in very good agreement with the flavor non-singlet results in Tab.~\ref{tab_results} and of course also reflect the connected only result in Tab.~\ref{tab_results}. As shown in Sec.~\ref{sec_GSmatrixelements} the flavor non-singlet contributions of the quark line disconnected diagrams in the extraction of the ground state matrix elements are small compared to the connected contributions. In leading order the structure function $F^q_1(x)$ corresponds to one half of the probability to find a quark of flavor $q$ with momentum fraction $x$. If we assume exact $\rm{SU}(3)$ flavor symmetry for the quark sea, the results in Tab.~\ref{tab_results} and~\ref{tab_structure_function_results} imply that in the pion the valence quarks carry about 35\% of the total momentum, while in the rho they carry about 40\% of the total momentum. It is remarkable that these values justify the assumption $F_1(x)^{\pi} \sim F_1(x)^{\rho}$, which is often used in phenomenological estimates. The structure functions $b_1(x)$ and $b_2(x)$ are sensitive to a possible dependence of the quark densities on the hadron polarization, i.e., they measure the difference in quark distributions of a spin projected $\lambda = 0$ and $\lambda=+/-$ rho meson. If the quarks were in a relative S-wave state, cf.\ the discussion in Sec.~\ref{sec_TwoPionStates}, one would expect $b_1 = b_2 = 0$. However, our results show a large contribution (compared to the scale of $a_2$) to the approximated valence quark contribution $d_2$ with a relative error of only $\sim$10\%. This confirms the conclusion in~\cite{Best:1997qp} that the quarks carry substantial angular momentum and also reflects the results of the various phenomenological studies cited in Sec.~\ref{sec_Introduction}.
\section{Summary and outlook\label{sec_Summary}}
In this article we have presented the computation of the first moments for the structure functions $F_1^{\pi}$, $F_1^{\rho}$, and $b_1$ including quark line disconnected contributions. Despite the fact that our final results are tainted with large statistical errors due to the flavor singlet disconnected contributions we are able to provide very accurate results for the flavor non-singlet combination $u+d-2s$. As an additional subtlety we had a closer look at possible two-pion contributions which might occur in our analysis. We do not find any evidence for the contribution of two-pion states. However, as discussed at the end of Sec.~\ref{sec_TwoPionStates}, using our analysis technique we cannot fully exclude them either. This is particularly true for two-pion states close to the resonance energy. To this end, the trustworthiness of our numerical results could be enhanced a posteriori, if future studies (e.g., by using the generalized eigenvalue method with two-pion interpolating currents, cf.\ refs.~\cite{Luscher:1990ck,Bali:2015gji,Erben:2019nmx,Fischer:2020fvl,Fischer:2020bgv}) can show that the overlap of smeared vector interpolating currents with the two-pion states is much smaller than for the local currents.\par%
\begin{table}
    \renewcommand{\arraystretch}{1.5}
    \centering
    \caption{\label{tab_structure_function_results}Estimated results for the first moments of the structure functions $F_1$ and $b_1$ exploiting Eqs.~\eqref{eq_f1pi}--\eqref{eq_b1rho_conn} and performing the extrapolations as discussed in Sec.~\ref{sec_Extrapolation}. Further we use the abbreviation $C \equiv 2 / C_2^{(1)}$.}%
    \begin{ruledtabular}
        \begin{tabular}{llc}
            Structure function & Final result & $\chi^2 / \rm{d.o.f.}$\\
            \hline
            $C M_2(F_1)_{\pi}^{}$ & 0.132 $( 33 )_s ( 32 )_m ( 57 )_a$ & 1.75 \\
            $C M_2(F_1)_{\rho}^{}$ & 0.156 $( 33 )_s ( 23 )_m ( 102 )_a$ & 1.84 \\
            $C M_2(b_1)_{\rho}^{}$ & 0.108 $( 41 )_s ( 1 )_m ( 13 )_a$ & 0.72 \\
            $C M_2(F_1)_{\pi}^{\text{conn.}}$ & 0.099 $( 5 )_s ( 0 )_m ( 4 )_a$ & 1.74 \\
            $C M_2(F_1)_{\rho}^{\text{conn.}}$ & 0.109 $( 8 )_s ( 2 )_m ( 9 )_a$ & 1.47 \\
            $C M_2(b_1)_{\rho}^{\text{conn.}}$ & 0.050 $( 10 )_s ( 1 )_m ( 2 )_a$ & 0.59 \\
        \end{tabular}
    \end{ruledtabular}
\end{table}
Despite the fact that we for the first time present comprehensive results including disconnected contributions we have reduced the statistical error considerably. This can be seen comparing the error of the connected contribution alone with earlier studies. However, to determine the phenomenologically important moments of the structure functions (at leading twist), one needs the flavor singlet combination, where the statistical error is still large. Future studies will have to aim at a further reduction of these statistical errors. Once this is achieved, also a nonperturbative calculation of the singlet renormalization factors and the inclusion of mixing with gluonic operators might be worthwhile.
\begin{acknowledgments}
Support of this project was granted by the German DFG (SFB/TRR 55). In addition this project has received funding from the European Union’s Horizon 2020 research and innovation program under the Marie Skłodowska-Curie grant agreement No. 813942. We are grateful to Gunnar S. Bali, Lorenzo Barca, Sara Collins, Vladimir Braun, Meinulf G{\"o}ckeler, and Christoph Lehner for the various fruitful discussions and to Wolfgang S{\"o}ldner for providing parts of the intermediate results to be published in~\cite{Bali:2023}. In addition we thank Benjamin Gl{\"a}{\ss}le and Simon Heybrock for co-developing some of the software used here and Daniel Richtmann, Peter Georg and Jakob Simeth for software development and software support and Enno E. Scholz for the support in data management. We gratefully acknowledge computing time granted by the John von Neumann Institute for Computing (NIC), provided on the Booster partition of the supercomputer JURECA~\cite{Juelich:2018} and use of computing time and services on the HDF Cloud~\cite{hdf}, funded as part of the Helmholtz Data Federation (HDF) strategic initiative, at J{\"u}lich Supercomputing Centre (\href{www.fz-juelich.de}{JSC}). Additional simulations were carried out at the QPACE2 and QPACE 3 Xeon Phi cluster of SFB/TRR 55 und the Regensburg computing cluster QPACE B and the Regensburg HPC cluster Athene2. We owe special thanks to Randy R{\"u}ckner for the software and runtime support concerning the Athene2 compute cluster. The authors also gratefully acknowledge the Gauss Centre for Supercomputing (GCS) for providing computing time for GCS Large-Scale Projects on SuperMUC and SuperMUC NG at Leibniz Supercomputing Centre (\href{https://www.lrz.de}{LRZ}). GCS is the alliance of the three national supercomputing centres HLRS (Universit{\"a}t Stuttgart), JSC (Forschungszentrum J{\"u}lich), and LRZ (BayerischeAkademie der Wissenschaften), funded by the German Federal Ministry of Education and Research (BMBF) and the German State Ministries for Research of Baden-W{\"u}rttemberg(MWK), Bayern (StMWFK) and Nordrhein-Westfalen (MIWF). We thank all our CLS colleagues for the joint generation of the gauge ensembles. The ensembles were generated using the \href{https://luscher.web.cern.ch/luscher/openQCD/)}{OpenQCD}~\cite{Luscher:2012av} software package.\par%
\end{acknowledgments}
\appendix
\section{Light-cone coordinates and polarization vectors\label{app_coordinates_and_polarization_vectors}}
We define the light-cone coordinates used in Sec.~\ref{sec_GeneralProperties} in such a way that the perpendicular (or transverse) part of the momentum always vanishes, i.e., $p_T = 0$ and $p_+$ is its large component. To achieve this let $v$ be any four-vector and $\hat{\vec{p}}$ the direction of the three-momentum. Then,%
\begin{align}%
    v^\mu &= v^+ n_+^\mu + v^- n_-^\mu +  v_T^\mu \,, & \text{ with }
    n_\pm^\mu &= \frac{1}{\sqrt2}   \begin{pmatrix}
                                        1 \\ \pm \hat{\vec{p}}
                                    \end{pmatrix}^\mu ,
\end{align}%
where $v^\pm = n_\mp \cdot v$ and%
\begin{align}%
    v_T^\mu &=  \begin{pmatrix}
                    0 \\ \vec{v_T}
                \end{pmatrix}^\mu\,, & \text{ with }
    \vec v_T &= \vec v -  (\hat{\vec{p}} \cdot \vec{v}) \hat{\vec{p}} \,,
\end{align}%
such that $\vec v_T \perp \hat{\vec{p}}$. For the momentum we then have $p^\pm=E\pm|\vec{p}|$ and $p_T=0$.\par%
For the polarization vectors we use a dimensionless definition. They obey the general transversality condition%
\begin{align}%
    \sum_{\lambda} \, \epsilon_{\mu}(\vec{p}, \lambda) \epsilon^{*}_{\nu}(\vec{p}, \lambda) = - \left(g_{\mu \nu} - \frac{p_{\mu} p_{\nu}}{m^2} \right) \,,
    \label{eq_trans-condition}
\end{align}%
where $m$ is the hadron mass. For momenta in $x$ direction the polarization vectors in the rest frame are given by $(0,\vec e_\lambda)$ with%
\begin{align}
    \vec e_0^x = \vec e_0 \biggr|_{\hat {\vec p} = \hat{\vec e}_1} &=   \begin{pmatrix}
                                                                            1 \\ 0 \\ 0
                                                                        \end{pmatrix} \,, &
    \vec e_\pm^x = \vec e_\pm \biggr|_{\hat {\vec p} = \hat{\vec e}_1} &= \frac{1}{\sqrt{2}}    \begin{pmatrix}
                                                                                                    0 \\ \mp 1 \\ -i
                                                                                                \end{pmatrix} \,,
\end{align}
where $\vec e_0^x$ corresponds to the longitudinal polarization, while $\vec e_\pm^x$ to the circular polarizations. For momenta in arbitrary direction, we have to rotate these vectors to%
\begin{align}
    \vec e_0 &= \hat{\vec p} \,, &
    \vec e_\pm &= \vec e_\pm^x - \frac{\hat {\vec p} \cdot \vec e_\pm^x}{1+ \hat {\vec p} \cdot \vec e_0^x } (\vec e_0^x + \hat {\vec p}) \,,
\end{align}
to obtain the longitudinal polarization vector $\vec e_0$ and the polarization vectors for the circular polarizations $\vec e_\pm$. Last but not least, we have to perform a boost to the laboratory frame. This only affects $\vec e_0$ ($\vec e_\pm$ are invariant because they are perpendicular to $\vec p$), and we obtain%
\begin{align}
    \epsilon^\mu(\vec p, 0) &=  \begin{pmatrix}
                                    \frac{| \vec p |}{m} \\ \frac{E}{m} \hat{\vec{p}}
                                \end{pmatrix}^\mu \,, &
    \epsilon^\mu(\vec p, \pm) &=    \begin{pmatrix}
                                        0 \\ \vec e_\pm
                                    \end{pmatrix}^\mu \,. &
\end{align}
In terms of the light-cone coordinates introduced above this yields%
\begin{align}
    \epsilon^{\pm}(\vec{p},0) &= \pm p^\pm/m \,, &
    \epsilon^{\mu}_T(\vec{p},0) &= 0 \,, \\
    \epsilon^{\pm}(\vec{p},+) &= 0 \,, &
    \epsilon^{\mu}_T(\vec{p},+) &= \epsilon^{\mu}(\vec{p},+) \,, \\
    \epsilon^{\pm}(\vec{p},-) &= 0\,, &
    \epsilon^{\mu}_T(\vec{p},-) &= \epsilon^{\mu}(\vec{p},-) \,.
\end{align}
\par
\section{\label{app_Operators}Operators and matrix element decomposition}
To avoid mixing as far as possible we use operators from suitably chosen multiplets that possess a definite C-parity and transform according to irreducible representations of~$\mathrm{H}(4)$, cf.\ refs.~\cite{Gockeler:1996mu,Best:1997qp}. To be specific, we will use the operators $\mathcal O_{\rm v2a}^i = \mathcal O^{0i}$ and $\mathcal O_{\rm v2b} = \frac{4}{3} \mathcal O^{00}$. For the special case of two indices the action of the symmetrizing and trace-subtracting operator $\mathcal S$ is defined by%
\begin{align}
    \mathcal S \mathcal O^{\mu\nu} &= \mathcal S^{\mu\nu}_{\rho\sigma} \mathcal O^{\rho\sigma} = \frac12 \Bigl( g^\mu_\rho g^\nu_\sigma + g^\mu_\sigma g^\nu_\rho - \frac{2}{d} g^{\mu\nu} g_{\rho\sigma} \Bigr) \mathcal O^{\rho\sigma} \,.
\end{align}
Plugging this into the Lorentz decomposition Eq.~\eqref{eq_pseudoscalar-matrix-element} for the pion one finds
\begin{align}%
    \bra{\vec p} \mathcal O_{\rm v2a}^i \ket{\vec p} &= 2 v^q_2 \, E p^i \,, \label{eq_ov2a_v2} \\
    \bra{\vec p} \mathcal O_{\rm v2b} \ket{\vec p} &= 2 v^q_2 \, \frac{4 E^2 - m^2}{3} \,. \label{eq_ov2b_v2}
\end{align}%
Using Eq.~\eqref{eq_vector-matrix-element}, we can show for the rho that%
\begin{align}
    \langle \bm p, \lambda | \mathcal O_{\rm v2a}^i | \bm p, \lambda \rangle
    &=  2 E p^i     \begin{cases}
                        a^q_2 + \frac{2}{3} d^q_2 & \text{ for } \lambda = 0 \\
                        a^q_2 - \frac{1}{3} d^q_2 & \text{ for } \lambda=\pm
                    \end{cases} \,, \label{eq_ov2a_a2d2} \\
    \begin{split}
        \langle \bm p, \lambda | \mathcal O_{\rm v2b} | \bm p, \lambda \rangle
        &= 2 \Bigl(a^q_2 - \frac{d^q_2}{3}\Bigr)\frac{4E^2 - m^2}{3} \taghere \\
        &\quad+ 2 \frac{d^q_2}{3}
        \begin{cases}
            4 E^2 - 3 m^2 & \text{ for } \lambda = 0 \\
            m^2 & \text{ for } \lambda=\pm
        \end{cases} \,. \label{eq_ov2b_a2d2}
    \end{split}
\end{align}
In the actual computations we use the explicit operators
\begin{align}
    \mathcal O_{\rm v2a}^i &= O^{\{0i\}}, \quad \quad \,\,\,\, \text{with } i = 1,2,3, \\
    \mathcal O_{\rm v2b} &= O^{00} + \frac{1}{3} \bigl( O^{11} + O^{22} + O^{33} \bigr),
\end{align}
where $O^{\mu \nu}$ is defined as
\begin{align}
    O_{\mu \nu} = \frac{i}{2} \, \bar{q} \, \gamma^{\mu} \, \overleftrightarrow{D}^{\nu} \, q.
\end{align}
The conversion between Minkowski and Euclidean convention is finally given by
\begin{align}
    \mathcal O_{\rm v2a}^{(E), i} = -i \, \mathcal O_{\rm v2a}^i\,\,, & & \mathcal O_{\rm v2b}^{(E)} = - \mathcal O_{\rm v2b}\,\,,
\end{align}
see ref.~\cite{Best:1997qp}.
\section{\label{sec_NumericalMethods}Numerical methods}%
\subsection{\label{sec_StochasticPropagator}Stochastic propagator}%
\begin{figure}%
    \includegraphics[scale=0.3]{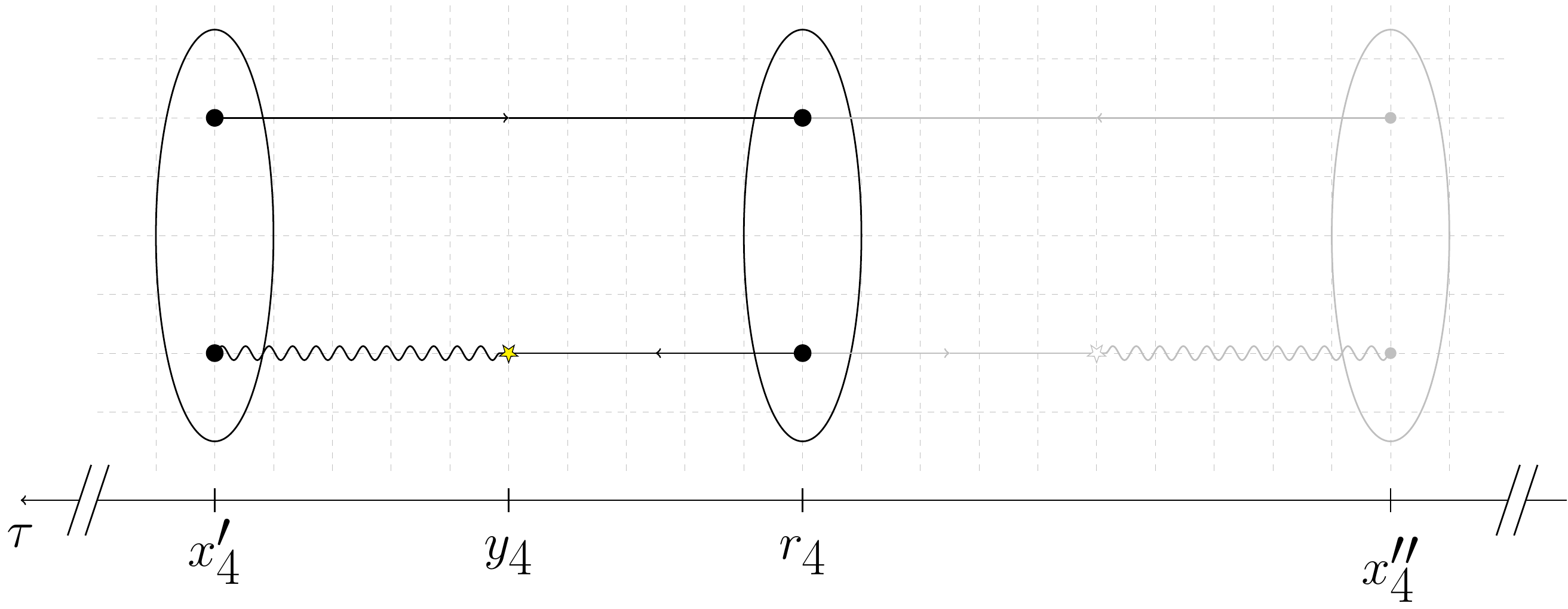}
    \caption{\label{fig_sto_threept_fb}Sketch of a generic meson three-point function in forward and backward direction. The source timeslice is $r_4$, the backward/forward sink timeslice $x_4^\prime$/$x_4^{\prime\prime}$ and the current is located at timeslice $y_4$. While the solid lines represent point-to-all propagators the wiggly line illustrates the stochastic timeslice-to-all propagator connecting the sink and the operator insertion.}
\end{figure}%
Using the common sequential source method~\cite{Martinelli:1988rr} the computational cost of evaluating three-point functions is high because a new inversion is necessary for each sink setup (timeslice, momentum and interpolating current). To reduce the computational cost and maximize synergies in calculating matrix elements we implemented a stochastic algorithm~\cite{Bali:2017mft,Rodl:2020ulu} which circumvents this limitations. The implementation we propose parallelizes the computations in such a way that those for multiple source positions and multiple insertion positions can be done simultaneously. A similar approach was already used in~\cite{Evans:2010tg, Alexandrou:2013xon, Bali:2013dpa, Yang:2015zja}, however, by storing the uncontracted data, with all spin indices open, on disk, our implementation enables the user to analyze any channel of interest at a later stage.\par%
First of all we factorize the three-point correlation function into two largely independent parts denoted as spectator $S$ and insertion $I$ part which can be computed separately. The generic expression of our factorized meson three-point function with open spin indices (Greek letters) as well as color, stochastic and flavor indices reads%
\begin{align}%
    \begin{split}
        \MoveEqLeft[1] C_{\rm{3pt}}(\vec{p}^\cprime, \vec{q}, x_4^\prime, y_4)_{f_1, f_2, f_3}^{\alpha^\cprime \beta^\cprime \Tilde{\alpha} \Tilde{\beta} \beta \alpha}
        = \Gamma_{\rm{snk}}^{\alpha^\cprime \beta^\cprime} \, \Gamma_{\rm{ins}}^{\tilde{\alpha} \tilde{\beta}} \, \Gamma_{\rm{src}}^{\beta \alpha} \\
        &\times \frac{1}{N_{\rm{sto}}} \sum_{i = 1}^{N_{\rm{sto}}} S_{i, f_1}(\vec{p}^\cprime,x^\prime_4)^{\beta^\cprime \alpha^\cprime \alpha}_{a}
        \, I_{i, f_2, f_3}(\vec{q},y_4)^{\tilde{\alpha} \tilde{\beta} \beta}_{a} \,. \taghere
    \end{split}
\end{align}%
$\Gamma_{\rm{src/snk}}$ corresponds to the interpolating currents of the (smeared) meson source or sink and $\Gamma_{\rm{ins}}$ corresponds to the local operator insertion which can contain additional derivatives (at the moment only first derivatives are implemented). We define the spectator $S_{i, f_1}(\vec{p}^\prime, x^\prime_4)^{\beta^\prime \alpha^\prime \alpha}_{a}$ and the insertion $I_{i, f_2, f_3}(\vec{q},y_4)^{\tilde{\alpha} \tilde{\beta} \beta}_{a}$ parts as%
\begin{align}%
    \begin{split}
        \MoveEqLeft[1] S_{i, f_1}(\vec{p}^\cprime, x^\prime_4)^{\beta^\cprime \alpha^\cprime \alpha}_{a} = \\
        &\sum_{\vec{x}^\prime} \delta_{a^\cprime b^\cprime} \Bigl[ \eta_i(x^\prime) \gamma_5 \Bigr]^{\beta^\cprime}_{b^\cprime}
        \, \Bigl[ \gamma_5 G_{f_1}^{\dagger}(x^\prime,r) \gamma_5 \Bigr]^{\alpha^\cprime \alpha}_{a^\cprime a}
        \, e^{-i \vec{p}^\cprime \cdot \vec{x}^\cprime} \,,\\
        \taghere
    \end{split}
    \label{eq_spectator} \\
    \begin{split}
        \MoveEqLeft[1] I_{i, f_2, f_3}(\vec{q},y_4)^{\tilde{\alpha} \tilde{\beta} \beta}_{a} = \\
        &\sum_{\vec{y}} \delta_{ab} \delta_{\tilde{a} \tilde{b}}  \Bigl[ \gamma_5 s_{i, \, f_2}(y)  \Bigr]^{*\tilde{\alpha}}_{\phantom{*}\tilde{a}}
        \, G_{f_3}(y,r)^{\tilde{\beta} \beta}_{\tilde{b} b}
        \, e^{i \vec{q} \cdot \vec{y}} \,, \taghere
    \end{split}
    \label{eq_insertion}
\end{align}%
where we assume that the spatial source is located at the origin without loss of generality. In Fig.~\ref{fig_sto_threept_fb} we show a sketch of a generic meson three-point function to further illustrate the factorization. We have two quark propagators $G_{f_i}$ in Eqs.~\eqref{eq_spectator} and~\eqref{eq_insertion} depicted as solid lines connecting the source position $r$ with all other points of the lattice. These point-to-all propagators are computed using the solver methods introduced in Sec.~\ref{sec_LatticeSetup}. The third propagator connecting the sink timeslice with the insertion current is plotted as a wiggly line and estimated by%
\begin{align}%
    G_{f_2}(y, x^\prime)^{\tilde{\alpha} \beta^\cprime}_{\tilde{a} b^\cprime} &\approx \frac{1}{N} \sum_{i=1}^{N_{\rm{sto}}} s_{i, \, f_2}(y)^{\tilde{\alpha}}_{\tilde{a}} \,\, \eta_i^{*} (x^\prime)^{\beta^\cprime}_{b^\cprime},
\end{align}%
where the sum runs over $N_{\rm{sto}}$ realizations of the noise vector~$\eta_i(x^\prime)$, with the properties%
\begin{align}%
    \frac{1}{N} \sum_{i=1}^{N} \eta_i(x)^{\alpha}_a
    &= 0 + \mathcal{O} \left(\frac{1}{\sqrt{N}} \right) \,,\\
    \frac{1}{N} \sum_{i=1}^{N} \eta_i (x)^{\alpha}_a \eta_i^{*} (x^{\prime})^{\alpha^\prime}_{a^\prime}
    &= \delta_{xx^\prime} \delta_{\alpha \alpha^\prime} \delta_{aa^\prime} + \mathcal{O} \left(\frac{1}{\sqrt{N}} \right) \,.
\end{align}%
In the implementation presented in this work we use time partitioned $\mathbb{Z}_2$~\cite{Dong:1993pk} noise vectors~$\eta_i(x)$ which are set to zero unless $x_4 = x_4^{\prime\prime}$ or $x_4 = x_4^\prime$. Seeding the noise vectors in forward and backward temporal direction enables us to increase statistics by a factor of two with only little computational overhead. Moreover, the insertion part of our factorization is constructed such that it can be reused in the calculation of baryon three-point functions by contracting an appropriate spectator part. The results in \cite{Bali:2019svt} for the baryon computation look very promising. In this work we showed that this holds for the meson computations as well and that the stochastic approach is a serious alternative to the sequential source method. For  more details about the implementation the interested reader is referred to~\cite{Bali:2017mft}.
\subsection{\label{sec_DisconnPropagator}Disconnected contributions}%
\begin{figure}%
    \includegraphics[scale=0.5]{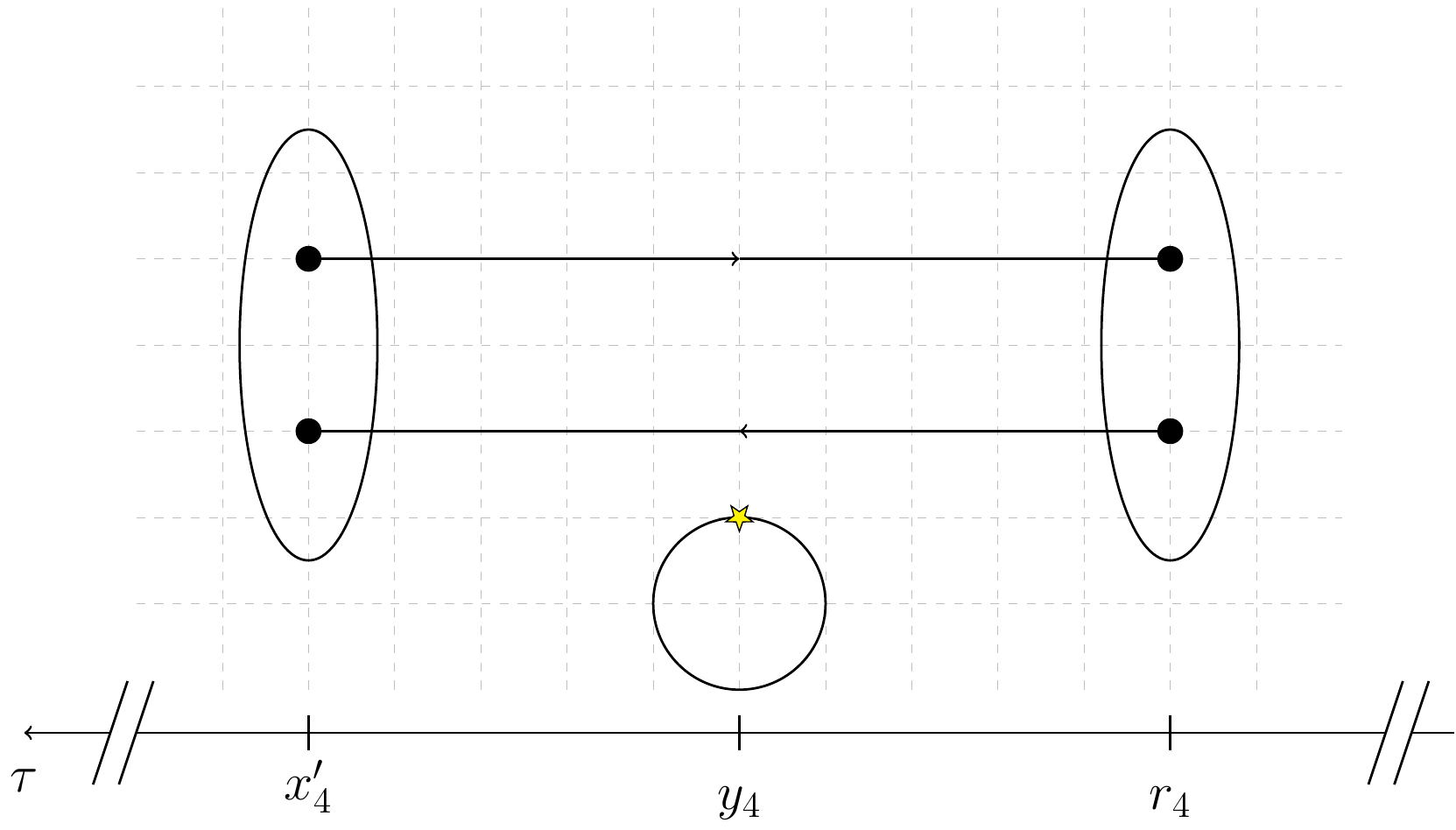}
    \caption{\label{fig_threept_discon}Sketch of a generic meson disconnected three-point function. The source timeslice is $r_4$, the sink timeslice $x_4^\prime$ and the current (loop) is located at timeslice $y_4$. While the solid lines represent point-to-all propagators the solid circle represents a quark loop.}
\end{figure}%
\begin{figure*}[tbp]
    \centering
    \includegraphics[width=0.9\textwidth,keepaspectratio]{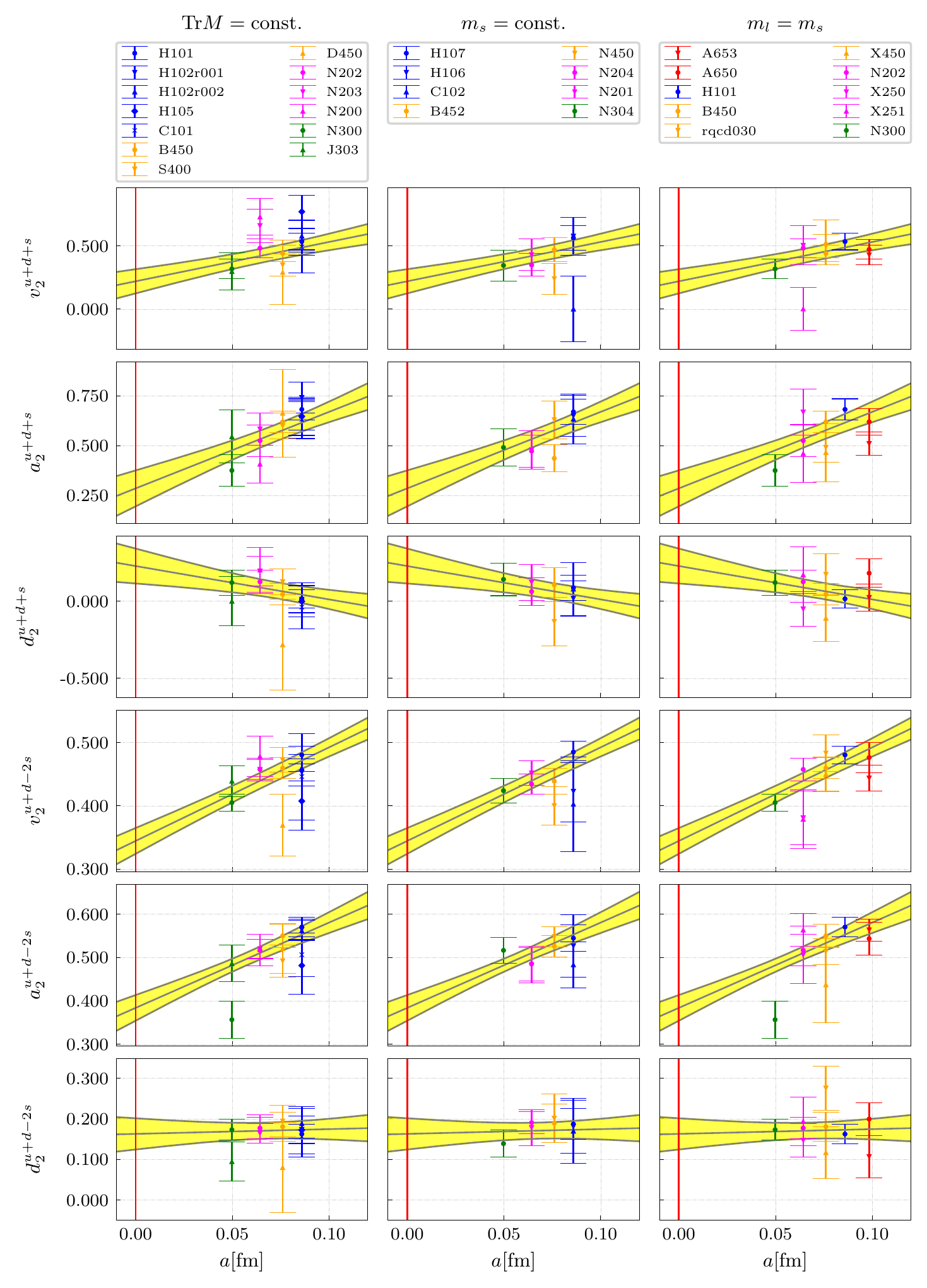}
    \caption{\label{fig_extrapolation_only_a}Extrapolation for the flavor singlet operator ($\bar{u} u + \bar{d} d + \bar{s} s$) and the flavor non-singlet operator ($\bar{u} u + \bar{d} d - 2 \bar{s} s$) using the fits shown in, e.g., Fig.~\ref{fig_N204_ratio_v2a}. The three different columns show the extrapolation for the lattice spacing dependence along the ${\rm Tr M} = \text{const.}$, $m_s = \text{const.}$, and the $m_l = m_s$ trajectories for the reduced matrix elements $v_2$, $a_2$, and $d_2$.}
\end{figure*}%
In addition to connected contributions to the three-point function, which we treated in Sec.~\ref{sec_StochasticPropagator}, we also compute the disconnected contributions illustrated in Fig.~\ref{fig_threept_discon}. To get the disconnected contribution we multiply the two-point function by a disonnected loop $L(\tau)$~\cite{Bali:2016lvx} which reads
\begin{align}%
    \MoveEqLeft[1] C^{\rm{discon}}_{\rm{3pt}}(\vec{p}^\cprime, \vec{q}, x_4^\prime, y_4) = \nonumber\\
    &\, \langle C^{\rm{c}}_{\rm{2pt}}(x_4^\prime, r_4) L^{\rm{c}}(y_4) \rangle_{\rm{c}} - \langle C^{\rm{c}}_{\rm{2pt}}(x_4^\prime, r_4) \rangle_{\rm{c}} \langle L^{\rm{c}}(y_4) \rangle_{\rm{c}},
\end{align}%
where $\langle \rangle_{\rm{c}}$ makes the configuration average explicit. The loop is given by
\begin{align}%
    L^{\rm{c}}(y_4) = \sum_{\vec{y}} \, \Tr[G_{f}(y, y) \Gamma_{\rm{ins}} ],
    \label{eq_loop}
\end{align}%
where $G_f$ denotes a quark propagator of flavor $f$ with $y = (y_4, \vec{y})$. Note that we set $r_4 = 0$ without loss of generality. To compute the propagator in Eq.~(\ref{eq_loop}) we use stochastic estimators similar to the approach presented in the last section and the corresponding solvers introduced in Sec.~\ref{sec_LatticeSetup}. More details on our approach of the computation of disconnected loops are given in~\cite{Bali:2009hu}.
\section{\label{sec_AdditionalPlots}Additional plots}%
In Fig.~\ref{fig_extrapolation_only_a} we explicitly show the $a$ dependence of the global extrapolation plots for the reduced matrix elements $v_2$, $a_2$, and $d_2$. We still use physical masses but in contrast to the main text we do not average over equal lattice spacings at this point.
\newpage

\bibliography{meson3pt}

\providecommand{\noopsort}[1]{}\providecommand{\singleletter}[1]{#1}%
\begin{thebibliography}{109}%
\makeatletter
\providecommand \@ifxundefined [1]{%
 \@ifx{#1\undefined}
}%
\providecommand \@ifnum [1]{%
 \ifnum #1\expandafter \@firstoftwo
 \else \expandafter \@secondoftwo
 \fi
}%
\providecommand \@ifx [1]{%
 \ifx #1\expandafter \@firstoftwo
 \else \expandafter \@secondoftwo
 \fi
}%
\providecommand \natexlab [1]{#1}%
\providecommand \enquote  [1]{``#1''}%
\providecommand \bibnamefont  [1]{#1}%
\providecommand \bibfnamefont [1]{#1}%
\providecommand \citenamefont [1]{#1}%
\providecommand \href@noop [0]{\@secondoftwo}%
\providecommand \href [0]{\begingroup \@sanitize@url \@href}%
\providecommand \@href[1]{\@@startlink{#1}\@@href}%
\providecommand \@@href[1]{\endgroup#1\@@endlink}%
\providecommand \@sanitize@url [0]{\catcode `\\12\catcode `\$12\catcode
  `\&12\catcode `\#12\catcode `\^12\catcode `\_12\catcode `\%12\relax}%
\providecommand \@@startlink[1]{}%
\providecommand \@@endlink[0]{}%
\providecommand \url  [0]{\begingroup\@sanitize@url \@url }%
\providecommand \@url [1]{\endgroup\@href {#1}{\urlprefix }}%
\providecommand \urlprefix  [0]{URL }%
\providecommand \Eprint [0]{\href }%
\providecommand \doibase [0]{https://doi.org/}%
\providecommand \selectlanguage [0]{\@gobble}%
\providecommand \bibinfo  [0]{\@secondoftwo}%
\providecommand \bibfield  [0]{\@secondoftwo}%
\providecommand \translation [1]{[#1]}%
\providecommand \BibitemOpen [0]{}%
\providecommand \bibitemStop [0]{}%
\providecommand \bibitemNoStop [0]{.\EOS\space}%
\providecommand \EOS [0]{\spacefactor3000\relax}%
\providecommand \BibitemShut  [1]{\csname bibitem#1\endcsname}%
\let\auto@bib@innerbib\@empty
\bibitem [{\citenamefont {Sullivan}(1972)}]{Sullivan:1971kd}%
  \BibitemOpen
  \bibfield  {author} {\bibinfo {author} {\bibfnamefont {J.~D.}\ \bibnamefont
  {Sullivan}},\ }\bibfield  {title} {\bibinfo {title} {{One-Pion Exchange and
  Deep-Inelastic Electron-Nucleon Scattering}},\ }\href
  {https://doi.org/10.1103/PhysRevD.5.1732} {\bibfield  {journal} {\bibinfo
  {journal} {Phys. Rev. D}\ }\textbf {\bibinfo {volume} {5}},\ \bibinfo {pages}
  {1732} (\bibinfo {year} {1972})}\BibitemShut {NoStop}%
\bibitem [{\citenamefont {Betev}\ \emph {et~al.}(1985)\citenamefont {Betev}
  \emph {et~al.}}]{Betev:1985pf}%
  \BibitemOpen
  \bibfield  {author} {\bibinfo {author} {\bibfnamefont {B.}~\bibnamefont
  {Betev}} \emph {et~al.} (\bibinfo {collaboration} {NA10}),\ }\bibfield
  {title} {\bibinfo {title} {{Differential Cross-Section of High Mass Muon
  Pairs Produced by a \unit{194}{\giga\electronvolt/c} \ $\pi^-$ Beam on a
  Tungsten Target}},\ }\href {https://doi.org/10.1007/BF01550243} {\bibfield
  {journal} {\bibinfo  {journal} {Z. Phys. C}\ }\textbf {\bibinfo {volume}
  {28}},\ \bibinfo {pages} {9} (\bibinfo {year} {1985})}\BibitemShut {NoStop}%
\bibitem [{\citenamefont {Greenlee}\ \emph {et~al.}(1985)\citenamefont
  {Greenlee} \emph {et~al.}}]{Greenlee:1985gd}%
  \BibitemOpen
  \bibfield  {author} {\bibinfo {author} {\bibfnamefont {H.~B.}\ \bibnamefont
  {Greenlee}} \emph {et~al.},\ }\bibfield  {title} {\bibinfo {title} {{The
  Production of Massive Muon Pairs in $\pi^-$-Nucleus Collisions}},\ }\href
  {https://doi.org/10.1103/PhysRevLett.55.1555} {\bibfield  {journal} {\bibinfo
   {journal} {Phys. Rev. Lett.}\ }\textbf {\bibinfo {volume} {55}},\ \bibinfo
  {pages} {1555} (\bibinfo {year} {1985})}\BibitemShut {NoStop}%
\bibitem [{\citenamefont {Conway}\ \emph {et~al.}(1989)\citenamefont {Conway}
  \emph {et~al.}}]{Conway:1989fs}%
  \BibitemOpen
  \bibfield  {author} {\bibinfo {author} {\bibfnamefont {J.~S.}\ \bibnamefont
  {Conway}} \emph {et~al.},\ }\bibfield  {title} {\bibinfo {title}
  {{Experimental study of muon pairs produced by 252-GeV pions on tungsten}},\
  }\href {https://doi.org/10.1103/PhysRevD.39.92} {\bibfield  {journal}
  {\bibinfo  {journal} {Phys. Rev. D}\ }\textbf {\bibinfo {volume} {39}},\
  \bibinfo {pages} {92} (\bibinfo {year} {1989})}\BibitemShut {NoStop}%
\bibitem [{\citenamefont {Aghasyan}\ \emph {et~al.}(2017)\citenamefont
  {Aghasyan} \emph {et~al.}}]{Aghasyan:2017jop}%
  \BibitemOpen
  \bibfield  {author} {\bibinfo {author} {\bibfnamefont {M.}~\bibnamefont
  {Aghasyan}} \emph {et~al.} (\bibinfo {collaboration} {COMPASS}),\ }\bibfield
  {title} {\bibinfo {title} {{First Measurement of Transverse-Spin-Dependent
  Azimuthal Asymmetries in the Drell-Yan Process}},\ }\href
  {https://doi.org/10.1103/PhysRevLett.119.112002} {\bibfield  {journal}
  {\bibinfo  {journal} {Phys. Rev. Lett.}\ }\textbf {\bibinfo {volume} {119}},\
  \bibinfo {pages} {112002} (\bibinfo {year} {2017})},\ \Eprint
  {https://arxiv.org/abs/1704.00488} {arXiv:1704.00488 [hep-ex]} \BibitemShut
  {NoStop}%
\bibitem [{\citenamefont {Adams}\ \emph {et~al.}(2018)\citenamefont {Adams}
  \emph {et~al.}}]{Denisov:2018unj}%
  \BibitemOpen
  \bibfield  {author} {\bibinfo {author} {\bibfnamefont {B.}~\bibnamefont
  {Adams}} \emph {et~al.},\ }\bibfield  {title} {\bibinfo {title} {{Letter of
  Intent: A New QCD facility at the M2 beam line of the CERN SPS
  (COMPASS++/AMBER)}},\ }\href@noop {} {\  (\bibinfo {year} {2018})},\ \Eprint
  {https://arxiv.org/abs/1808.00848} {arXiv:1808.00848 [hep-ex]} \BibitemShut
  {NoStop}%
\bibitem [{\citenamefont {{CERN}}(2021)}]{Amber:2021}%
  \BibitemOpen
  \bibfield  {author} {\bibinfo {author} {\bibnamefont {{CERN}}},\ }\href@noop
  {} {\bibinfo {title} {Amber -- a new qcd facility at the m2 beam line of the
  cern sps}},\ \bibinfo {howpublished} {\url{https://nqf-m2.web.cern.ch/}}
  (\bibinfo {year} {2021}),\ \bibinfo {note} {[Online; accessed
  29-July-2021]}\BibitemShut {NoStop}%
\bibitem [{\citenamefont {Chekanov}\ \emph {et~al.}(2002)\citenamefont
  {Chekanov} \emph {et~al.}}]{Chekanov:2002pf}%
  \BibitemOpen
  \bibfield  {author} {\bibinfo {author} {\bibfnamefont {S.}~\bibnamefont
  {Chekanov}} \emph {et~al.} (\bibinfo {collaboration} {ZEUS}),\ }\bibfield
  {title} {\bibinfo {title} {{Leading neutron production in $e^+$ $p$
  collisions at HERA}},\ }\href {https://doi.org/10.1016/S0550-3213(02)00439-X}
  {\bibfield  {journal} {\bibinfo  {journal} {Nucl. Phys. B}\ }\textbf
  {\bibinfo {volume} {637}},\ \bibinfo {pages} {3} (\bibinfo {year} {2002})},\
  \Eprint {https://arxiv.org/abs/hep-ex/0205076} {arXiv:hep-ex/0205076}
  \BibitemShut {NoStop}%
\bibitem [{\citenamefont {Aaron}\ \emph {et~al.}(2010)\citenamefont {Aaron}
  \emph {et~al.}}]{Aaron:2010ab}%
  \BibitemOpen
  \bibfield  {author} {\bibinfo {author} {\bibfnamefont {F.~D.}\ \bibnamefont
  {Aaron}} \emph {et~al.} (\bibinfo {collaboration} {H1}),\ }\bibfield  {title}
  {\bibinfo {title} {{Measurement of leading neutron production in
  deep-inelastic scattering at HERA}},\ }\href
  {https://doi.org/10.1140/epjc/s10052-010-1369-4} {\bibfield  {journal}
  {\bibinfo  {journal} {Eur. Phys. J. C}\ }\textbf {\bibinfo {volume} {68}},\
  \bibinfo {pages} {381} (\bibinfo {year} {2010})},\ \Eprint
  {https://arxiv.org/abs/1001.0532} {arXiv:1001.0532 [hep-ex]} \BibitemShut
  {NoStop}%
\bibitem [{\citenamefont {Holtmann}\ \emph {et~al.}(1994)\citenamefont
  {Holtmann}, \citenamefont {Levman}, \citenamefont {Nikolaev}, \citenamefont
  {Szczurek},\ and\ \citenamefont {Speth}}]{Holtmann:1994rs}%
  \BibitemOpen
  \bibfield  {author} {\bibinfo {author} {\bibfnamefont {H.}~\bibnamefont
  {Holtmann}}, \bibinfo {author} {\bibfnamefont {G.}~\bibnamefont {Levman}},
  \bibinfo {author} {\bibfnamefont {N.~N.}\ \bibnamefont {Nikolaev}}, \bibinfo
  {author} {\bibfnamefont {A.}~\bibnamefont {Szczurek}},\ and\ \bibinfo
  {author} {\bibfnamefont {J.}~\bibnamefont {Speth}},\ }\bibfield  {title}
  {\bibinfo {title} {{How to measure the pion structure function at HERA}},\
  }\href {https://doi.org/10.1016/0370-2693(94)91392-7} {\bibfield  {journal}
  {\bibinfo  {journal} {Phys. Lett. B}\ }\textbf {\bibinfo {volume} {338}},\
  \bibinfo {pages} {363} (\bibinfo {year} {1994})}\BibitemShut {NoStop}%
\bibitem [{\citenamefont {McKenney}\ \emph {et~al.}(2016)\citenamefont
  {McKenney}, \citenamefont {Sato}, \citenamefont {Melnitchouk},\ and\
  \citenamefont {Ji}}]{McKenney:2015xis}%
  \BibitemOpen
  \bibfield  {author} {\bibinfo {author} {\bibfnamefont {J.~R.}\ \bibnamefont
  {McKenney}}, \bibinfo {author} {\bibfnamefont {N.}~\bibnamefont {Sato}},
  \bibinfo {author} {\bibfnamefont {W.}~\bibnamefont {Melnitchouk}},\ and\
  \bibinfo {author} {\bibfnamefont {C.-R.}\ \bibnamefont {Ji}},\ }\bibfield
  {title} {\bibinfo {title} {{Pion structure function from leading neutron
  electroproduction and SU(2) flavor asymmetry}},\ }\href
  {https://doi.org/10.1103/PhysRevD.93.054011} {\bibfield  {journal} {\bibinfo
  {journal} {Phys. Rev. D}\ }\textbf {\bibinfo {volume} {93}},\ \bibinfo
  {pages} {054011} (\bibinfo {year} {2016})},\ \Eprint
  {https://arxiv.org/abs/1512.04459} {arXiv:1512.04459 [hep-ph]} \BibitemShut
  {NoStop}%
\bibitem [{\citenamefont {Barry}\ \emph {et~al.}(2018)\citenamefont {Barry},
  \citenamefont {Sato}, \citenamefont {Melnitchouk},\ and\ \citenamefont
  {Ji}}]{Barry:2018ort}%
  \BibitemOpen
  \bibfield  {author} {\bibinfo {author} {\bibfnamefont {P.~C.}\ \bibnamefont
  {Barry}}, \bibinfo {author} {\bibfnamefont {N.}~\bibnamefont {Sato}},
  \bibinfo {author} {\bibfnamefont {W.}~\bibnamefont {Melnitchouk}},\ and\
  \bibinfo {author} {\bibfnamefont {C.-R.}\ \bibnamefont {Ji}},\ }\bibfield
  {title} {\bibinfo {title} {{First Monte Carlo Global QCD Analysis of Pion
  Parton Distributions}},\ }\href
  {https://doi.org/10.1103/PhysRevLett.121.152001} {\bibfield  {journal}
  {\bibinfo  {journal} {Phys. Rev. Lett.}\ }\textbf {\bibinfo {volume} {121}},\
  \bibinfo {pages} {152001} (\bibinfo {year} {2018})},\ \Eprint
  {https://arxiv.org/abs/1804.01965} {arXiv:1804.01965 [hep-ph]} \BibitemShut
  {NoStop}%
\bibitem [{\citenamefont {Montgomery}\ \emph {et~al.}(2017)\citenamefont
  {Montgomery}, \citenamefont {Annand}, \citenamefont {Dutta}, \citenamefont
  {Keppel}, \citenamefont {King}, \citenamefont {Wojtsekhowski},\ and\
  \citenamefont {Zhang}}]{Montgomery:2017hab}%
  \BibitemOpen
  \bibfield  {author} {\bibinfo {author} {\bibfnamefont {R.~A.}\ \bibnamefont
  {Montgomery}}, \bibinfo {author} {\bibfnamefont {J.~R.~M.}\ \bibnamefont
  {Annand}}, \bibinfo {author} {\bibfnamefont {D.}~\bibnamefont {Dutta}},
  \bibinfo {author} {\bibfnamefont {C.~E.}\ \bibnamefont {Keppel}}, \bibinfo
  {author} {\bibfnamefont {P.}~\bibnamefont {King}}, \bibinfo {author}
  {\bibfnamefont {B.}~\bibnamefont {Wojtsekhowski}},\ and\ \bibinfo {author}
  {\bibfnamefont {J.}~\bibnamefont {Zhang}} (\bibinfo {collaboration} {TDIS,
  SBS}),\ }\bibfield  {title} {\bibinfo {title} {{Proposed Measurement of
  Tagged Deep Inelastic Scattering in Hall A of Jefferson lab}},\ }\href
  {https://doi.org/10.1063/1.4977122} {\bibfield  {journal} {\bibinfo
  {journal} {AIP Conf. Proc.}\ }\textbf {\bibinfo {volume} {1819}},\ \bibinfo
  {pages} {030004} (\bibinfo {year} {2017})}\BibitemShut {NoStop}%
\bibitem [{\citenamefont {Keppel}\ \emph {et~al.}(2015)\citenamefont {Keppel}
  \emph {et~al.}}]{Keppel:2015}%
  \BibitemOpen
  \bibfield  {author} {\bibinfo {author} {\bibfnamefont {C.~E.}\ \bibnamefont
  {Keppel}} \emph {et~al.},\ }\href@noop {} {\bibinfo {title} {{Measurement of
  Tagged Deep Inelastic Scattering (TDIS): Hall A and SBS Collaboration
  Proposal}}} (\bibinfo {year} {2015}),\ \bibinfo {note}
  {{PR12-15-006}}\BibitemShut {NoStop}%
\bibitem [{\citenamefont {Aguilar}\ \emph {et~al.}(2019)\citenamefont {Aguilar}
  \emph {et~al.}}]{Aguilar:2019teb}%
  \BibitemOpen
  \bibfield  {author} {\bibinfo {author} {\bibfnamefont {A.~C.}\ \bibnamefont
  {Aguilar}} \emph {et~al.},\ }\bibfield  {title} {\bibinfo {title} {{Pion and
  kaon structure at the electron-ion collider}},\ }\href
  {https://doi.org/10.1140/epja/i2019-12885-0} {\bibfield  {journal} {\bibinfo
  {journal} {Eur. Phys. J. A}\ }\textbf {\bibinfo {volume} {55}},\ \bibinfo
  {pages} {190} (\bibinfo {year} {2019})},\ \Eprint
  {https://arxiv.org/abs/1907.08218} {arXiv:1907.08218 [nucl-ex]} \BibitemShut
  {NoStop}%
\bibitem [{\citenamefont {Zyla}\ \emph {et~al.}(2020)\citenamefont {Zyla} \emph
  {et~al.}}]{Zyla:2020zbs}%
  \BibitemOpen
  \bibfield  {author} {\bibinfo {author} {\bibfnamefont {P.~A.}\ \bibnamefont
  {Zyla}} \emph {et~al.} (\bibinfo {collaboration} {PDG}),\ }\bibfield  {title}
  {\bibinfo {title} {{Review of Particle Physics}},\ }\href
  {https://doi.org/10.1093/ptep/ptaa104} {\bibfield  {journal} {\bibinfo
  {journal} {Prog. Theor. Exp. Phys.}\ }\textbf {\bibinfo {volume} {2020}},\
  \bibinfo {pages} {083C01} (\bibinfo {year} {2020})}\BibitemShut {NoStop}%
\bibitem [{\citenamefont {Hoodbhoy}\ \emph {et~al.}(1989)\citenamefont
  {Hoodbhoy}, \citenamefont {Jaffe},\ and\ \citenamefont
  {Manohar}}]{Hoodbhoy:1988am}%
  \BibitemOpen
  \bibfield  {author} {\bibinfo {author} {\bibfnamefont {P.}~\bibnamefont
  {Hoodbhoy}}, \bibinfo {author} {\bibfnamefont {R.~L.}\ \bibnamefont
  {Jaffe}},\ and\ \bibinfo {author} {\bibfnamefont {A.}~\bibnamefont
  {Manohar}},\ }\bibfield  {title} {\bibinfo {title} {{Novel effects in deep
  inelastic scattering from spin-one hadrons}},\ }\href
  {https://doi.org/10.1016/0550-3213(89)90572-5} {\bibfield  {journal}
  {\bibinfo  {journal} {Nucl. Phys. B}\ }\textbf {\bibinfo {volume} {312}},\
  \bibinfo {pages} {571} (\bibinfo {year} {1989})}\BibitemShut {NoStop}%
\bibitem [{\citenamefont {Best}\ \emph
  {et~al.}(1997{\natexlab{a}})\citenamefont {Best}, \citenamefont
  {G{\"o}ckeler}, \citenamefont {Horsley}, \citenamefont {Ilgenfritz},
  \citenamefont {Perlt}, \citenamefont {Rakow}, \citenamefont {Sch{\"a}fer},
  \citenamefont {Schierholz}, \citenamefont {Schiller},\ and\ \citenamefont
  {Schramm}}]{Best:1997qp}%
  \BibitemOpen
  \bibfield  {author} {\bibinfo {author} {\bibfnamefont {C.}~\bibnamefont
  {Best}}, \bibinfo {author} {\bibfnamefont {M.}~\bibnamefont {G{\"o}ckeler}},
  \bibinfo {author} {\bibfnamefont {R.}~\bibnamefont {Horsley}}, \bibinfo
  {author} {\bibfnamefont {E.-M.}\ \bibnamefont {Ilgenfritz}}, \bibinfo
  {author} {\bibfnamefont {H.}~\bibnamefont {Perlt}}, \bibinfo {author}
  {\bibfnamefont {P.}~\bibnamefont {Rakow}}, \bibinfo {author} {\bibfnamefont
  {A.}~\bibnamefont {Sch{\"a}fer}}, \bibinfo {author} {\bibfnamefont
  {G.}~\bibnamefont {Schierholz}}, \bibinfo {author} {\bibfnamefont
  {A.}~\bibnamefont {Schiller}},\ and\ \bibinfo {author} {\bibfnamefont
  {S.}~\bibnamefont {Schramm}},\ }\bibfield  {title} {\bibinfo {title} {{$\pi$
  and $\rho$ structure functions from lattice QCD}},\ }\href
  {https://doi.org/10.1103/PhysRevD.56.2743} {\bibfield  {journal} {\bibinfo
  {journal} {Phys. Rev. D}\ }\textbf {\bibinfo {volume} {56}},\ \bibinfo
  {pages} {2743} (\bibinfo {year} {1997}{\natexlab{a}})},\ \Eprint
  {https://arxiv.org/abs/hep-lat/9703014} {arXiv:hep-lat/9703014} \BibitemShut
  {NoStop}%
\bibitem [{\citenamefont {Ji}(2021)}]{Ji:private}%
  \BibitemOpen
  \bibfield  {author} {\bibinfo {author} {\bibfnamefont {X.-D.}\ \bibnamefont
  {Ji}},\ }\href@noop {} {}\bibinfo {howpublished} {private communication}
  (\bibinfo {year} {2021}),\ \bibinfo {note} {[July-2021]}\BibitemShut
  {NoStop}%
\bibitem [{\citenamefont {Airapetian}\ \emph {et~al.}(2005)\citenamefont
  {Airapetian} \emph {et~al.}}]{Airapetian:2005cb}%
  \BibitemOpen
  \bibfield  {author} {\bibinfo {author} {\bibfnamefont {A.}~\bibnamefont
  {Airapetian}} \emph {et~al.} (\bibinfo {collaboration} {HERMES}),\ }\bibfield
   {title} {\bibinfo {title} {{Measurement of the Tensor Structure Function
  b(1) of the Deuteron}},\ }\href
  {https://doi.org/10.1103/PhysRevLett.95.242001} {\bibfield  {journal}
  {\bibinfo  {journal} {Phys. Rev. Lett.}\ }\textbf {\bibinfo {volume} {95}},\
  \bibinfo {pages} {242001} (\bibinfo {year} {2005})},\ \Eprint
  {https://arxiv.org/abs/hep-ex/0506018} {arXiv:hep-ex/0506018} \BibitemShut
  {NoStop}%
\bibitem [{\citenamefont {Close}\ and\ \citenamefont
  {Kumano}(1990)}]{Close:1990zw}%
  \BibitemOpen
  \bibfield  {author} {\bibinfo {author} {\bibfnamefont {F.~E.}\ \bibnamefont
  {Close}}\ and\ \bibinfo {author} {\bibfnamefont {S.}~\bibnamefont {Kumano}},\
  }\bibfield  {title} {\bibinfo {title} {{Sum rule for the spin-dependent
  structure function $b_1(x)$ for spin-one hadrons}},\ }\href
  {https://doi.org/10.1103/PhysRevD.42.2377} {\bibfield  {journal} {\bibinfo
  {journal} {Phys. Rev. D}\ }\textbf {\bibinfo {volume} {42}},\ \bibinfo
  {pages} {2377} (\bibinfo {year} {1990})}\BibitemShut {NoStop}%
\bibitem [{\citenamefont {Kumano}\ and\ \citenamefont
  {Song}(2017)}]{Kumano:2016cqs}%
  \BibitemOpen
  \bibfield  {author} {\bibinfo {author} {\bibfnamefont {S.}~\bibnamefont
  {Kumano}}\ and\ \bibinfo {author} {\bibfnamefont {Q.-T.}\ \bibnamefont
  {Song}},\ }\bibfield  {title} {\bibinfo {title} {{Estimate on Spin Asymmetry
  for Drell-Yan Process at Fermilab with Tensor-Polarized Deuteron}},\ }\href
  {https://doi.org/10.7566/JPSCP.13.020048} {\bibfield  {journal} {\bibinfo
  {journal} {JPS Conf. Proc.}\ }\textbf {\bibinfo {volume} {13}},\ \bibinfo
  {pages} {020048} (\bibinfo {year} {2017})},\ \Eprint
  {https://arxiv.org/abs/1611.00474} {arXiv:1611.00474 [hep-ph]} \BibitemShut
  {NoStop}%
\bibitem [{\citenamefont {Alleda}\ \emph {et~al.}(2013)\citenamefont {Alleda}
  \emph {et~al.}}]{Alleda:2013}%
  \BibitemOpen
  \bibfield  {author} {\bibinfo {author} {\bibfnamefont {K.}~\bibnamefont
  {Alleda}} \emph {et~al.},\ }\href@noop {} {\bibinfo {title} {{The Deuteron
  Tensor Structure Function~$b_1$: A Proposal to Jefferson Lab PAC-40}}}
  (\bibinfo {year} {2013}),\ \bibinfo {note} {{PR12-13-011}}\BibitemShut
  {NoStop}%
\bibitem [{\citenamefont {Song}(2019)}]{Song:2019awx}%
  \BibitemOpen
  \bibfield  {author} {\bibinfo {author} {\bibfnamefont {Q.-T.}\ \bibnamefont
  {Song}},\ }\bibfield  {title} {\bibinfo {title} {{Structure of Deuteron by
  Polarized Proton-Deuteron Drell-Yan Process}},\ }\href
  {https://doi.org/10.7566/JPSCP.26.031001} {\bibfield  {journal} {\bibinfo
  {journal} {JPS Conf. Proc.}\ }\textbf {\bibinfo {volume} {26}},\ \bibinfo
  {pages} {031001} (\bibinfo {year} {2019})}\BibitemShut {NoStop}%
\bibitem [{\citenamefont {Boer}(2019)}]{Boer:2019pzo}%
  \BibitemOpen
  \bibfield  {author} {\bibinfo {author} {\bibfnamefont {D.}~\bibnamefont
  {Boer}},\ }\bibfield  {title} {\bibinfo {title} {{Overview of Spin Physics at
  EIC}},\ }\href {https://doi.org/10.22323/1.346.0167} {\bibfield  {journal}
  {\bibinfo  {journal} {PoS}\ }\textbf {\bibinfo {volume} {SPIN2018}},\
  \bibinfo {pages} {167} (\bibinfo {year} {2019})},\ \Eprint
  {https://arxiv.org/abs/1903.01119} {arXiv:1903.01119 [hep-ph]} \BibitemShut
  {NoStop}%
\bibitem [{\citenamefont {Martinelli}\ and\ \citenamefont
  {Sachrajda}(1987)}]{Martinelli:1987zd}%
  \BibitemOpen
  \bibfield  {author} {\bibinfo {author} {\bibfnamefont {G.}~\bibnamefont
  {Martinelli}}\ and\ \bibinfo {author} {\bibfnamefont {C.~T.}\ \bibnamefont
  {Sachrajda}},\ }\bibfield  {title} {\bibinfo {title} {{Pion structure
  functions from lattice QCD}},\ }\href
  {https://doi.org/10.1016/0370-2693(87)90601-0} {\bibfield  {journal}
  {\bibinfo  {journal} {Phys. Lett. B}\ }\textbf {\bibinfo {volume} {196}},\
  \bibinfo {pages} {184} (\bibinfo {year} {1987})}\BibitemShut {NoStop}%
\bibitem [{\citenamefont {Martinelli}\ and\ \citenamefont
  {Sachrajda}(1988)}]{Martinelli:1987bh}%
  \BibitemOpen
  \bibfield  {author} {\bibinfo {author} {\bibfnamefont {G.}~\bibnamefont
  {Martinelli}}\ and\ \bibinfo {author} {\bibfnamefont {C.~T.}\ \bibnamefont
  {Sachrajda}},\ }\bibfield  {title} {\bibinfo {title} {{A lattice calculation
  of the pion's form-factor and structure function}},\ }\href
  {https://doi.org/10.1016/0550-3213(88)90445-2} {\bibfield  {journal}
  {\bibinfo  {journal} {Nucl. Phys. B}\ }\textbf {\bibinfo {volume} {306}},\
  \bibinfo {pages} {865} (\bibinfo {year} {1988})}\BibitemShut {NoStop}%
\bibitem [{\citenamefont {Best}\ \emph
  {et~al.}(1997{\natexlab{b}})\citenamefont {Best} \emph
  {et~al.}}]{Best:1997ab}%
  \BibitemOpen
  \bibfield  {author} {\bibinfo {author} {\bibfnamefont {C.}~\bibnamefont
  {Best}} \emph {et~al.},\ }\bibfield  {title} {\bibinfo {title} {{Hadron
  structure functions from lattice QCD - 1997}},\ }\href
  {https://doi.org/10.1063/1.53699} {\bibfield  {journal} {\bibinfo  {journal}
  {AIP Conf. Proc.}\ }\textbf {\bibinfo {volume} {407}},\ \bibinfo {pages}
  {997} (\bibinfo {year} {1997}{\natexlab{b}})},\ \Eprint
  {https://arxiv.org/abs/hep-ph/9706502} {arXiv:hep-ph/9706502} \BibitemShut
  {NoStop}%
\bibitem [{\citenamefont {Best}\ \emph {et~al.}(1998)\citenamefont {Best},
  \citenamefont {G{\"o}ckeler}, \citenamefont {Horsley}, \citenamefont {Perlt},
  \citenamefont {Rakow}, \citenamefont {Sch{\"a}fer}, \citenamefont
  {Schierholz}, \citenamefont {Schiller},\ and\ \citenamefont
  {Schramm}}]{Best:1997rm}%
  \BibitemOpen
  \bibfield  {author} {\bibinfo {author} {\bibfnamefont {C.}~\bibnamefont
  {Best}}, \bibinfo {author} {\bibfnamefont {M.}~\bibnamefont {G{\"o}ckeler}},
  \bibinfo {author} {\bibfnamefont {R.}~\bibnamefont {Horsley}}, \bibinfo
  {author} {\bibfnamefont {H.}~\bibnamefont {Perlt}}, \bibinfo {author}
  {\bibfnamefont {P.}~\bibnamefont {Rakow}}, \bibinfo {author} {\bibfnamefont
  {A.}~\bibnamefont {Sch{\"a}fer}}, \bibinfo {author} {\bibfnamefont
  {G.}~\bibnamefont {Schierholz}}, \bibinfo {author} {\bibfnamefont
  {A.}~\bibnamefont {Schiller}},\ and\ \bibinfo {author} {\bibfnamefont
  {S.}~\bibnamefont {Schramm}},\ }\bibfield  {title} {\bibinfo {title} {{The
  Deep-Inelastic Structure Functions of $\pi$ and $\rho$ Mesons}},\ }\href
  {https://doi.org/10.1016/S0920-5632(97)00731-7} {\bibfield  {journal}
  {\bibinfo  {journal} {Nucl. Phys. B (Proc. Suppl.)}\ }\textbf {\bibinfo
  {volume} {63A-C}},\ \bibinfo {pages} {236} (\bibinfo {year} {1998})},\
  \Eprint {https://arxiv.org/abs/hep-lat/9710037} {arXiv:hep-lat/9710037}
  \BibitemShut {NoStop}%
\bibitem [{\citenamefont {Guagnelli}\ \emph {et~al.}(2005)\citenamefont
  {Guagnelli}, \citenamefont {Jansen}, \citenamefont {Palombi}, \citenamefont
  {Petronzio}, \citenamefont {Shindler},\ and\ \citenamefont
  {Wetzorke}}]{Guagnelli:2004ga}%
  \BibitemOpen
  \bibfield  {author} {\bibinfo {author} {\bibfnamefont {M.}~\bibnamefont
  {Guagnelli}}, \bibinfo {author} {\bibfnamefont {K.}~\bibnamefont {Jansen}},
  \bibinfo {author} {\bibfnamefont {F.}~\bibnamefont {Palombi}}, \bibinfo
  {author} {\bibfnamefont {R.}~\bibnamefont {Petronzio}}, \bibinfo {author}
  {\bibfnamefont {A.}~\bibnamefont {Shindler}},\ and\ \bibinfo {author}
  {\bibfnamefont {I.}~\bibnamefont {Wetzorke}} (\bibinfo {collaboration}
  {ZeRo}),\ }\bibfield  {title} {\bibinfo {title} {{Non-perturbative pion
  matrix element of a twist-2 operator from the lattice}},\ }\href
  {https://doi.org/10.1140/epjc/s2005-02121-5} {\bibfield  {journal} {\bibinfo
  {journal} {Eur. Phys. J. C}\ }\textbf {\bibinfo {volume} {40}},\ \bibinfo
  {pages} {69} (\bibinfo {year} {2005})},\ \Eprint
  {https://arxiv.org/abs/hep-lat/0405027} {arXiv:hep-lat/0405027} \BibitemShut
  {NoStop}%
\bibitem [{\citenamefont {Capitani}\ \emph {et~al.}(2006)\citenamefont
  {Capitani}, \citenamefont {Jansen}, \citenamefont {Papinutto}, \citenamefont
  {Shindler}, \citenamefont {Urbach},\ and\ \citenamefont
  {Wetzorke}}]{Capitani:2005jp}%
  \BibitemOpen
  \bibfield  {author} {\bibinfo {author} {\bibfnamefont {S.}~\bibnamefont
  {Capitani}}, \bibinfo {author} {\bibfnamefont {K.}~\bibnamefont {Jansen}},
  \bibinfo {author} {\bibfnamefont {M.}~\bibnamefont {Papinutto}}, \bibinfo
  {author} {\bibfnamefont {A.}~\bibnamefont {Shindler}}, \bibinfo {author}
  {\bibfnamefont {C.}~\bibnamefont {Urbach}},\ and\ \bibinfo {author}
  {\bibfnamefont {I.}~\bibnamefont {Wetzorke}},\ }\bibfield  {title} {\bibinfo
  {title} {{Parton distribution functions with twisted mass fermions}},\ }\href
  {https://doi.org/10.1016/j.physletb.2006.02.047} {\bibfield  {journal}
  {\bibinfo  {journal} {Phys. Lett. B}\ }\textbf {\bibinfo {volume} {639}},\
  \bibinfo {pages} {520} (\bibinfo {year} {2006})},\ \Eprint
  {https://arxiv.org/abs/hep-lat/0511013} {arXiv:hep-lat/0511013} \BibitemShut
  {NoStop}%
\bibitem [{\citenamefont {Br{\"o}mmel}\ \emph {et~al.}(2008)\citenamefont
  {Br{\"o}mmel} \emph {et~al.}}]{Brommel:2006zz}%
  \BibitemOpen
  \bibfield  {author} {\bibinfo {author} {\bibfnamefont {D.}~\bibnamefont
  {Br{\"o}mmel}} \emph {et~al.} (\bibinfo {collaboration} {QCDSF/UKQCD}),\
  }\bibfield  {title} {\bibinfo {title} {{Quark distributions in the pion}},\
  }\href {https://doi.org/10.22323/1.042.0140} {\bibfield  {journal} {\bibinfo
  {journal} {PoS}\ }\textbf {\bibinfo {volume} {LATTICE2007}},\ \bibinfo
  {pages} {140} (\bibinfo {year} {2008})}\BibitemShut {NoStop}%
\bibitem [{\citenamefont {Bali}\ \emph
  {et~al.}(2014{\natexlab{a}})\citenamefont {Bali}, \citenamefont {Collins},
  \citenamefont {Gl{\"a}ssle}, \citenamefont {G{\"o}ckeler}, \citenamefont
  {Javadi-Motaghi}, \citenamefont {Najjar}, \citenamefont {S{\"o}ldner},\ and\
  \citenamefont {Sternbeck}}]{Bali:2013gya}%
  \BibitemOpen
  \bibfield  {author} {\bibinfo {author} {\bibfnamefont {G.~S.}\ \bibnamefont
  {Bali}}, \bibinfo {author} {\bibfnamefont {S.}~\bibnamefont {Collins}},
  \bibinfo {author} {\bibfnamefont {B.}~\bibnamefont {Gl{\"a}ssle}}, \bibinfo
  {author} {\bibfnamefont {M.}~\bibnamefont {G{\"o}ckeler}}, \bibinfo {author}
  {\bibfnamefont {N.}~\bibnamefont {Javadi-Motaghi}}, \bibinfo {author}
  {\bibfnamefont {J.}~\bibnamefont {Najjar}}, \bibinfo {author} {\bibfnamefont
  {W.}~\bibnamefont {S{\"o}ldner}},\ and\ \bibinfo {author} {\bibfnamefont
  {A.}~\bibnamefont {Sternbeck}},\ }\bibfield  {title} {\bibinfo {title} {{Pion
  structure from lattice QCD}},\ }\href {https://doi.org/10.22323/1.187.0447}
  {\bibfield  {journal} {\bibinfo  {journal} {PoS}\ }\textbf {\bibinfo {volume}
  {LATTICE2013}},\ \bibinfo {pages} {447} (\bibinfo {year}
  {2014}{\natexlab{a}})},\ \Eprint {https://arxiv.org/abs/1311.7639}
  {arXiv:1311.7639 [hep-lat]} \BibitemShut {NoStop}%
\bibitem [{\citenamefont {Abdel-Rehim}\ \emph {et~al.}(2015)\citenamefont
  {Abdel-Rehim} \emph {et~al.}}]{Abdel-Rehim:2015owa}%
  \BibitemOpen
  \bibfield  {author} {\bibinfo {author} {\bibfnamefont {A.}~\bibnamefont
  {Abdel-Rehim}} \emph {et~al.},\ }\bibfield  {title} {\bibinfo {title}
  {{Nucleon and pion structure with lattice QCD simulations at physical value
  of the pion mass}},\ }\href {https://doi.org/10.1103/PhysRevD.92.114513}
  {\bibfield  {journal} {\bibinfo  {journal} {Phys. Rev. D}\ }\textbf {\bibinfo
  {volume} {92}},\ \bibinfo {pages} {114513} (\bibinfo {year} {2015})},\
  \bibinfo {note} {[Erratum: Phys.~Rev.~D~{\bf 93}, 039904(E) (2016)]},\
  \Eprint {https://arxiv.org/abs/1507.04936} {arXiv:1507.04936 [hep-lat]}
  \BibitemShut {NoStop}%
\bibitem [{\citenamefont {Oehm}\ \emph {et~al.}(2019)\citenamefont {Oehm},
  \citenamefont {Alexandrou}, \citenamefont {Constantinou}, \citenamefont
  {Jansen}, \citenamefont {Koutsou}, \citenamefont {Kostrzewa}, \citenamefont
  {Steffens}, \citenamefont {Urbach},\ and\ \citenamefont
  {Zafeiropoulos}}]{Oehm:2018jvm}%
  \BibitemOpen
  \bibfield  {author} {\bibinfo {author} {\bibfnamefont {M.}~\bibnamefont
  {Oehm}}, \bibinfo {author} {\bibfnamefont {C.}~\bibnamefont {Alexandrou}},
  \bibinfo {author} {\bibfnamefont {M.}~\bibnamefont {Constantinou}}, \bibinfo
  {author} {\bibfnamefont {K.}~\bibnamefont {Jansen}}, \bibinfo {author}
  {\bibfnamefont {G.}~\bibnamefont {Koutsou}}, \bibinfo {author} {\bibfnamefont
  {B.}~\bibnamefont {Kostrzewa}}, \bibinfo {author} {\bibfnamefont
  {F.}~\bibnamefont {Steffens}}, \bibinfo {author} {\bibfnamefont
  {C.}~\bibnamefont {Urbach}},\ and\ \bibinfo {author} {\bibfnamefont
  {S.}~\bibnamefont {Zafeiropoulos}},\ }\bibfield  {title} {\bibinfo {title}
  {{$\langle x\rangle$ and $\langle x^2\rangle$ of the pion PDF from lattice
  QCD with $N_f=2+1+1$ dynamical quark flavors}},\ }\href
  {https://doi.org/10.1103/PhysRevD.99.014508} {\bibfield  {journal} {\bibinfo
  {journal} {Phys. Rev. D}\ }\textbf {\bibinfo {volume} {99}},\ \bibinfo
  {pages} {014508} (\bibinfo {year} {2019})},\ \Eprint
  {https://arxiv.org/abs/1810.09743} {arXiv:1810.09743 [hep-lat]} \BibitemShut
  {NoStop}%
\bibitem [{\citenamefont {Alexandrou}\ \emph {et~al.}(2020)\citenamefont
  {Alexandrou}, \citenamefont {Bacchio}, \citenamefont {Cl{\"o}et},
  \citenamefont {Constantinou}, \citenamefont {Hadjiyiannakou}, \citenamefont
  {Koutsou},\ and\ \citenamefont {Lauer}}]{Alexandrou:2020gxs}%
  \BibitemOpen
  \bibfield  {author} {\bibinfo {author} {\bibfnamefont {C.}~\bibnamefont
  {Alexandrou}}, \bibinfo {author} {\bibfnamefont {S.}~\bibnamefont {Bacchio}},
  \bibinfo {author} {\bibfnamefont {I.}~\bibnamefont {Cl{\"o}et}}, \bibinfo
  {author} {\bibfnamefont {M.}~\bibnamefont {Constantinou}}, \bibinfo {author}
  {\bibfnamefont {K.}~\bibnamefont {Hadjiyiannakou}}, \bibinfo {author}
  {\bibfnamefont {G.}~\bibnamefont {Koutsou}},\ and\ \bibinfo {author}
  {\bibfnamefont {C.}~\bibnamefont {Lauer}},\ }\bibfield  {title} {\bibinfo
  {title} {{The Mellin moments $\langle x \rangle$ and $\langle x^2 \rangle$
  for the pion and kaon from lattice QCD}},\ }\href@noop {} {\  (\bibinfo
  {year} {2020})},\ \Eprint {https://arxiv.org/abs/2010.03495}
  {arXiv:2010.03495 [hep-lat]} \BibitemShut {NoStop}%
\bibitem [{\citenamefont {Alexandrou}\ \emph {et~al.}(2021)\citenamefont
  {Alexandrou}, \citenamefont {Bacchio}, \citenamefont {Clo\"et}, \citenamefont
  {Constantinou}, \citenamefont {Hadjiyiannakou}, \citenamefont {Koutsou},\
  and\ \citenamefont {Lauer}}]{Alexandrou:2021mmi}%
  \BibitemOpen
  \bibfield  {author} {\bibinfo {author} {\bibfnamefont {C.}~\bibnamefont
  {Alexandrou}}, \bibinfo {author} {\bibfnamefont {S.}~\bibnamefont {Bacchio}},
  \bibinfo {author} {\bibfnamefont {I.}~\bibnamefont {Clo\"et}}, \bibinfo
  {author} {\bibfnamefont {M.}~\bibnamefont {Constantinou}}, \bibinfo {author}
  {\bibfnamefont {K.}~\bibnamefont {Hadjiyiannakou}}, \bibinfo {author}
  {\bibfnamefont {G.}~\bibnamefont {Koutsou}},\ and\ \bibinfo {author}
  {\bibfnamefont {C.}~\bibnamefont {Lauer}} (\bibinfo {collaboration} {ETM}),\
  }\bibfield  {title} {\bibinfo {title} {{The pion and kaon $\langle x^3
  \rangle$ from lattice QCD and PDF reconstruction from Mellin moments}},\
  }\href@noop {} {\  (\bibinfo {year} {2021})},\ \Eprint
  {https://arxiv.org/abs/2104.02247} {arXiv:2104.02247 [hep-lat]} \BibitemShut
  {NoStop}%
\bibitem [{\citenamefont {Braun}\ and\ \citenamefont
  {M\"uller}(2008)}]{Braun:2007wv}%
  \BibitemOpen
  \bibfield  {author} {\bibinfo {author} {\bibfnamefont {V.}~\bibnamefont
  {Braun}}\ and\ \bibinfo {author} {\bibfnamefont {D.}~\bibnamefont
  {M\"uller}},\ }\bibfield  {title} {\bibinfo {title} {{Exclusive processes in
  position space and the pion distribution amplitude}},\ }\href
  {https://doi.org/10.1140/epjc/s10052-008-0608-4} {\bibfield  {journal}
  {\bibinfo  {journal} {Eur. Phys. J. C}\ }\textbf {\bibinfo {volume} {55}},\
  \bibinfo {pages} {349} (\bibinfo {year} {2008})},\ \Eprint
  {https://arxiv.org/abs/0709.1348} {arXiv:0709.1348 [hep-ph]} \BibitemShut
  {NoStop}%
\bibitem [{\citenamefont {Ji}(2013)}]{Ji:2013dva}%
  \BibitemOpen
  \bibfield  {author} {\bibinfo {author} {\bibfnamefont {X.}~\bibnamefont
  {Ji}},\ }\bibfield  {title} {\bibinfo {title} {{Parton Physics on a Euclidean
  Lattice}},\ }\href {https://doi.org/10.1103/PhysRevLett.110.262002}
  {\bibfield  {journal} {\bibinfo  {journal} {Phys. Rev. Lett.}\ }\textbf
  {\bibinfo {volume} {110}},\ \bibinfo {pages} {262002} (\bibinfo {year}
  {2013})},\ \Eprint {https://arxiv.org/abs/1305.1539} {arXiv:1305.1539
  [hep-ph]} \BibitemShut {NoStop}%
\bibitem [{\citenamefont {Ma}\ and\ \citenamefont {Qiu}(2018)}]{Ma:2014jla}%
  \BibitemOpen
  \bibfield  {author} {\bibinfo {author} {\bibfnamefont {Y.-Q.}\ \bibnamefont
  {Ma}}\ and\ \bibinfo {author} {\bibfnamefont {J.-W.}\ \bibnamefont {Qiu}},\
  }\bibfield  {title} {\bibinfo {title} {{Extracting Parton Distribution
  Functions from Lattice QCD Calculations}},\ }\href
  {https://doi.org/10.1103/PhysRevD.98.074021} {\bibfield  {journal} {\bibinfo
  {journal} {Phys. Rev. D}\ }\textbf {\bibinfo {volume} {98}},\ \bibinfo
  {pages} {074021} (\bibinfo {year} {2018})},\ \Eprint
  {https://arxiv.org/abs/1404.6860} {arXiv:1404.6860 [hep-ph]} \BibitemShut
  {NoStop}%
\bibitem [{\citenamefont {Radyushkin}(2017)}]{Radyushkin:2017cyf}%
  \BibitemOpen
  \bibfield  {author} {\bibinfo {author} {\bibfnamefont {A.~V.}\ \bibnamefont
  {Radyushkin}},\ }\bibfield  {title} {\bibinfo {title} {{Quasi-parton
  distribution functions, momentum distributions, and pseudo-parton
  distribution functions}},\ }\href
  {https://doi.org/10.1103/PhysRevD.96.034025} {\bibfield  {journal} {\bibinfo
  {journal} {Phys. Rev. D}\ }\textbf {\bibinfo {volume} {96}},\ \bibinfo
  {pages} {034025} (\bibinfo {year} {2017})},\ \Eprint
  {https://arxiv.org/abs/1705.01488} {arXiv:1705.01488 [hep-ph]} \BibitemShut
  {NoStop}%
\bibitem [{\citenamefont {Ji}\ \emph {et~al.}(2021)\citenamefont {Ji},
  \citenamefont {Liu}, \citenamefont {Liu}, \citenamefont {Zhang},\ and\
  \citenamefont {Zhao}}]{Ji:2020ect}%
  \BibitemOpen
  \bibfield  {author} {\bibinfo {author} {\bibfnamefont {X.}~\bibnamefont
  {Ji}}, \bibinfo {author} {\bibfnamefont {Y.-S.}\ \bibnamefont {Liu}},
  \bibinfo {author} {\bibfnamefont {Y.}~\bibnamefont {Liu}}, \bibinfo {author}
  {\bibfnamefont {J.-H.}\ \bibnamefont {Zhang}},\ and\ \bibinfo {author}
  {\bibfnamefont {Y.}~\bibnamefont {Zhao}},\ }\bibfield  {title} {\bibinfo
  {title} {{Large-momentum effective theory}},\ }\href
  {https://doi.org/10.1103/RevModPhys.93.035005} {\bibfield  {journal}
  {\bibinfo  {journal} {Rev. Mod. Phys.}\ }\textbf {\bibinfo {volume} {93}},\
  \bibinfo {pages} {035005} (\bibinfo {year} {2021})},\ \Eprint
  {https://arxiv.org/abs/2004.03543} {arXiv:2004.03543 [hep-ph]} \BibitemShut
  {NoStop}%
\bibitem [{\citenamefont {Hua}\ \emph {et~al.}(2021)\citenamefont {Hua},
  \citenamefont {Chu}, \citenamefont {Sun}, \citenamefont {Wang}, \citenamefont
  {Xu}, \citenamefont {Yang}, \citenamefont {Zhang},\ and\ \citenamefont
  {Zhang}}]{Hua:2020gnw}%
  \BibitemOpen
  \bibfield  {author} {\bibinfo {author} {\bibfnamefont {J.}~\bibnamefont
  {Hua}}, \bibinfo {author} {\bibfnamefont {M.-H.}\ \bibnamefont {Chu}},
  \bibinfo {author} {\bibfnamefont {P.}~\bibnamefont {Sun}}, \bibinfo {author}
  {\bibfnamefont {W.}~\bibnamefont {Wang}}, \bibinfo {author} {\bibfnamefont
  {J.}~\bibnamefont {Xu}}, \bibinfo {author} {\bibfnamefont {Y.-B.}\
  \bibnamefont {Yang}}, \bibinfo {author} {\bibfnamefont {J.-H.}\ \bibnamefont
  {Zhang}},\ and\ \bibinfo {author} {\bibfnamefont {Q.-A.}\ \bibnamefont
  {Zhang}} (\bibinfo {collaboration} {Lattice Parton}),\ }\bibfield  {title}
  {\bibinfo {title} {{Distribution Amplitudes of K* and \ensuremath{\phi} at
  the Physical Pion Mass from Lattice QCD}},\ }\href
  {https://doi.org/10.1103/PhysRevLett.127.062002} {\bibfield  {journal}
  {\bibinfo  {journal} {Phys. Rev. Lett.}\ }\textbf {\bibinfo {volume} {127}},\
  \bibinfo {pages} {062002} (\bibinfo {year} {2021})},\ \Eprint
  {https://arxiv.org/abs/2011.09788} {arXiv:2011.09788 [hep-lat]} \BibitemShut
  {NoStop}%
\bibitem [{\citenamefont {Braun}\ \emph {et~al.}(2017)\citenamefont {Braun}
  \emph {et~al.}}]{Braun:2016wnx}%
  \BibitemOpen
  \bibfield  {author} {\bibinfo {author} {\bibfnamefont {V.~M.}\ \bibnamefont
  {Braun}} \emph {et~al.},\ }\bibfield  {title} {\bibinfo {title} {{The
  \ensuremath{\rho}-meson light-cone distribution amplitudes from lattice
  QCD}},\ }\href {https://doi.org/10.1007/JHEP04(2017)082} {\bibfield
  {journal} {\bibinfo  {journal} {JHEP}\ }\textbf {\bibinfo {volume} {04}},\
  \bibinfo {pages} {082}},\ \Eprint {https://arxiv.org/abs/1612.02955}
  {arXiv:1612.02955 [hep-lat]} \BibitemShut {NoStop}%
\bibitem [{\citenamefont {Cichy}(2021)}]{Cichy:2021lih}%
  \BibitemOpen
  \bibfield  {author} {\bibinfo {author} {\bibfnamefont {K.}~\bibnamefont
  {Cichy}},\ }\bibfield  {title} {\bibinfo {title} {{Progress in $x$-dependent
  partonic distributions from lattice QCD}},\ }in\ \href@noop {} {\emph
  {\bibinfo {booktitle} {{38th International Symposium on Lattice Field
  Theory}}}}\ (\bibinfo {year} {2021})\ \Eprint
  {https://arxiv.org/abs/2110.07440} {arXiv:2110.07440 [hep-lat]} \BibitemShut
  {NoStop}%
\bibitem [{\citenamefont {Sufian}\ \emph {et~al.}(2019)\citenamefont {Sufian},
  \citenamefont {Karpie}, \citenamefont {Egerer}, \citenamefont {Orginos},
  \citenamefont {Qiu},\ and\ \citenamefont {Richards}}]{Sufian:2019bol}%
  \BibitemOpen
  \bibfield  {author} {\bibinfo {author} {\bibfnamefont {R.~S.}\ \bibnamefont
  {Sufian}}, \bibinfo {author} {\bibfnamefont {J.}~\bibnamefont {Karpie}},
  \bibinfo {author} {\bibfnamefont {C.}~\bibnamefont {Egerer}}, \bibinfo
  {author} {\bibfnamefont {K.}~\bibnamefont {Orginos}}, \bibinfo {author}
  {\bibfnamefont {J.-W.}\ \bibnamefont {Qiu}},\ and\ \bibinfo {author}
  {\bibfnamefont {D.~G.}\ \bibnamefont {Richards}},\ }\bibfield  {title}
  {\bibinfo {title} {{Pion valence quark distribution from matrix element
  calculated in lattice QCD}},\ }\href
  {https://doi.org/10.1103/PhysRevD.99.074507} {\bibfield  {journal} {\bibinfo
  {journal} {Phys. Rev. D}\ }\textbf {\bibinfo {volume} {99}},\ \bibinfo
  {pages} {074507} (\bibinfo {year} {2019})},\ \Eprint
  {https://arxiv.org/abs/1901.03921} {arXiv:1901.03921 [hep-lat]} \BibitemShut
  {NoStop}%
\bibitem [{\citenamefont {Sufian}\ \emph {et~al.}(2020)\citenamefont {Sufian},
  \citenamefont {Egerer}, \citenamefont {Karpie}, \citenamefont {Edwards},
  \citenamefont {Jo{\'o}}, \citenamefont {Ma}, \citenamefont {Orginos},
  \citenamefont {Qiu},\ and\ \citenamefont {Richards}}]{Sufian:2020vzb}%
  \BibitemOpen
  \bibfield  {author} {\bibinfo {author} {\bibfnamefont {R.~S.}\ \bibnamefont
  {Sufian}}, \bibinfo {author} {\bibfnamefont {C.}~\bibnamefont {Egerer}},
  \bibinfo {author} {\bibfnamefont {J.}~\bibnamefont {Karpie}}, \bibinfo
  {author} {\bibfnamefont {R.~G.}\ \bibnamefont {Edwards}}, \bibinfo {author}
  {\bibfnamefont {B.}~\bibnamefont {Jo{\'o}}}, \bibinfo {author} {\bibfnamefont
  {Y.-Q.}\ \bibnamefont {Ma}}, \bibinfo {author} {\bibfnamefont
  {K.}~\bibnamefont {Orginos}}, \bibinfo {author} {\bibfnamefont {J.-W.}\
  \bibnamefont {Qiu}},\ and\ \bibinfo {author} {\bibfnamefont {D.~G.}\
  \bibnamefont {Richards}},\ }\bibfield  {title} {\bibinfo {title} {{Pion
  valence quark distribution from current-current correlation in lattice
  QCD}},\ }\href {https://doi.org/10.1103/PhysRevD.102.054508} {\bibfield
  {journal} {\bibinfo  {journal} {Phys. Rev. D}\ }\textbf {\bibinfo {volume}
  {102}},\ \bibinfo {pages} {054508} (\bibinfo {year} {2020})},\ \Eprint
  {https://arxiv.org/abs/2001.04960} {arXiv:2001.04960 [hep-lat]} \BibitemShut
  {NoStop}%
\bibitem [{\citenamefont {Zhang}\ \emph {et~al.}(2019)\citenamefont {Zhang},
  \citenamefont {Chen}, \citenamefont {Jin}, \citenamefont {Lin}, \citenamefont
  {Sch{\"a}fer},\ and\ \citenamefont {Zhao}}]{Chen:2018fwa}%
  \BibitemOpen
  \bibfield  {author} {\bibinfo {author} {\bibfnamefont {J.-H.}\ \bibnamefont
  {Zhang}}, \bibinfo {author} {\bibfnamefont {J.-W.}\ \bibnamefont {Chen}},
  \bibinfo {author} {\bibfnamefont {L.}~\bibnamefont {Jin}}, \bibinfo {author}
  {\bibfnamefont {H.-W.}\ \bibnamefont {Lin}}, \bibinfo {author} {\bibfnamefont
  {A.}~\bibnamefont {Sch{\"a}fer}},\ and\ \bibinfo {author} {\bibfnamefont
  {Y.}~\bibnamefont {Zhao}},\ }\bibfield  {title} {\bibinfo {title} {{First
  direct lattice-QCD calculation of the $x$-dependence of the pion parton
  distribution function}},\ }\href
  {https://doi.org/10.1103/PhysRevD.100.034505} {\bibfield  {journal} {\bibinfo
   {journal} {Phys. Rev. D}\ }\textbf {\bibinfo {volume} {100}},\ \bibinfo
  {pages} {034505} (\bibinfo {year} {2019})},\ \Eprint
  {https://arxiv.org/abs/1804.01483} {arXiv:1804.01483 [hep-lat]} \BibitemShut
  {NoStop}%
\bibitem [{\citenamefont {Izubuchi}\ \emph {et~al.}(2019)\citenamefont
  {Izubuchi}, \citenamefont {Jin}, \citenamefont {Kallidonis}, \citenamefont
  {Karthik}, \citenamefont {Mukherjee}, \citenamefont {Petreczky},
  \citenamefont {Shugert},\ and\ \citenamefont {Syritsyn}}]{Izubuchi:2019lyk}%
  \BibitemOpen
  \bibfield  {author} {\bibinfo {author} {\bibfnamefont {T.}~\bibnamefont
  {Izubuchi}}, \bibinfo {author} {\bibfnamefont {L.}~\bibnamefont {Jin}},
  \bibinfo {author} {\bibfnamefont {C.}~\bibnamefont {Kallidonis}}, \bibinfo
  {author} {\bibfnamefont {N.}~\bibnamefont {Karthik}}, \bibinfo {author}
  {\bibfnamefont {S.}~\bibnamefont {Mukherjee}}, \bibinfo {author}
  {\bibfnamefont {P.}~\bibnamefont {Petreczky}}, \bibinfo {author}
  {\bibfnamefont {C.}~\bibnamefont {Shugert}},\ and\ \bibinfo {author}
  {\bibfnamefont {S.}~\bibnamefont {Syritsyn}},\ }\bibfield  {title} {\bibinfo
  {title} {{Valence parton distribution function of pion from fine lattice}},\
  }\href {https://doi.org/10.1103/PhysRevD.100.034516} {\bibfield  {journal}
  {\bibinfo  {journal} {Phys. Rev. D}\ }\textbf {\bibinfo {volume} {100}},\
  \bibinfo {pages} {034516} (\bibinfo {year} {2019})},\ \Eprint
  {https://arxiv.org/abs/1905.06349} {arXiv:1905.06349 [hep-lat]} \BibitemShut
  {NoStop}%
\bibitem [{\citenamefont {Jo{\'o}}\ \emph {et~al.}(2019)\citenamefont
  {Jo{\'o}}, \citenamefont {Karpie}, \citenamefont {Orginos}, \citenamefont
  {Radyushkin}, \citenamefont {Richards}, \citenamefont {Sufian},\ and\
  \citenamefont {Zafeiropoulos}}]{Joo:2019bzr}%
  \BibitemOpen
  \bibfield  {author} {\bibinfo {author} {\bibfnamefont {B.}~\bibnamefont
  {Jo{\'o}}}, \bibinfo {author} {\bibfnamefont {J.}~\bibnamefont {Karpie}},
  \bibinfo {author} {\bibfnamefont {K.}~\bibnamefont {Orginos}}, \bibinfo
  {author} {\bibfnamefont {A.~V.}\ \bibnamefont {Radyushkin}}, \bibinfo
  {author} {\bibfnamefont {D.~G.}\ \bibnamefont {Richards}}, \bibinfo {author}
  {\bibfnamefont {R.~S.}\ \bibnamefont {Sufian}},\ and\ \bibinfo {author}
  {\bibfnamefont {S.}~\bibnamefont {Zafeiropoulos}},\ }\bibfield  {title}
  {\bibinfo {title} {{Pion valence structure from Ioffe-time parton
  pseudodistribution functions}},\ }\href
  {https://doi.org/10.1103/PhysRevD.100.114512} {\bibfield  {journal} {\bibinfo
   {journal} {Phys. Rev. D}\ }\textbf {\bibinfo {volume} {100}},\ \bibinfo
  {pages} {114512} (\bibinfo {year} {2019})},\ \Eprint
  {https://arxiv.org/abs/1909.08517} {arXiv:1909.08517 [hep-lat]} \BibitemShut
  {NoStop}%
\bibitem [{\citenamefont {Manohar}(1992)}]{Manohar:1992tz}%
  \BibitemOpen
  \bibfield  {author} {\bibinfo {author} {\bibfnamefont {A.~V.}\ \bibnamefont
  {Manohar}},\ }\bibfield  {title} {\bibinfo {title} {{An introduction to spin
  dependent deep inelastic scattering}},\ }\href@noop {} {\  (\bibinfo {year}
  {1992})},\ \Eprint {https://arxiv.org/abs/hep-ph/9204208}
  {arXiv:hep-ph/9204208} \BibitemShut {NoStop}%
\bibitem [{\citenamefont {Ji}(1998)}]{Ji:1998pc}%
  \BibitemOpen
  \bibfield  {author} {\bibinfo {author} {\bibfnamefont {X.-D.}\ \bibnamefont
  {Ji}},\ }\bibfield  {title} {\bibinfo {title} {{Off forward parton
  distributions}},\ }\href {https://doi.org/10.1088/0954-3899/24/7/002}
  {\bibfield  {journal} {\bibinfo  {journal} {J. Phys. G}\ }\textbf {\bibinfo
  {volume} {24}},\ \bibinfo {pages} {1181} (\bibinfo {year} {1998})},\ \Eprint
  {https://arxiv.org/abs/hep-ph/9807358} {arXiv:hep-ph/9807358} \BibitemShut
  {NoStop}%
\bibitem [{\citenamefont {Dokshitzer}(1977)}]{Dokshitzer:1977sg}%
  \BibitemOpen
  \bibfield  {author} {\bibinfo {author} {\bibfnamefont {Y.~L.}\ \bibnamefont
  {Dokshitzer}},\ }\bibfield  {title} {\bibinfo {title} {{Calculation of the
  structure functions for deep inelastic scattering and $e^+ e^-$ annihilation
  by perturbation theory in quantum chromodynamics}},\ }\href@noop {}
  {\bibfield  {journal} {\bibinfo  {journal} {Sov. Phys. JETP}\ }\textbf
  {\bibinfo {volume} {46}},\ \bibinfo {pages} {641} (\bibinfo {year}
  {1977})}\BibitemShut {NoStop}%
\bibitem [{\citenamefont {Gribov}\ and\ \citenamefont
  {Lipatov}(1972)}]{Gribov:1972ri}%
  \BibitemOpen
  \bibfield  {author} {\bibinfo {author} {\bibfnamefont {V.~N.}\ \bibnamefont
  {Gribov}}\ and\ \bibinfo {author} {\bibfnamefont {L.~N.}\ \bibnamefont
  {Lipatov}},\ }\bibfield  {title} {\bibinfo {title} {{Deep inelastic $e p$
  scattering in perturbation theory}},\ }\href@noop {} {\bibfield  {journal}
  {\bibinfo  {journal} {Sov. J. Nucl. Phys.}\ }\textbf {\bibinfo {volume}
  {15}},\ \bibinfo {pages} {438} (\bibinfo {year} {1972})}\BibitemShut
  {NoStop}%
\bibitem [{\citenamefont {Altarelli}\ and\ \citenamefont
  {Parisi}(1977)}]{Altarelli:1977zs}%
  \BibitemOpen
  \bibfield  {author} {\bibinfo {author} {\bibfnamefont {G.}~\bibnamefont
  {Altarelli}}\ and\ \bibinfo {author} {\bibfnamefont {G.}~\bibnamefont
  {Parisi}},\ }\bibfield  {title} {\bibinfo {title} {{Asymptotic freedom in
  parton language}},\ }\href {https://doi.org/10.1016/0550-3213(77)90384-4}
  {\bibfield  {journal} {\bibinfo  {journal} {Nucl. Phys. B}\ }\textbf
  {\bibinfo {volume} {126}},\ \bibinfo {pages} {298} (\bibinfo {year}
  {1977})}\BibitemShut {NoStop}%
\bibitem [{\citenamefont {Diehl}(2003)}]{Diehl:2003ny}%
  \BibitemOpen
  \bibfield  {author} {\bibinfo {author} {\bibfnamefont {M.}~\bibnamefont
  {Diehl}},\ }\bibfield  {title} {\bibinfo {title} {{Generalized parton
  distributions}},\ }\href {https://doi.org/10.1016/j.physrep.2003.08.002}
  {\bibfield  {journal} {\bibinfo  {journal} {Phys. Rept.}\ }\textbf {\bibinfo
  {volume} {388}},\ \bibinfo {pages} {41} (\bibinfo {year} {2003})},\ \Eprint
  {https://arxiv.org/abs/hep-ph/0307382} {arXiv:hep-ph/0307382} \BibitemShut
  {NoStop}%
\bibitem [{\citenamefont {Callan}\ and\ \citenamefont
  {Gross}(1969)}]{Callan:1969uq}%
  \BibitemOpen
  \bibfield  {author} {\bibinfo {author} {\bibfnamefont {J.}~\bibnamefont
  {Callan}, \bibfnamefont {C.~G.}}\ and\ \bibinfo {author} {\bibfnamefont
  {D.~J.}\ \bibnamefont {Gross}},\ }\bibfield  {title} {\bibinfo {title}
  {{High-energy electroproduction and the constitution of the electric
  current}},\ }\href {https://doi.org/10.1103/PhysRevLett.22.156} {\bibfield
  {journal} {\bibinfo  {journal} {Phys. Rev. Lett.}\ }\textbf {\bibinfo
  {volume} {22}},\ \bibinfo {pages} {156} (\bibinfo {year} {1969})}\BibitemShut
  {NoStop}%
\bibitem [{\citenamefont {Ji}(1997)}]{Ji:1996ek}%
  \BibitemOpen
  \bibfield  {author} {\bibinfo {author} {\bibfnamefont {X.}~\bibnamefont
  {Ji}},\ }\bibfield  {title} {\bibinfo {title} {{Gauge-Invariant Decomposition
  of Nucleon Spin}},\ }\href {https://doi.org/10.1103/PhysRevLett.78.610}
  {\bibfield  {journal} {\bibinfo  {journal} {Phys. Rev. Lett.}\ }\textbf
  {\bibinfo {volume} {78}},\ \bibinfo {pages} {610} (\bibinfo {year} {1997})},\
  \Eprint {https://arxiv.org/abs/hep-ph/9603249} {arXiv:hep-ph/9603249}
  \BibitemShut {NoStop}%
\bibitem [{\citenamefont {Bruno}\ \emph {et~al.}(2015)\citenamefont {Bruno}
  \emph {et~al.}}]{Bruno:2014jqa}%
  \BibitemOpen
  \bibfield  {author} {\bibinfo {author} {\bibfnamefont {M.}~\bibnamefont
  {Bruno}} \emph {et~al.},\ }\bibfield  {title} {\bibinfo {title} {{Simulation
  of QCD with $N_f=2+1$ flavors of non-perturbatively improved Wilson
  fermions}},\ }\href {https://doi.org/10.1007/JHEP02(2015)043} {\bibfield
  {journal} {\bibinfo  {journal} {JHEP}\ }\textbf {\bibinfo {volume} {02}},\
  \bibinfo {pages} {043}},\ \Eprint {https://arxiv.org/abs/1411.3982}
  {arXiv:1411.3982 [hep-lat]} \BibitemShut {NoStop}%
\bibitem [{\citenamefont {L{\"u}scher}\ and\ \citenamefont
  {Schaefer}(2013)}]{Luscher:2012av}%
  \BibitemOpen
  \bibfield  {author} {\bibinfo {author} {\bibfnamefont {M.}~\bibnamefont
  {L{\"u}scher}}\ and\ \bibinfo {author} {\bibfnamefont {S.}~\bibnamefont
  {Schaefer}},\ }\bibfield  {title} {\bibinfo {title} {{Lattice QCD with open
  boundary conditions and twisted-mass reweighting}},\ }\href
  {https://doi.org/10.1016/j.cpc.2012.10.003} {\bibfield  {journal} {\bibinfo
  {journal} {Comput. Phys. Commun.}\ }\textbf {\bibinfo {volume} {184}},\
  \bibinfo {pages} {519} (\bibinfo {year} {2013})},\ \Eprint
  {https://arxiv.org/abs/1206.2809} {arXiv:1206.2809 [hep-lat]} \BibitemShut
  {NoStop}%
\bibitem [{\citenamefont {L{\"u}scher}\ and\ \citenamefont
  {Schaefer}(2011)}]{Luscher:2011kk}%
  \BibitemOpen
  \bibfield  {author} {\bibinfo {author} {\bibfnamefont {M.}~\bibnamefont
  {L{\"u}scher}}\ and\ \bibinfo {author} {\bibfnamefont {S.}~\bibnamefont
  {Schaefer}},\ }\bibfield  {title} {\bibinfo {title} {{Lattice QCD without
  topology barriers}},\ }\href {https://doi.org/10.1007/JHEP07(2011)036}
  {\bibfield  {journal} {\bibinfo  {journal} {JHEP}\ }\textbf {\bibinfo
  {volume} {07}},\ \bibinfo {pages} {036}},\ \Eprint
  {https://arxiv.org/abs/1105.4749} {arXiv:1105.4749 [hep-lat]} \BibitemShut
  {NoStop}%
\bibitem [{\citenamefont {Bali}\ \emph
  {et~al.}(2016{\natexlab{a}})\citenamefont {Bali}, \citenamefont {Scholz},
  \citenamefont {Simeth},\ and\ \citenamefont {S{\"o}ldner}}]{Bali:2016umi}%
  \BibitemOpen
  \bibfield  {author} {\bibinfo {author} {\bibfnamefont {G.~S.}\ \bibnamefont
  {Bali}}, \bibinfo {author} {\bibfnamefont {E.~E.}\ \bibnamefont {Scholz}},
  \bibinfo {author} {\bibfnamefont {J.}~\bibnamefont {Simeth}},\ and\ \bibinfo
  {author} {\bibfnamefont {W.}~\bibnamefont {S{\"o}ldner}} (\bibinfo
  {collaboration} {RQCD}),\ }\bibfield  {title} {\bibinfo {title} {{Lattice
  simulations with $N_f=2+1$ improved Wilson fermions at a fixed strange quark
  mass}},\ }\href {https://doi.org/10.1103/PhysRevD.94.074501} {\bibfield
  {journal} {\bibinfo  {journal} {Phys. Rev. D}\ }\textbf {\bibinfo {volume}
  {94}},\ \bibinfo {pages} {074501} (\bibinfo {year} {2016}{\natexlab{a}})},\
  \Eprint {https://arxiv.org/abs/1606.09039} {arXiv:1606.09039 [hep-lat]}
  \BibitemShut {NoStop}%
\bibitem [{\citenamefont {Bali}\ \emph
  {et~al.}(2020{\natexlab{a}})\citenamefont {Bali}, \citenamefont {Barca},
  \citenamefont {Collins}, \citenamefont {Gruber}, \citenamefont {L{\"o}ffler},
  \citenamefont {Sch{\"a}fer}, \citenamefont {S{\"o}ldner}, \citenamefont
  {Wein}, \citenamefont {Weish{\"a}upl},\ and\ \citenamefont
  {Wurm}}]{Bali:2019yiy}%
  \BibitemOpen
  \bibfield  {author} {\bibinfo {author} {\bibfnamefont {G.~S.}\ \bibnamefont
  {Bali}}, \bibinfo {author} {\bibfnamefont {L.}~\bibnamefont {Barca}},
  \bibinfo {author} {\bibfnamefont {S.}~\bibnamefont {Collins}}, \bibinfo
  {author} {\bibfnamefont {M.}~\bibnamefont {Gruber}}, \bibinfo {author}
  {\bibfnamefont {M.}~\bibnamefont {L{\"o}ffler}}, \bibinfo {author}
  {\bibfnamefont {A.}~\bibnamefont {Sch{\"a}fer}}, \bibinfo {author}
  {\bibfnamefont {W.}~\bibnamefont {S{\"o}ldner}}, \bibinfo {author}
  {\bibfnamefont {P.}~\bibnamefont {Wein}}, \bibinfo {author} {\bibfnamefont
  {S.}~\bibnamefont {Weish{\"a}upl}},\ and\ \bibinfo {author} {\bibfnamefont
  {T.}~\bibnamefont {Wurm}} (\bibinfo {collaboration} {RQCD}),\ }\bibfield
  {title} {\bibinfo {title} {{Nucleon axial structure from lattice QCD}},\
  }\href {https://doi.org/10.1007/JHEP05(2020)126} {\bibfield  {journal}
  {\bibinfo  {journal} {JHEP}\ }\textbf {\bibinfo {volume} {05}},\ \bibinfo
  {pages} {126}},\ \Eprint {https://arxiv.org/abs/1911.13150} {arXiv:1911.13150
  [hep-lat]} \BibitemShut {NoStop}%
\bibitem [{\citenamefont {Babich}\ \emph {et~al.}(2010)\citenamefont {Babich},
  \citenamefont {Brannick}, \citenamefont {Brower}, \citenamefont {Clark},
  \citenamefont {Manteuffel}, \citenamefont {McCormick}, \citenamefont
  {Osborn},\ and\ \citenamefont {Rebbi}}]{Babich:2010qb}%
  \BibitemOpen
  \bibfield  {author} {\bibinfo {author} {\bibfnamefont {R.}~\bibnamefont
  {Babich}}, \bibinfo {author} {\bibfnamefont {J.}~\bibnamefont {Brannick}},
  \bibinfo {author} {\bibfnamefont {R.~C.}\ \bibnamefont {Brower}}, \bibinfo
  {author} {\bibfnamefont {M.~A.}\ \bibnamefont {Clark}}, \bibinfo {author}
  {\bibfnamefont {T.~A.}\ \bibnamefont {Manteuffel}}, \bibinfo {author}
  {\bibfnamefont {S.~F.}\ \bibnamefont {McCormick}}, \bibinfo {author}
  {\bibfnamefont {J.~C.}\ \bibnamefont {Osborn}},\ and\ \bibinfo {author}
  {\bibfnamefont {C.}~\bibnamefont {Rebbi}},\ }\bibfield  {title} {\bibinfo
  {title} {{Adaptive multigrid algorithm for the lattice Wilson-Dirac
  operator}},\ }\href {https://doi.org/10.1103/PhysRevLett.105.201602}
  {\bibfield  {journal} {\bibinfo  {journal} {Phys. Rev. Lett.}\ }\textbf
  {\bibinfo {volume} {105}},\ \bibinfo {pages} {201602} (\bibinfo {year}
  {2010})},\ \Eprint {https://arxiv.org/abs/1005.3043} {arXiv:1005.3043
  [hep-lat]} \BibitemShut {NoStop}%
\bibitem [{\citenamefont {Frommer}\ \emph {et~al.}(2014)\citenamefont
  {Frommer}, \citenamefont {Kahl}, \citenamefont {Krieg}, \citenamefont
  {Leder},\ and\ \citenamefont {Rottmann}}]{Frommer:2013fsa}%
  \BibitemOpen
  \bibfield  {author} {\bibinfo {author} {\bibfnamefont {A.}~\bibnamefont
  {Frommer}}, \bibinfo {author} {\bibfnamefont {K.}~\bibnamefont {Kahl}},
  \bibinfo {author} {\bibfnamefont {S.}~\bibnamefont {Krieg}}, \bibinfo
  {author} {\bibfnamefont {B.}~\bibnamefont {Leder}},\ and\ \bibinfo {author}
  {\bibfnamefont {M.}~\bibnamefont {Rottmann}},\ }\bibfield  {title} {\bibinfo
  {title} {{Adaptive aggregation based domain decomposition multigrid for the
  lattice Wilson Dirac operator}},\ }\href {https://doi.org/10.1137/130919507}
  {\bibfield  {journal} {\bibinfo  {journal} {SIAM J. Sci. Comput.}\ }\textbf
  {\bibinfo {volume} {36}},\ \bibinfo {pages} {A1581} (\bibinfo {year}
  {2014})},\ \Eprint {https://arxiv.org/abs/1303.1377} {arXiv:1303.1377
  [hep-lat]} \BibitemShut {NoStop}%
\bibitem [{\citenamefont {Heybrock}\ \emph {et~al.}(2016)\citenamefont
  {Heybrock}, \citenamefont {Rottmann}, \citenamefont {Georg},\ and\
  \citenamefont {Wettig}}]{Heybrock:2015kpy}%
  \BibitemOpen
  \bibfield  {author} {\bibinfo {author} {\bibfnamefont {S.}~\bibnamefont
  {Heybrock}}, \bibinfo {author} {\bibfnamefont {M.}~\bibnamefont {Rottmann}},
  \bibinfo {author} {\bibfnamefont {P.}~\bibnamefont {Georg}},\ and\ \bibinfo
  {author} {\bibfnamefont {T.}~\bibnamefont {Wettig}},\ }\bibfield  {title}
  {\bibinfo {title} {{Adaptive algebraic multigrid on SIMD architectures}},\
  }\href {https://doi.org/10.22323/1.251.0036} {\bibfield  {journal} {\bibinfo
  {journal} {PoS}\ }\textbf {\bibinfo {volume} {LATTICE2015}},\ \bibinfo
  {pages} {036} (\bibinfo {year} {2016})},\ \Eprint
  {https://arxiv.org/abs/1512.04506} {arXiv:1512.04506 [physics.comp-ph]}
  \BibitemShut {NoStop}%
\bibitem [{\citenamefont {Richtmann}\ \emph {et~al.}(2016)\citenamefont
  {Richtmann}, \citenamefont {Heybrock},\ and\ \citenamefont
  {Wettig}}]{Richtmann:2016kcq}%
  \BibitemOpen
  \bibfield  {author} {\bibinfo {author} {\bibfnamefont {D.}~\bibnamefont
  {Richtmann}}, \bibinfo {author} {\bibfnamefont {S.}~\bibnamefont
  {Heybrock}},\ and\ \bibinfo {author} {\bibfnamefont {T.}~\bibnamefont
  {Wettig}},\ }\bibfield  {title} {\bibinfo {title} {{Multiple right-hand-side
  setup for the DD-$\alpha$AMG}},\ }\href {https://doi.org/10.22323/1.251.0035}
  {\bibfield  {journal} {\bibinfo  {journal} {PoS}\ }\textbf {\bibinfo {volume}
  {LATTICE2015}},\ \bibinfo {pages} {035} (\bibinfo {year} {2016})},\ \Eprint
  {https://arxiv.org/abs/1601.03184} {arXiv:1601.03184 [hep-lat]} \BibitemShut
  {NoStop}%
\bibitem [{\citenamefont {Georg}\ \emph {et~al.}(2017)\citenamefont {Georg},
  \citenamefont {Richtmann},\ and\ \citenamefont {Wettig}}]{Georg:2017diz}%
  \BibitemOpen
  \bibfield  {author} {\bibinfo {author} {\bibfnamefont {P.}~\bibnamefont
  {Georg}}, \bibinfo {author} {\bibfnamefont {D.}~\bibnamefont {Richtmann}},\
  and\ \bibinfo {author} {\bibfnamefont {T.}~\bibnamefont {Wettig}},\
  }\bibfield  {title} {\bibinfo {title} {{pMR: A high-performance communication
  library}},\ }\href {https://doi.org/10.22323/1.256.0361} {\bibfield
  {journal} {\bibinfo  {journal} {PoS}\ }\textbf {\bibinfo {volume}
  {LATTICE2016}},\ \bibinfo {pages} {361} (\bibinfo {year} {2017})},\ \Eprint
  {https://arxiv.org/abs/1701.08521} {arXiv:1701.08521 [hep-lat]} \BibitemShut
  {NoStop}%
\bibitem [{\citenamefont {Georg}\ \emph {et~al.}(2018)\citenamefont {Georg},
  \citenamefont {Richtmann},\ and\ \citenamefont {Wettig}}]{Georg:2017zua}%
  \BibitemOpen
  \bibfield  {author} {\bibinfo {author} {\bibfnamefont {P.}~\bibnamefont
  {Georg}}, \bibinfo {author} {\bibfnamefont {D.}~\bibnamefont {Richtmann}},\
  and\ \bibinfo {author} {\bibfnamefont {T.}~\bibnamefont {Wettig}},\
  }\bibfield  {title} {\bibinfo {title} {{DD-$\alpha$AMG on QPACE~3}},\ }\href
  {https://doi.org/10.1051/epjconf/201817502007} {\bibfield  {journal}
  {\bibinfo  {journal} {EPJ Web Conf.}\ }\textbf {\bibinfo {volume} {175}},\
  \bibinfo {pages} {02007} (\bibinfo {year} {2018})},\ \Eprint
  {https://arxiv.org/abs/1710.07041} {arXiv:1710.07041 [hep-lat]} \BibitemShut
  {NoStop}%
\bibitem [{\citenamefont {L{\"u}scher}(2007{\natexlab{a}})}]{Luscher:2007se}%
  \BibitemOpen
  \bibfield  {author} {\bibinfo {author} {\bibfnamefont {M.}~\bibnamefont
  {L{\"u}scher}},\ }\bibfield  {title} {\bibinfo {title} {{Local coherence and
  deflation of the low quark modes in lattice QCD}},\ }\href
  {https://doi.org/10.1088/1126-6708/2007/07/081} {\bibfield  {journal}
  {\bibinfo  {journal} {JHEP}\ }\textbf {\bibinfo {volume} {07}},\ \bibinfo
  {pages} {081}},\ \Eprint {https://arxiv.org/abs/0706.2298} {arXiv:0706.2298
  [hep-lat]} \BibitemShut {NoStop}%
\bibitem [{\citenamefont {L{\"u}scher}(2007{\natexlab{b}})}]{Luscher:2007es}%
  \BibitemOpen
  \bibfield  {author} {\bibinfo {author} {\bibfnamefont {M.}~\bibnamefont
  {L{\"u}scher}},\ }\bibfield  {title} {\bibinfo {title} {{Deflation
  acceleration of lattice QCD simulations}},\ }\href
  {https://doi.org/10.1088/1126-6708/2007/12/011} {\bibfield  {journal}
  {\bibinfo  {journal} {JHEP}\ }\textbf {\bibinfo {volume} {12}},\ \bibinfo
  {pages} {011}},\ \Eprint {https://arxiv.org/abs/0710.5417} {arXiv:0710.5417
  [hep-lat]} \BibitemShut {NoStop}%
\bibitem [{\citenamefont {G{\"u}sken}(1990)}]{Gusken:1989qx}%
  \BibitemOpen
  \bibfield  {author} {\bibinfo {author} {\bibfnamefont {S.}~\bibnamefont
  {G{\"u}sken}},\ }\bibfield  {title} {\bibinfo {title} {{A study of smearing
  techniques for hadron correlation functions}},\ }\href
  {https://doi.org/10.1016/0920-5632(90)90273-W} {\bibfield  {journal}
  {\bibinfo  {journal} {Nucl. Phys. B (Proc. Suppl.)}\ }\textbf {\bibinfo
  {volume} {17}},\ \bibinfo {pages} {361} (\bibinfo {year} {1990})}\BibitemShut
  {NoStop}%
\bibitem [{\citenamefont {Falcioni}\ \emph {et~al.}(1985)\citenamefont
  {Falcioni}, \citenamefont {Paciello}, \citenamefont {Parisi},\ and\
  \citenamefont {Taglienti}}]{Falcioni:1984ei}%
  \BibitemOpen
  \bibfield  {author} {\bibinfo {author} {\bibfnamefont {M.}~\bibnamefont
  {Falcioni}}, \bibinfo {author} {\bibfnamefont {M.~L.}\ \bibnamefont
  {Paciello}}, \bibinfo {author} {\bibfnamefont {G.}~\bibnamefont {Parisi}},\
  and\ \bibinfo {author} {\bibfnamefont {B.}~\bibnamefont {Taglienti}},\
  }\bibfield  {title} {\bibinfo {title} {{Again on SU(3) glueball mass}},\
  }\href {https://doi.org/10.1016/0550-3213(85)90280-9} {\bibfield  {journal}
  {\bibinfo  {journal} {Nucl. Phys. B}\ }\textbf {\bibinfo {volume} {251}},\
  \bibinfo {pages} {624} (\bibinfo {year} {1985})}\BibitemShut {NoStop}%
\bibitem [{\citenamefont {Edwards}\ and\ \citenamefont
  {Jo{\'o}}(2005)}]{Edwards:2004sx}%
  \BibitemOpen
  \bibfield  {author} {\bibinfo {author} {\bibfnamefont {R.~G.}\ \bibnamefont
  {Edwards}}\ and\ \bibinfo {author} {\bibfnamefont {B.}~\bibnamefont
  {Jo{\'o}}} (\bibinfo {collaboration} {SciDAC, LHPC, UKQCD}),\ }\bibfield
  {title} {\bibinfo {title} {{The Chroma Software System for Lattice QCD}},\
  }\href {https://doi.org/10.1016/j.nuclphysbps.2004.11.254} {\bibfield
  {journal} {\bibinfo  {journal} {Nucl. Phys. B (Proc. Suppl.)}\ }\textbf
  {\bibinfo {volume} {140}},\ \bibinfo {pages} {832} (\bibinfo {year}
  {2005})},\ \Eprint {https://arxiv.org/abs/hep-lat/0409003}
  {arXiv:hep-lat/0409003} \BibitemShut {NoStop}%
\bibitem [{\citenamefont {Alexandrou}\ \emph {et~al.}(2017)\citenamefont
  {Alexandrou}, \citenamefont {Constantinou}, \citenamefont {Hadjiyiannakou},
  \citenamefont {Jansen}, \citenamefont {Panagopoulos},\ and\ \citenamefont
  {Wiese}}]{Alexandrou:2016ekb}%
  \BibitemOpen
  \bibfield  {author} {\bibinfo {author} {\bibfnamefont {C.}~\bibnamefont
  {Alexandrou}}, \bibinfo {author} {\bibfnamefont {M.}~\bibnamefont
  {Constantinou}}, \bibinfo {author} {\bibfnamefont {K.}~\bibnamefont
  {Hadjiyiannakou}}, \bibinfo {author} {\bibfnamefont {K.}~\bibnamefont
  {Jansen}}, \bibinfo {author} {\bibfnamefont {H.}~\bibnamefont
  {Panagopoulos}},\ and\ \bibinfo {author} {\bibfnamefont {C.}~\bibnamefont
  {Wiese}},\ }\bibfield  {title} {\bibinfo {title} {{Gluon momentum fraction of
  the nucleon from lattice QCD}},\ }\href
  {https://doi.org/10.1103/PhysRevD.96.054503} {\bibfield  {journal} {\bibinfo
  {journal} {Phys. Rev. D}\ }\textbf {\bibinfo {volume} {96}},\ \bibinfo
  {pages} {054503} (\bibinfo {year} {2017})},\ \Eprint
  {https://arxiv.org/abs/1611.06901} {arXiv:1611.06901 [hep-lat]} \BibitemShut
  {NoStop}%
\bibitem [{\citenamefont {Shanahan}\ and\ \citenamefont
  {Detmold}(2019)}]{Shanahan:2018pib}%
  \BibitemOpen
  \bibfield  {author} {\bibinfo {author} {\bibfnamefont {P.~E.}\ \bibnamefont
  {Shanahan}}\ and\ \bibinfo {author} {\bibfnamefont {W.}~\bibnamefont
  {Detmold}},\ }\bibfield  {title} {\bibinfo {title} {{Gluon gravitational form
  factors of the nucleon and the pion from lattice QCD}},\ }\href
  {https://doi.org/10.1103/PhysRevD.99.014511} {\bibfield  {journal} {\bibinfo
  {journal} {Phys. Rev. D}\ }\textbf {\bibinfo {volume} {99}},\ \bibinfo
  {pages} {014511} (\bibinfo {year} {2019})},\ \Eprint
  {https://arxiv.org/abs/1810.04626} {arXiv:1810.04626 [hep-lat]} \BibitemShut
  {NoStop}%
\bibitem [{\citenamefont {G{\"o}ckeler}\ \emph {et~al.}(1996)\citenamefont
  {G{\"o}ckeler}, \citenamefont {Horsley}, \citenamefont {Ilgenfritz},
  \citenamefont {Perlt}, \citenamefont {Rakow}, \citenamefont {Schierholz},\
  and\ \citenamefont {Schiller}}]{Gockeler:1996mu}%
  \BibitemOpen
  \bibfield  {author} {\bibinfo {author} {\bibfnamefont {M.}~\bibnamefont
  {G{\"o}ckeler}}, \bibinfo {author} {\bibfnamefont {R.}~\bibnamefont
  {Horsley}}, \bibinfo {author} {\bibfnamefont {E.-M.}\ \bibnamefont
  {Ilgenfritz}}, \bibinfo {author} {\bibfnamefont {H.}~\bibnamefont {Perlt}},
  \bibinfo {author} {\bibfnamefont {P.~E.~L.}\ \bibnamefont {Rakow}}, \bibinfo
  {author} {\bibfnamefont {G.}~\bibnamefont {Schierholz}},\ and\ \bibinfo
  {author} {\bibfnamefont {A.}~\bibnamefont {Schiller}},\ }\bibfield  {title}
  {\bibinfo {title} {{Lattice operators for moments of the structure functions
  and their transformation under the hypercubic group}},\ }\href
  {https://doi.org/10.1103/PhysRevD.54.5705} {\bibfield  {journal} {\bibinfo
  {journal} {Phys. Rev. D}\ }\textbf {\bibinfo {volume} {54}},\ \bibinfo
  {pages} {5705} (\bibinfo {year} {1996})},\ \Eprint
  {https://arxiv.org/abs/hep-lat/9602029} {arXiv:hep-lat/9602029} \BibitemShut
  {NoStop}%
\bibitem [{\citenamefont {Bali}\ \emph {et~al.}(2021)\citenamefont {Bali},
  \citenamefont {B\"urger}, \citenamefont {Collins}, \citenamefont
  {G\"ockeler}, \citenamefont {Gruber}, \citenamefont {Piemonte}, \citenamefont
  {Sch\"afer}, \citenamefont {Sternbeck},\ and\ \citenamefont
  {Wein}}]{Bali:2020lwx}%
  \BibitemOpen
  \bibfield  {author} {\bibinfo {author} {\bibfnamefont {G.~S.}\ \bibnamefont
  {Bali}}, \bibinfo {author} {\bibfnamefont {S.}~\bibnamefont {B\"urger}},
  \bibinfo {author} {\bibfnamefont {S.}~\bibnamefont {Collins}}, \bibinfo
  {author} {\bibfnamefont {M.}~\bibnamefont {G\"ockeler}}, \bibinfo {author}
  {\bibfnamefont {M.}~\bibnamefont {Gruber}}, \bibinfo {author} {\bibfnamefont
  {S.}~\bibnamefont {Piemonte}}, \bibinfo {author} {\bibfnamefont
  {A.}~\bibnamefont {Sch\"afer}}, \bibinfo {author} {\bibfnamefont
  {A.}~\bibnamefont {Sternbeck}},\ and\ \bibinfo {author} {\bibfnamefont
  {P.}~\bibnamefont {Wein}},\ }\bibfield  {title} {\bibinfo {title}
  {{Nonperturbative Renormalization in Lattice QCD with three Flavors of Clover
  Fermions: Using Periodic and Open Boundary Conditions}},\ }\href
  {https://doi.org/10.1103/PhysRevD.103.094511} {\bibfield  {journal} {\bibinfo
   {journal} {Phys. Rev. D}\ }\textbf {\bibinfo {volume} {103}},\ \bibinfo
  {pages} {094511} (\bibinfo {year} {2021})},\ \Eprint
  {https://arxiv.org/abs/2012.06284} {arXiv:2012.06284 [hep-lat]} \BibitemShut
  {NoStop}%
\bibitem [{\citenamefont {Werner}\ \emph {et~al.}(2020)\citenamefont {Werner}
  \emph {et~al.}}]{Werner:2019hxc}%
  \BibitemOpen
  \bibfield  {author} {\bibinfo {author} {\bibfnamefont {M.}~\bibnamefont
  {Werner}} \emph {et~al.} (\bibinfo {collaboration} {ETMC}),\ }\bibfield
  {title} {\bibinfo {title} {{Hadron--Hadron interactions from $N_f=2+1+1$
  lattice QCD: the $\rho$-resonance}},\ }\href
  {https://doi.org/10.1140/epja/s10050-020-00057-4} {\bibfield  {journal}
  {\bibinfo  {journal} {Eur. Phys. J.}\ }\textbf {\bibinfo {volume} {A56}},\
  \bibinfo {pages} {61} (\bibinfo {year} {2020})},\ \Eprint
  {https://arxiv.org/abs/1907.01237} {arXiv:1907.01237 [hep-lat]} \BibitemShut
  {NoStop}%
\bibitem [{\citenamefont {G{\"o}ckeler}\ \emph {et~al.}(2012)\citenamefont
  {G{\"o}ckeler}, \citenamefont {Horsley}, \citenamefont {Lage}, \citenamefont
  {Mei{\ss}ner}, \citenamefont {Rakow}, \citenamefont {Rusetsky}, \citenamefont
  {Schierholz},\ and\ \citenamefont {Zanotti}}]{Gockeler:2012yj}%
  \BibitemOpen
  \bibfield  {author} {\bibinfo {author} {\bibfnamefont {M.}~\bibnamefont
  {G{\"o}ckeler}}, \bibinfo {author} {\bibfnamefont {R.}~\bibnamefont
  {Horsley}}, \bibinfo {author} {\bibfnamefont {M.}~\bibnamefont {Lage}},
  \bibinfo {author} {\bibfnamefont {U.-G.}\ \bibnamefont {Mei{\ss}ner}},
  \bibinfo {author} {\bibfnamefont {P.~E.~L.}\ \bibnamefont {Rakow}}, \bibinfo
  {author} {\bibfnamefont {A.}~\bibnamefont {Rusetsky}}, \bibinfo {author}
  {\bibfnamefont {G.}~\bibnamefont {Schierholz}},\ and\ \bibinfo {author}
  {\bibfnamefont {J.~M.}\ \bibnamefont {Zanotti}},\ }\bibfield  {title}
  {\bibinfo {title} {{Scattering phases for meson and baryon resonances on
  general moving-frame lattices}},\ }\href
  {https://doi.org/10.1103/PhysRevD.86.094513} {\bibfield  {journal} {\bibinfo
  {journal} {Phys. Rev.}\ }\textbf {\bibinfo {volume} {D86}},\ \bibinfo {pages}
  {094513} (\bibinfo {year} {2012})},\ \Eprint
  {https://arxiv.org/abs/1206.4141} {arXiv:1206.4141 [hep-lat]} \BibitemShut
  {NoStop}%
\bibitem [{\citenamefont {Erben}\ \emph {et~al.}(2020)\citenamefont {Erben},
  \citenamefont {Green}, \citenamefont {Mohler},\ and\ \citenamefont
  {Wittig}}]{Erben:2019nmx}%
  \BibitemOpen
  \bibfield  {author} {\bibinfo {author} {\bibfnamefont {F.}~\bibnamefont
  {Erben}}, \bibinfo {author} {\bibfnamefont {J.~R.}\ \bibnamefont {Green}},
  \bibinfo {author} {\bibfnamefont {D.}~\bibnamefont {Mohler}},\ and\ \bibinfo
  {author} {\bibfnamefont {H.}~\bibnamefont {Wittig}},\ }\bibfield  {title}
  {\bibinfo {title} {{Rho resonance, timelike pion form factor, and
  implications for lattice studies of the hadronic vacuum polarization}},\
  }\href {https://doi.org/10.1103/PhysRevD.101.054504} {\bibfield  {journal}
  {\bibinfo  {journal} {Phys. Rev.}\ }\textbf {\bibinfo {volume} {D101}},\
  \bibinfo {pages} {054504} (\bibinfo {year} {2020})},\ \Eprint
  {https://arxiv.org/abs/1910.01083} {arXiv:1910.01083 [hep-lat]} \BibitemShut
  {NoStop}%
\bibitem [{\citenamefont {L{\"u}scher}(1986)}]{Luscher:1986pf}%
  \BibitemOpen
  \bibfield  {author} {\bibinfo {author} {\bibfnamefont {M.}~\bibnamefont
  {L{\"u}scher}},\ }\bibfield  {title} {\bibinfo {title} {{Volume Dependence of
  the Energy Spectrum in Massive Quantum Field Theories. 2. Scattering
  States}},\ }\href {https://doi.org/10.1007/BF01211097} {\bibfield  {journal}
  {\bibinfo  {journal} {Commun. Math. Phys.}\ }\textbf {\bibinfo {volume}
  {105}},\ \bibinfo {pages} {153} (\bibinfo {year} {1986})}\BibitemShut
  {NoStop}%
\bibitem [{\citenamefont {L{\"u}scher}(1991)}]{Luscher:1990ux}%
  \BibitemOpen
  \bibfield  {author} {\bibinfo {author} {\bibfnamefont {M.}~\bibnamefont
  {L{\"u}scher}},\ }\bibfield  {title} {\bibinfo {title} {{Two-particle states
  on a torus and their relation to the scattering matrix}},\ }\href
  {https://doi.org/10.1016/0550-3213(91)90366-6} {\bibfield  {journal}
  {\bibinfo  {journal} {Nucl. Phys.}\ }\textbf {\bibinfo {volume} {B354}},\
  \bibinfo {pages} {531} (\bibinfo {year} {1991})}\BibitemShut {NoStop}%
\bibitem [{\citenamefont {Dudek}\ \emph {et~al.}(2013)\citenamefont {Dudek},
  \citenamefont {Edwards},\ and\ \citenamefont {Thomas}}]{Dudek:2012xn}%
  \BibitemOpen
  \bibfield  {author} {\bibinfo {author} {\bibfnamefont {J.~J.}\ \bibnamefont
  {Dudek}}, \bibinfo {author} {\bibfnamefont {R.~G.}\ \bibnamefont {Edwards}},\
  and\ \bibinfo {author} {\bibfnamefont {C.~E.}\ \bibnamefont {Thomas}}
  (\bibinfo {collaboration} {HSC}),\ }\bibfield  {title} {\bibinfo {title}
  {{Energy dependence of the $\rho$ resonance in $\pi\pi$ elastic scattering
  from lattice QCD}},\ }\href {https://doi.org/10.1103/PhysRevD.87.034505}
  {\bibfield  {journal} {\bibinfo  {journal} {Phys. Rev.}\ }\textbf {\bibinfo
  {volume} {D87}},\ \bibinfo {pages} {034505} (\bibinfo {year} {2013})},\
  \bibinfo {note} {[Erratum: Phys.~Rev.~{\bf{D90}}~(2014)~099902(E)]},\ \Eprint
  {https://arxiv.org/abs/1212.0830} {arXiv:1212.0830 [hep-ph]} \BibitemShut
  {NoStop}%
\bibitem [{\citenamefont {Wilson}\ \emph {et~al.}(2015)\citenamefont {Wilson},
  \citenamefont {Brice\~no}, \citenamefont {Dudek}, \citenamefont {Edwards},\
  and\ \citenamefont {Thomas}}]{Wilson:2015dqa}%
  \BibitemOpen
  \bibfield  {author} {\bibinfo {author} {\bibfnamefont {D.~J.}\ \bibnamefont
  {Wilson}}, \bibinfo {author} {\bibfnamefont {R.~A.}\ \bibnamefont
  {Brice\~no}}, \bibinfo {author} {\bibfnamefont {J.~J.}\ \bibnamefont
  {Dudek}}, \bibinfo {author} {\bibfnamefont {R.~G.}\ \bibnamefont {Edwards}},\
  and\ \bibinfo {author} {\bibfnamefont {C.~E.}\ \bibnamefont {Thomas}},\
  }\bibfield  {title} {\bibinfo {title} {{Coupled $\pi\pi, K\bar{K}$ scattering
  in $P$-wave and the $\rho$ resonance from lattice QCD}},\ }\href
  {https://doi.org/10.1103/PhysRevD.92.094502} {\bibfield  {journal} {\bibinfo
  {journal} {Phys. Rev. D}\ }\textbf {\bibinfo {volume} {92}},\ \bibinfo
  {pages} {094502} (\bibinfo {year} {2015})},\ \Eprint
  {https://arxiv.org/abs/1507.02599} {arXiv:1507.02599 [hep-ph]} \BibitemShut
  {NoStop}%
\bibitem [{\citenamefont {Gounaris}\ and\ \citenamefont
  {Sakurai}(1968)}]{Gounaris:1968mw}%
  \BibitemOpen
  \bibfield  {author} {\bibinfo {author} {\bibfnamefont {G.~J.}\ \bibnamefont
  {Gounaris}}\ and\ \bibinfo {author} {\bibfnamefont {J.~J.}\ \bibnamefont
  {Sakurai}},\ }\bibfield  {title} {\bibinfo {title} {{Finite-width corrections
  to the vector-meson-dominance prediction for $\rho \to e^+ e^-$}},\ }\href
  {https://doi.org/10.1103/PhysRevLett.21.244} {\bibfield  {journal} {\bibinfo
  {journal} {Phys. Rev. Lett.}\ }\textbf {\bibinfo {volume} {21}},\ \bibinfo
  {pages} {244} (\bibinfo {year} {1968})}\BibitemShut {NoStop}%
\bibitem [{\citenamefont {Kawarabayashi}\ and\ \citenamefont
  {Suzuki}(1966)}]{Kawarabayashi:1966kd}%
  \BibitemOpen
  \bibfield  {author} {\bibinfo {author} {\bibfnamefont {K.}~\bibnamefont
  {Kawarabayashi}}\ and\ \bibinfo {author} {\bibfnamefont {M.}~\bibnamefont
  {Suzuki}},\ }\bibfield  {title} {\bibinfo {title} {{Partially conserved
  axial-vector current and the decays of vector mesons}},\ }\href
  {https://doi.org/10.1103/PhysRevLett.16.255} {\bibfield  {journal} {\bibinfo
  {journal} {Phys. Rev. Lett.}\ }\textbf {\bibinfo {volume} {16}},\ \bibinfo
  {pages} {255} (\bibinfo {year} {1966})}\BibitemShut {NoStop}%
\bibitem [{\citenamefont {Riazuddin}\ and\ \citenamefont
  {Fayyazuddin}(1966)}]{Riazuddin:1966sw}%
  \BibitemOpen
  \bibfield  {author} {\bibinfo {author} {\bibnamefont {Riazuddin}}\ and\
  \bibinfo {author} {\bibnamefont {Fayyazuddin}},\ }\bibfield  {title}
  {\bibinfo {title} {{Algebra of current components and decay widths of $\rho$
  and $K^*$ mesons}},\ }\href {https://doi.org/10.1103/PhysRev.147.1071}
  {\bibfield  {journal} {\bibinfo  {journal} {Phys. Rev.}\ }\textbf {\bibinfo
  {volume} {147}},\ \bibinfo {pages} {1071} (\bibinfo {year}
  {1966})}\BibitemShut {NoStop}%
\bibitem [{\citenamefont {Djukanovic}\ \emph {et~al.}(2004)\citenamefont
  {Djukanovic}, \citenamefont {Schindler}, \citenamefont {Gegelia},
  \citenamefont {Japaridze},\ and\ \citenamefont
  {Scherer}}]{Djukanovic:2004mm}%
  \BibitemOpen
  \bibfield  {author} {\bibinfo {author} {\bibfnamefont {D.}~\bibnamefont
  {Djukanovic}}, \bibinfo {author} {\bibfnamefont {M.~R.}\ \bibnamefont
  {Schindler}}, \bibinfo {author} {\bibfnamefont {J.}~\bibnamefont {Gegelia}},
  \bibinfo {author} {\bibfnamefont {G.}~\bibnamefont {Japaridze}},\ and\
  \bibinfo {author} {\bibfnamefont {S.}~\bibnamefont {Scherer}},\ }\bibfield
  {title} {\bibinfo {title} {{Universality of the $\rho$ meson coupling in
  effective field theory}},\ }\href
  {https://doi.org/10.1103/PhysRevLett.93.122002} {\bibfield  {journal}
  {\bibinfo  {journal} {Phys. Rev. Lett.}\ }\textbf {\bibinfo {volume} {93}},\
  \bibinfo {pages} {122002} (\bibinfo {year} {2004})},\ \Eprint
  {https://arxiv.org/abs/hep-ph/0407239} {arXiv:hep-ph/0407239} \BibitemShut
  {NoStop}%
\bibitem [{\citenamefont {Feng}\ \emph {et~al.}(2015)\citenamefont {Feng},
  \citenamefont {Aoki}, \citenamefont {Hashimoto},\ and\ \citenamefont
  {Kaneko}}]{Feng:2014gba}%
  \BibitemOpen
  \bibfield  {author} {\bibinfo {author} {\bibfnamefont {X.}~\bibnamefont
  {Feng}}, \bibinfo {author} {\bibfnamefont {S.}~\bibnamefont {Aoki}}, \bibinfo
  {author} {\bibfnamefont {S.}~\bibnamefont {Hashimoto}},\ and\ \bibinfo
  {author} {\bibfnamefont {T.}~\bibnamefont {Kaneko}},\ }\bibfield  {title}
  {\bibinfo {title} {{Timelike pion form factor in lattice QCD}},\ }\href
  {https://doi.org/10.1103/PhysRevD.91.054504} {\bibfield  {journal} {\bibinfo
  {journal} {Phys. Rev.}\ }\textbf {\bibinfo {volume} {D91}},\ \bibinfo {pages}
  {054504} (\bibinfo {year} {2015})},\ \Eprint
  {https://arxiv.org/abs/1412.6319} {arXiv:1412.6319 [hep-lat]} \BibitemShut
  {NoStop}%
\bibitem [{\citenamefont {Meyer}(2011)}]{Meyer:2011um}%
  \BibitemOpen
  \bibfield  {author} {\bibinfo {author} {\bibfnamefont {H.~B.}\ \bibnamefont
  {Meyer}},\ }\bibfield  {title} {\bibinfo {title} {{Lattice QCD and the
  Timelike Pion Form Factor}},\ }\href
  {https://doi.org/10.1103/PhysRevLett.107.072002} {\bibfield  {journal}
  {\bibinfo  {journal} {Phys. Rev. Lett.}\ }\textbf {\bibinfo {volume} {107}},\
  \bibinfo {pages} {072002} (\bibinfo {year} {2011})},\ \Eprint
  {https://arxiv.org/abs/1105.1892} {arXiv:1105.1892 [hep-lat]} \BibitemShut
  {NoStop}%
\bibitem [{\citenamefont {L{\"u}scher}\ and\ \citenamefont
  {Wolff}(1990)}]{Luscher:1990ck}%
  \BibitemOpen
  \bibfield  {author} {\bibinfo {author} {\bibfnamefont {M.}~\bibnamefont
  {L{\"u}scher}}\ and\ \bibinfo {author} {\bibfnamefont {U.}~\bibnamefont
  {Wolff}},\ }\bibfield  {title} {\bibinfo {title} {{How to calculate the
  elastic scattering matrix in two-dimensional quantum field theories by
  numerical simulation}},\ }\href
  {https://doi.org/10.1016/0550-3213(90)90540-T} {\bibfield  {journal}
  {\bibinfo  {journal} {Nucl. Phys. B}\ }\textbf {\bibinfo {volume} {339}},\
  \bibinfo {pages} {222} (\bibinfo {year} {1990})}\BibitemShut {NoStop}%
\bibitem [{\citenamefont {Bali}\ \emph
  {et~al.}(2016{\natexlab{b}})\citenamefont {Bali}, \citenamefont {Collins},
  \citenamefont {Cox}, \citenamefont {Donald}, \citenamefont {G{\"o}ckeler},
  \citenamefont {Lang},\ and\ \citenamefont {Sch{\"a}fer}}]{Bali:2015gji}%
  \BibitemOpen
  \bibfield  {author} {\bibinfo {author} {\bibfnamefont {G.~S.}\ \bibnamefont
  {Bali}}, \bibinfo {author} {\bibfnamefont {S.}~\bibnamefont {Collins}},
  \bibinfo {author} {\bibfnamefont {A.}~\bibnamefont {Cox}}, \bibinfo {author}
  {\bibfnamefont {G.}~\bibnamefont {Donald}}, \bibinfo {author} {\bibfnamefont
  {M.}~\bibnamefont {G{\"o}ckeler}}, \bibinfo {author} {\bibfnamefont {C.~B.}\
  \bibnamefont {Lang}},\ and\ \bibinfo {author} {\bibfnamefont
  {A.}~\bibnamefont {Sch{\"a}fer}} (\bibinfo {collaboration} {RQCD}),\
  }\bibfield  {title} {\bibinfo {title} {{$\rho$ and $K^*$ resonances on the
  lattice at nearly physical quark masses and $N_f=2$}},\ }\href
  {https://doi.org/10.1103/PhysRevD.93.054509} {\bibfield  {journal} {\bibinfo
  {journal} {Phys. Rev. D}\ }\textbf {\bibinfo {volume} {93}},\ \bibinfo
  {pages} {054509} (\bibinfo {year} {2016}{\natexlab{b}})},\ \Eprint
  {https://arxiv.org/abs/1512.08678} {arXiv:1512.08678 [hep-lat]} \BibitemShut
  {NoStop}%
\bibitem [{\citenamefont {Fischer}\ \emph
  {et~al.}(2020{\natexlab{a}})\citenamefont {Fischer}, \citenamefont
  {Kostrzewa}, \citenamefont {Mai}, \citenamefont {Petschlies}, \citenamefont
  {Pittler}, \citenamefont {Ueding}, \citenamefont {Urbach},\ and\
  \citenamefont {Werner}}]{Fischer:2020fvl}%
  \BibitemOpen
  \bibfield  {author} {\bibinfo {author} {\bibfnamefont {M.}~\bibnamefont
  {Fischer}}, \bibinfo {author} {\bibfnamefont {B.}~\bibnamefont {Kostrzewa}},
  \bibinfo {author} {\bibfnamefont {M.}~\bibnamefont {Mai}}, \bibinfo {author}
  {\bibfnamefont {M.}~\bibnamefont {Petschlies}}, \bibinfo {author}
  {\bibfnamefont {F.}~\bibnamefont {Pittler}}, \bibinfo {author} {\bibfnamefont
  {M.}~\bibnamefont {Ueding}}, \bibinfo {author} {\bibfnamefont
  {C.}~\bibnamefont {Urbach}},\ and\ \bibinfo {author} {\bibfnamefont
  {M.}~\bibnamefont {Werner}} (\bibinfo {collaboration} {ETM}),\ }\bibfield
  {title} {\bibinfo {title} {{The $\rho$-resonance with physical pion mass from
  $N_f=2$ lattice QCD}},\ }\href@noop {} {\  (\bibinfo {year}
  {2020}{\natexlab{a}})},\ \Eprint {https://arxiv.org/abs/2006.13805}
  {arXiv:2006.13805 [hep-lat]} \BibitemShut {NoStop}%
\bibitem [{\citenamefont {Fischer}\ \emph
  {et~al.}(2020{\natexlab{b}})\citenamefont {Fischer}, \citenamefont
  {Kostrzewa}, \citenamefont {Ostmeyer}, \citenamefont {Ottnad}, \citenamefont
  {Ueding},\ and\ \citenamefont {Urbach}}]{Fischer:2020bgv}%
  \BibitemOpen
  \bibfield  {author} {\bibinfo {author} {\bibfnamefont {M.}~\bibnamefont
  {Fischer}}, \bibinfo {author} {\bibfnamefont {B.}~\bibnamefont {Kostrzewa}},
  \bibinfo {author} {\bibfnamefont {J.}~\bibnamefont {Ostmeyer}}, \bibinfo
  {author} {\bibfnamefont {K.}~\bibnamefont {Ottnad}}, \bibinfo {author}
  {\bibfnamefont {M.}~\bibnamefont {Ueding}},\ and\ \bibinfo {author}
  {\bibfnamefont {C.}~\bibnamefont {Urbach}},\ }\bibfield  {title} {\bibinfo
  {title} {{On the generalised eigenvalue method and its relation to Prony and
  generalised pencil of function methods}},\ }\href
  {https://doi.org/10.1140/epja/s10050-020-00205-w} {\bibfield  {journal}
  {\bibinfo  {journal} {Eur. Phys. J. A}\ }\textbf {\bibinfo {volume} {56}},\
  \bibinfo {pages} {206} (\bibinfo {year} {2020}{\natexlab{b}})},\ \Eprint
  {https://arxiv.org/abs/2004.10472} {arXiv:2004.10472 [hep-lat]} \BibitemShut
  {NoStop}%
\bibitem [{\citenamefont {Bali}\ \emph {et~al.}()\citenamefont {Bali} \emph
  {et~al.}}]{Bali:2023}%
  \BibitemOpen
  \bibfield  {author} {\bibinfo {author} {\bibfnamefont {G.~S.}\ \bibnamefont
  {Bali}} \emph {et~al.},\ }\bibfield  {title} {\bibinfo {title} {{Scale
  setting and the light hadron spectrum in $N_f = 2 + 1$ QCD with Wilson
  fermions}},\ }\href@noop {} {\bibfield  {journal} {\bibinfo  {journal} {to be
  published}\ }}\bibinfo {note} {{in preparation}}\BibitemShut {NoStop}%
\bibitem [{\citenamefont {{J{\"u}lich Supercomputing
  Centre}}(2018)}]{Juelich:2018}%
  \BibitemOpen
  \bibfield  {author} {\bibinfo {author} {\bibnamefont {{J{\"u}lich
  Supercomputing Centre}}},\ }\bibfield  {title} {\bibinfo {title} {{JURECA:
  Modular supercomputer at J{\"u}lich Supercomputing Centre}},\ }\href
  {https://doi.org/http://dx.doi.org/10.17815/jlsrf-4-121-1} {\bibfield
  {journal} {\bibinfo  {journal} {Journal of large-scale research facilities}\
  }\textbf {\bibinfo {volume} {4}},\ \bibinfo {pages} {A132} (\bibinfo {year}
  {2018})}\BibitemShut {NoStop}%
\bibitem [{\citenamefont {{{Jülich Supercomputing Centre}}}(2019)}]{hdf}%
  \BibitemOpen
  \bibfield  {author} {\bibinfo {author} {\bibnamefont {{{Jülich
  Supercomputing Centre}}}},\ }\bibfield  {title} {\bibinfo {title} {{HDF Cloud
  – Helmholtz Data Federation Cloud Resources at the Jülich Supercomputing
  Centre}},\ }\href {https://doi.org/10.17815/jlsrf-5-173} {\bibfield
  {journal} {\bibinfo  {journal} {Journal of Large-Scale Research Facilities}\
  }\textbf {\bibinfo {volume} {5}},\ \bibinfo {pages} {A137} (\bibinfo {year}
  {2019})}\BibitemShut {NoStop}%
\bibitem [{\citenamefont {Martinelli}\ and\ \citenamefont
  {Sachrajda}(1989)}]{Martinelli:1988rr}%
  \BibitemOpen
  \bibfield  {author} {\bibinfo {author} {\bibfnamefont {G.}~\bibnamefont
  {Martinelli}}\ and\ \bibinfo {author} {\bibfnamefont {C.~T.}\ \bibnamefont
  {Sachrajda}},\ }\bibfield  {title} {\bibinfo {title} {{A lattice study of
  nucleon structure}},\ }\href {https://doi.org/10.1016/0550-3213(89)90035-7}
  {\bibfield  {journal} {\bibinfo  {journal} {Nucl. Phys. B}\ }\textbf
  {\bibinfo {volume} {316}},\ \bibinfo {pages} {355} (\bibinfo {year}
  {1989})}\BibitemShut {NoStop}%
\bibitem [{\citenamefont {Bali}\ \emph {et~al.}(2018)\citenamefont {Bali},
  \citenamefont {Collins}, \citenamefont {Gl{\"a}{\ss}le}, \citenamefont
  {Heybrock}, \citenamefont {Korcyl}, \citenamefont {L{\"o}ffler},
  \citenamefont {R{\"o}dl},\ and\ \citenamefont {Sch{\"a}fer}}]{Bali:2017mft}%
  \BibitemOpen
  \bibfield  {author} {\bibinfo {author} {\bibfnamefont {G.~S.}\ \bibnamefont
  {Bali}}, \bibinfo {author} {\bibfnamefont {S.}~\bibnamefont {Collins}},
  \bibinfo {author} {\bibfnamefont {B.}~\bibnamefont {Gl{\"a}{\ss}le}},
  \bibinfo {author} {\bibfnamefont {S.}~\bibnamefont {Heybrock}}, \bibinfo
  {author} {\bibfnamefont {P.}~\bibnamefont {Korcyl}}, \bibinfo {author}
  {\bibfnamefont {M.}~\bibnamefont {L{\"o}ffler}}, \bibinfo {author}
  {\bibfnamefont {R.}~\bibnamefont {R{\"o}dl}},\ and\ \bibinfo {author}
  {\bibfnamefont {A.}~\bibnamefont {Sch{\"a}fer}},\ }\bibfield  {title}
  {\bibinfo {title} {{Baryonic and mesonic 3-point functions with open spin
  indices}},\ }\href {https://doi.org/10.1051/epjconf/201817506014} {\bibfield
  {journal} {\bibinfo  {journal} {EPJ Web Conf.}\ }\textbf {\bibinfo {volume}
  {175}},\ \bibinfo {pages} {06014} (\bibinfo {year} {2018})},\ \Eprint
  {https://arxiv.org/abs/1711.02384} {arXiv:1711.02384 [hep-lat]} \BibitemShut
  {NoStop}%
\bibitem [{\citenamefont {R\"odl}(2020)}]{Rodl:2020ulu}%
  \BibitemOpen
  \bibfield  {author} {\bibinfo {author} {\bibfnamefont {R.~H.}\ \bibnamefont
  {R\"odl}},\ }\emph {\bibinfo {title} {{Nucleon generalized form factors from
  two-flavor lattice QCD and stochastic three-point functions with open
  indices}}},\ \href@noop {} {Ph.D. thesis},\ \bibinfo  {school} {Regensburg
  U.} (\bibinfo {year} {2020})\BibitemShut {NoStop}%
\bibitem [{\citenamefont {Evans}\ \emph {et~al.}(2010)\citenamefont {Evans},
  \citenamefont {Bali},\ and\ \citenamefont {Collins}}]{Evans:2010tg}%
  \BibitemOpen
  \bibfield  {author} {\bibinfo {author} {\bibfnamefont {R.}~\bibnamefont
  {Evans}}, \bibinfo {author} {\bibfnamefont {G.~S.}\ \bibnamefont {Bali}},\
  and\ \bibinfo {author} {\bibfnamefont {S.}~\bibnamefont {Collins}},\
  }\bibfield  {title} {\bibinfo {title} {{Improved semileptonic form factor
  calculations in lattice QCD}},\ }\href
  {https://doi.org/10.1103/PhysRevD.82.094501} {\bibfield  {journal} {\bibinfo
  {journal} {Phys. Rev. D}\ }\textbf {\bibinfo {volume} {82}},\ \bibinfo
  {pages} {094501} (\bibinfo {year} {2010})},\ \Eprint
  {https://arxiv.org/abs/1008.3293} {arXiv:1008.3293 [hep-lat]} \BibitemShut
  {NoStop}%
\bibitem [{\citenamefont {Alexandrou}\ \emph {et~al.}(2014)\citenamefont
  {Alexandrou}, \citenamefont {Constantinou}, \citenamefont {Dinter},
  \citenamefont {Drach}, \citenamefont {Jansen}, \citenamefont
  {Hadjiyiannakou},\ and\ \citenamefont {Renner}}]{Alexandrou:2013xon}%
  \BibitemOpen
  \bibfield  {author} {\bibinfo {author} {\bibfnamefont {C.}~\bibnamefont
  {Alexandrou}}, \bibinfo {author} {\bibfnamefont {M.}~\bibnamefont
  {Constantinou}}, \bibinfo {author} {\bibfnamefont {S.}~\bibnamefont
  {Dinter}}, \bibinfo {author} {\bibfnamefont {V.}~\bibnamefont {Drach}},
  \bibinfo {author} {\bibfnamefont {K.}~\bibnamefont {Jansen}}, \bibinfo
  {author} {\bibfnamefont {K.}~\bibnamefont {Hadjiyiannakou}},\ and\ \bibinfo
  {author} {\bibfnamefont {D.~B.}\ \bibnamefont {Renner}} (\bibinfo
  {collaboration} {ETM}),\ }\bibfield  {title} {\bibinfo {title} {{A stochastic
  method for computing hadronic matrix elements}},\ }\href
  {https://doi.org/10.1140/epjc/s10052-013-2692-3} {\bibfield  {journal}
  {\bibinfo  {journal} {Eur. Phys. J. C}\ }\textbf {\bibinfo {volume} {74}},\
  \bibinfo {pages} {2692} (\bibinfo {year} {2014})},\ \Eprint
  {https://arxiv.org/abs/1302.2608} {arXiv:1302.2608 [hep-lat]} \BibitemShut
  {NoStop}%
\bibitem [{\citenamefont {Bali}\ \emph
  {et~al.}(2014{\natexlab{b}})\citenamefont {Bali} \emph
  {et~al.}}]{Bali:2013dpa}%
  \BibitemOpen
  \bibfield  {author} {\bibinfo {author} {\bibfnamefont {G.~S.}\ \bibnamefont
  {Bali}} \emph {et~al.},\ }\bibfield  {title} {\bibinfo {title} {{Nucleon
  generalized form factors and sigma term from lattice QCD near the physical
  quark mass}},\ }\href {https://doi.org/10.22323/1.187.0291} {\bibfield
  {journal} {\bibinfo  {journal} {PoS}\ }\textbf {\bibinfo {volume}
  {Lattice2013}},\ \bibinfo {pages} {291} (\bibinfo {year}
  {2014}{\natexlab{b}})},\ \Eprint {https://arxiv.org/abs/1312.0828}
  {arXiv:1312.0828 [hep-lat]} \BibitemShut {NoStop}%
\bibitem [{\citenamefont {Yang}\ \emph {et~al.}(2016)\citenamefont {Yang},
  \citenamefont {Alexandru}, \citenamefont {Draper}, \citenamefont {Gong},\
  and\ \citenamefont {Liu}}]{Yang:2015zja}%
  \BibitemOpen
  \bibfield  {author} {\bibinfo {author} {\bibfnamefont {Y.-B.}\ \bibnamefont
  {Yang}}, \bibinfo {author} {\bibfnamefont {A.}~\bibnamefont {Alexandru}},
  \bibinfo {author} {\bibfnamefont {T.}~\bibnamefont {Draper}}, \bibinfo
  {author} {\bibfnamefont {M.}~\bibnamefont {Gong}},\ and\ \bibinfo {author}
  {\bibfnamefont {K.-F.}\ \bibnamefont {Liu}},\ }\bibfield  {title} {\bibinfo
  {title} {{Stochastic method with low mode substitution for nucleon isovector
  matrix elements}},\ }\href {https://doi.org/10.1103/PhysRevD.93.034503}
  {\bibfield  {journal} {\bibinfo  {journal} {Phys. Rev. D}\ }\textbf {\bibinfo
  {volume} {93}},\ \bibinfo {pages} {034503} (\bibinfo {year} {2016})},\
  \Eprint {https://arxiv.org/abs/1509.04616} {arXiv:1509.04616 [hep-lat]}
  \BibitemShut {NoStop}%
\bibitem [{\citenamefont {Dong}\ and\ \citenamefont {Liu}(1994)}]{Dong:1993pk}%
  \BibitemOpen
  \bibfield  {author} {\bibinfo {author} {\bibfnamefont {S.-J.}\ \bibnamefont
  {Dong}}\ and\ \bibinfo {author} {\bibfnamefont {K.-F.}\ \bibnamefont {Liu}},\
  }\bibfield  {title} {\bibinfo {title} {{Stochastic estimation with $Z_2$
  noise}},\ }\href {https://doi.org/10.1016/0370-2693(94)90440-5} {\bibfield
  {journal} {\bibinfo  {journal} {Phys. Lett. B}\ }\textbf {\bibinfo {volume}
  {328}},\ \bibinfo {pages} {130} (\bibinfo {year} {1994})},\ \Eprint
  {https://arxiv.org/abs/hep-lat/9308015} {arXiv:hep-lat/9308015} \BibitemShut
  {NoStop}%
\bibitem [{\citenamefont {Bali}\ \emph
  {et~al.}(2020{\natexlab{b}})\citenamefont {Bali}, \citenamefont {Collins},
  \citenamefont {Korcyl}, \citenamefont {R{\"o}dl}, \citenamefont
  {Weish{\"a}upl},\ and\ \citenamefont {Wurm}}]{Bali:2019svt}%
  \BibitemOpen
  \bibfield  {author} {\bibinfo {author} {\bibfnamefont {G.~S.}\ \bibnamefont
  {Bali}}, \bibinfo {author} {\bibfnamefont {S.}~\bibnamefont {Collins}},
  \bibinfo {author} {\bibfnamefont {P.}~\bibnamefont {Korcyl}}, \bibinfo
  {author} {\bibfnamefont {R.}~\bibnamefont {R{\"o}dl}}, \bibinfo {author}
  {\bibfnamefont {S.}~\bibnamefont {Weish{\"a}upl}},\ and\ \bibinfo {author}
  {\bibfnamefont {T.}~\bibnamefont {Wurm}},\ }\bibfield  {title} {\bibinfo
  {title} {{Hyperon couplings from $N_f=2+1$ lattice QCD}},\ }\href@noop {}
  {\bibfield  {journal} {\bibinfo  {journal} {PoS}\ }\textbf {\bibinfo {volume}
  {LATTICE2019}},\ \bibinfo {pages} {099} (\bibinfo {year}
  {2020}{\natexlab{b}})},\ \Eprint {https://arxiv.org/abs/1907.13454}
  {arXiv:1907.13454 [hep-lat]} \BibitemShut {NoStop}%
\bibitem [{\citenamefont {Bali}\ \emph
  {et~al.}(2016{\natexlab{c}})\citenamefont {Bali}, \citenamefont {Collins},
  \citenamefont {Richtmann}, \citenamefont {Sch{\"a}fer}, \citenamefont
  {S{\"o}ldner},\ and\ \citenamefont {Sternbeck}}]{Bali:2016lvx}%
  \BibitemOpen
  \bibfield  {author} {\bibinfo {author} {\bibfnamefont {G.~S.}\ \bibnamefont
  {Bali}}, \bibinfo {author} {\bibfnamefont {S.}~\bibnamefont {Collins}},
  \bibinfo {author} {\bibfnamefont {D.}~\bibnamefont {Richtmann}}, \bibinfo
  {author} {\bibfnamefont {A.}~\bibnamefont {Sch{\"a}fer}}, \bibinfo {author}
  {\bibfnamefont {W.}~\bibnamefont {S{\"o}ldner}},\ and\ \bibinfo {author}
  {\bibfnamefont {A.}~\bibnamefont {Sternbeck}} (\bibinfo {collaboration}
  {RQCD}),\ }\bibfield  {title} {\bibinfo {title} {{Direct determinations of
  the nucleon and pion $\sigma$ terms at nearly physical quark masses}},\
  }\href {https://doi.org/10.1103/PhysRevD.93.094504} {\bibfield  {journal}
  {\bibinfo  {journal} {Phys. Rev. D}\ }\textbf {\bibinfo {volume} {93}},\
  \bibinfo {pages} {094504} (\bibinfo {year} {2016}{\natexlab{c}})},\ \Eprint
  {https://arxiv.org/abs/1603.00827} {arXiv:1603.00827 [hep-lat]} \BibitemShut
  {NoStop}%
\bibitem [{\citenamefont {Bali}\ \emph {et~al.}(2010)\citenamefont {Bali},
  \citenamefont {Collins},\ and\ \citenamefont {Sch{\"a}fer}}]{Bali:2009hu}%
  \BibitemOpen
  \bibfield  {author} {\bibinfo {author} {\bibfnamefont {G.~S.}\ \bibnamefont
  {Bali}}, \bibinfo {author} {\bibfnamefont {S.}~\bibnamefont {Collins}},\ and\
  \bibinfo {author} {\bibfnamefont {A.}~\bibnamefont {Sch{\"a}fer}},\
  }\bibfield  {title} {\bibinfo {title} {{Effective noise reduction techniques
  for disconnected loops in Lattice QCD}},\ }\href
  {https://doi.org/10.1016/j.cpc.2010.05.008} {\bibfield  {journal} {\bibinfo
  {journal} {Comput. Phys. Commun.}\ }\textbf {\bibinfo {volume} {181}},\
  \bibinfo {pages} {1570} (\bibinfo {year} {2010})},\ \Eprint
  {https://arxiv.org/abs/0910.3970} {arXiv:0910.3970 [hep-lat]} \BibitemShut
  {NoStop}%
\end{thebibliography}%
\end{document}